\documentclass[twocolumn]{article} 
\usepackage[utf8]{inputenc}
\usepackage{amsmath}
\usepackage{graphicx}
\usepackage[a4paper, total={6in, 8in}]{geometry}
\usepackage{listings}
\usepackage{float}
\usepackage{placeins}
\usepackage[nottoc,numbib]{tocbibind}
\usepackage{xcolor}
\usepackage{hyperref}
\hypersetup{colorlinks,linkcolor={blue},citecolor={blue},urlcolor={blue}}
\usepackage[]{lineno}
\usepackage{bm}
\usepackage{url}
\usepackage[affil-it]{authblk}

\title{3D PIC Study of Magnetic Field Effects
on Hall Thruster Electron Drift Instability}

\author[1]{Kunpeng Zhong}
\author[1]{Demai Zeng}
\author[1,*]{Yinjian Zhao}
\author[1]{Daren Yu}

\affil[1]{School of Energy Science and Engineering, Harbin Institute of Technology, Harbin 150001, People’s Republic of China}
\affil[*]{Corresponding author: Yinjian Zhao,
zhaoyinjian@hit.edu.cn}

\date{\today}

\begin{document}

\twocolumn[
  \begin{@twocolumnfalse}
    \maketitle
    \begin{abstract}

To fully characterize electron drift instability,
a critical phenomenon governing electron transport in Hall thrusters, large-scale three-dimensional (3D) particle-in-cell (PIC) simulations are essential, as the instability inherently exhibits 3D features. While prior 3D PIC studies of this instability exist, their setups remain oversimplified to mitigate computational costs, often employing analytical approximations for ionization and magnetic fields. Notably, these models typically assume a purely radial magnetic field, significantly deviating from real thruster configurations.
This work presents the first 3D PIC study incorporating realistic magnetic fields with both radial and axial components, coupled with a Monte Carlo collision model for ionization and a self-consistent fluid solver for neutral gas density. These advancements enable a systematic investigation of magnetic field effects on electron drift instability. Results demonstrate that both the spatial configuration and strength of the magnetic field profoundly influence instability dynamics.

    \end{abstract}
  \end{@twocolumnfalse}
]

\section{Introduction}

%Human quests know no boundaries. From man taking his first upright steps, to the completion of the first circumnavigation, and onward to the footprints on the Moon, we have never ceased to explore. Now we crave celestial steeds to embark on odysseys, with Hall thrusters emerging as the preferred solution due to the high specific impulse and long operational lifespan.
The Hall thruster is among the most promising electric propulsion technologies for space applications. Although it offers numerous advantages, there is still potential to further enhance its efficiency and performance.
To improve the performance of Hall thrusters,
it is crucial to understand the physics hiding in the electron drift instability (EDI). Numerous studies to date have been conducted on
two-dimensional simulations, such as 2D azimuthal-axial simulations and 2D azimuthal-radial simulations. However, when it comes to EDI, since the instabilities are intrinsically three-dimensional, these 2D simulations contain fundamental deficiencies that render them incapable of providing accurate physical descriptions. Consequently, implementing three-dimensional simulations becomes imperative.

While multiple factors govern the development of EDI, our previous research has focused on investigating the effect of plasma initialization \cite{Xie_2024}, this study specifically addresses the magnetic field effects on EDI. Lafleur et al. (2016) developed a 1D simulation model to investigate the effects of a uniform radial magnetic field configuration on azimuthal instability development.
They found that an increased magnetic field or a decreased electron density, will reduce the effect of the instability on the macroscopic plasma transport, and if the magnetic field is increased too far, the instability becomes too weak and the associated electron–ion friction can no longer enhance the electron cross-field transport \cite{Lafleur1}.
Reza et al. (2023) used an axial-azimuthal quasi-2D simulation framework to analyze the impact of radial magnetic field configurations on the evolution of azimuthal instabilities, and they noticed that the evolution and the characteristics of the azimuthal instabilities were majorly affected by the radial B-field gradient \cite{Reza_2023}.
Reza et al. (2023) employed a quasi-2D simulation model incorporating partially real radial magnetic field configurations, and demonstrated that the gradients in the magnetic field configuration affect the spectrum of the azimuthal instabilities, which consequently changes the dominant mechanism behind the electron axial mobility \cite{R2}.
Currently, no 3D PIC simulations with real
magnetic fields exist for Hall thruster azimuthal instabilities, which defines the main goal of this work.

This paper utilizes a self-developed
3D PIC code named
PMSL-PIC-HET-3D to conduct three-dimensional simulations of both the channel region and plume region in Hall thrusters. By optimizing the initial particle distribution and neutral flow velocity configurations, the convergence speed and numerical accuracy of the simulations are significantly improved. In engineering design, we predominantly utilize the tool such as FEMM to generate the real magnetic field, so in our work, two real magnetic field configurations and one analytical are implemented. Through monitoring time-dependent variations in electric field intensity and averaged electron current, a qualitative analysis is conducted to assess the influence of magnetic field distributions on EDI.

The paper is organized as follows: In Sec.\ref{sec:2}, the 3D basic simulation model including the domain setup, the simulation parameters, the plasma initialization, the cathode electron injection, neutral gas fluid solver, magnetic field configuration, and the simulation cases are introduced. In Sec.\ref{sec:3}, three simulation cases are tested and compared. At last, conclusions and discussions are drawn in Sec.\ref{sec:4}.

\section{Simulation Method}
\label{sec:2}

All the simulations in
this paper are carried out
using the self-developed code PMSL-PIC-HET-3D,
whose introduction and validation
are published somewhere else.

\subsection{Simulation Domain and Parameters}

\begin{figure}[!ht]
\centering
\includegraphics[width=0.45\textwidth]{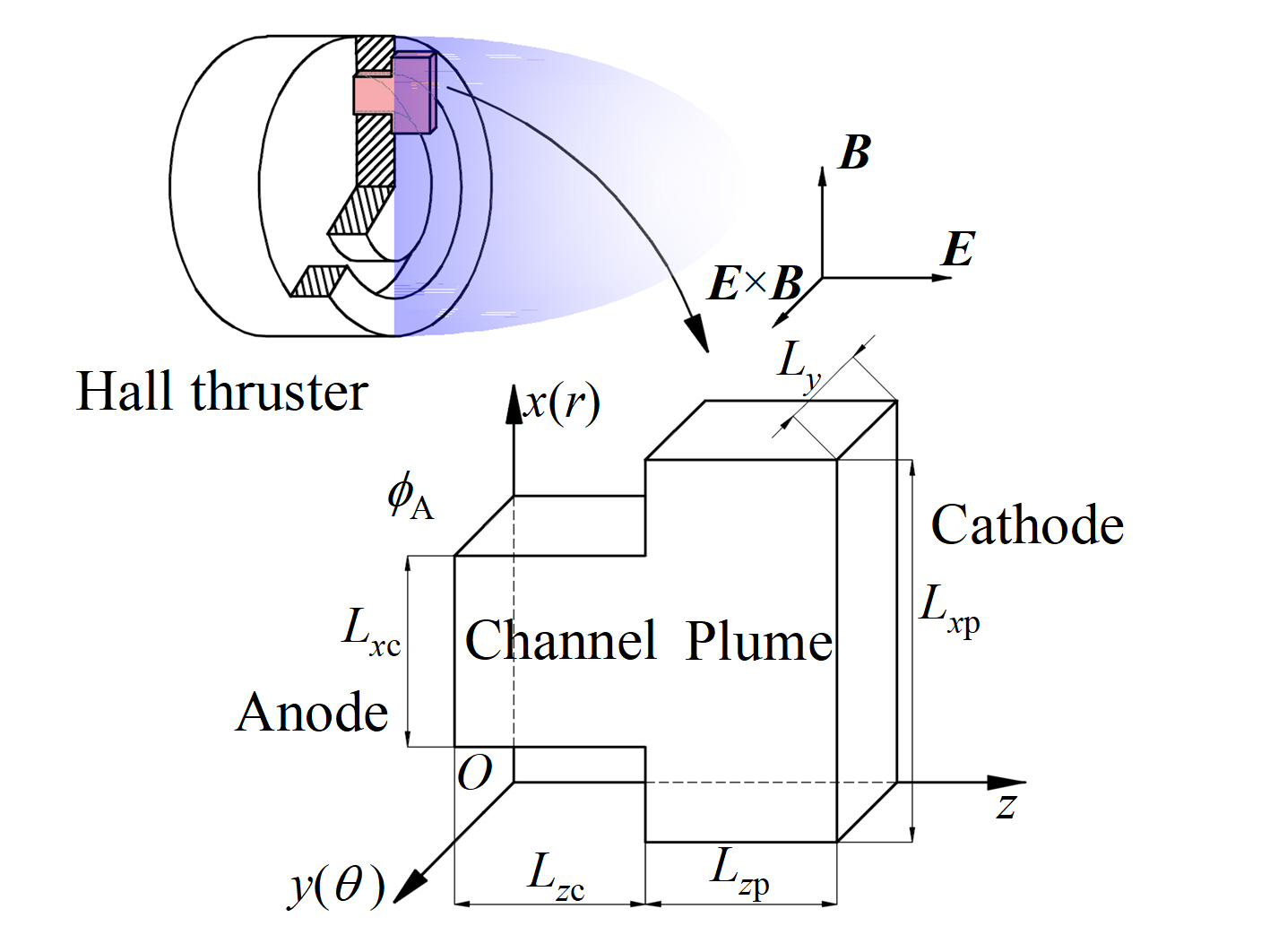}
\caption{
Illustration of the
simulation domain.
}
\label{fig:domain}
\end{figure}

The simulation domain contains a channel
and a plume as shown in Fig.\ref{fig:domain}.
$x$, $y$, $z$ represent the radial ($r$), azimuthal
($\theta$), and axial directions in the Hall thruster,
respectively. The surface at $z=0$ is set to be
the anode with fixed potential boundary conditions $\phi(z=0)=\phi_A$;
other surfaces in $x$ and $z$
are considered to be either the wall boundaries
or the vacuum boundaries with fixed zero potential;
the two surfaces in $y$ are set to be periodic.
Particles are initially sampled
according to the given profiles
that are introduced later in
Sec.\ref{sec:initialization}.
For those particles who reach the anode, cathode, wall,
or vacuum boundaries are removed;
electron-ion pairs due to ionization are generated
through the classic Monte Carlo Collision (MCC)
method\cite{VAHEDI1995179}.
Only the ionization collision type is
considered.
Free electrons are injected through the
cathode boundary
to serve as ionization sources
and maintain the overall neutrality,
which is introduced in Sec.\ref{sec:cathode}.

The simulation parameters are chosen based on
a recent work
with pure 3D cubic simulation domain
\cite{Xie_2024}.
Uniform grid cells are applied and the
curvature in the azimuthal direction is
ignored as applied in many previous numerical research
\cite{Charoy2019,Villafana2021,Villafana2023}.
The anode potential is set to be $\phi_A=200$ V.
The ion mass is chosen to be
that of Xenon.
The cell sizes are chosen as
$\Delta x = \Delta y = \Delta z =
0.1$ mm,
smaller than the typical Debye length
of the simulated plasma,
and the timestep is set to be
$\Delta t = 1.5 \times 10^{-11}$ s,
both of which are
the same as the previously applied value
in
\cite{Xie_2024}.
The number of cells is chosen to be
$N_{xc} = 128$ and 
$N_{zc} = 72$ for the channel,
the corresponding physical domain sizes are
$L_{xc} = 1.28$ cm and
$L_{zc} = 0.72$ cm,
as illustrated in Fig.\ref{fig:domain}.
The number of cells is chosen to be
$N_{xp} = 256$ and 
$N_{zp} = 184$ for the plume,
the corresponding physical domain sizes are
$L_{xp} = 2.56$ cm and
$L_{zp} = 1.84$ cm.
The channel region is located in the middle of the
plume region along $x$.
Both the channel and the plume
have the same azimuthal size for all cases
with $N_y=64$ and $L_{yc}=L_{yp}=L_y=0.64$ cm.
Note that in the previous study\cite{Xie_2024}
we have shown
that using a relatively small size ($N_y=64$)
in the azimuthal direction,
does not affect the dominant
modes and the evolution of the instability,
but can greatly save the computational cost.
The macro-particle weight is set to be
$w_0 \approx 19555$.
Electrons and ions share the same $w_0$.
The maximum total number of macro-particles
of the simulations is about
$1.5 \times 10^{8}$.

\subsection{The Plasma Initialization}

\begin{figure}[!ht]
\centering
(a)
\includegraphics[width=0.45\textwidth]{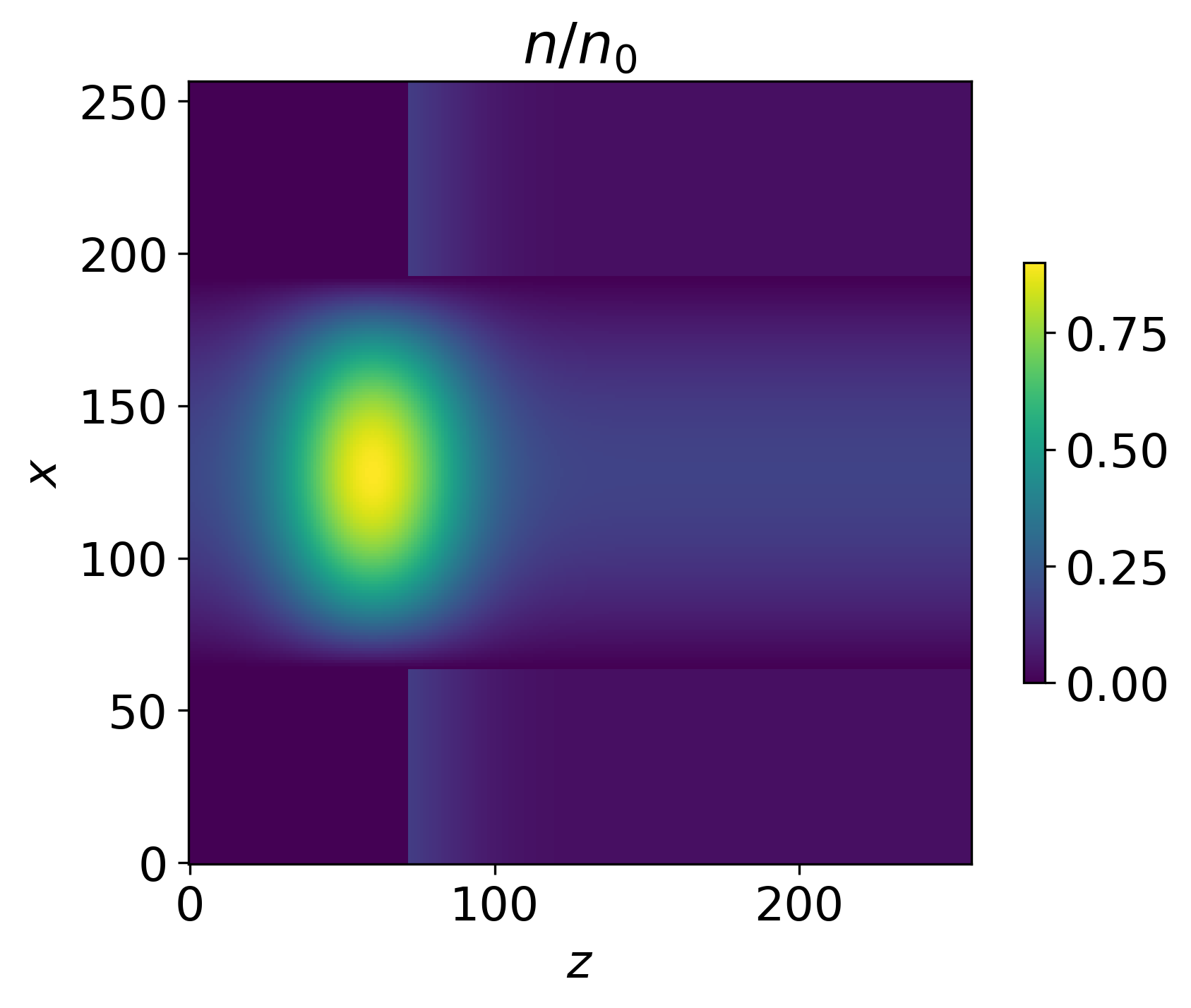}\\
(b)
\includegraphics[width=0.45\textwidth]{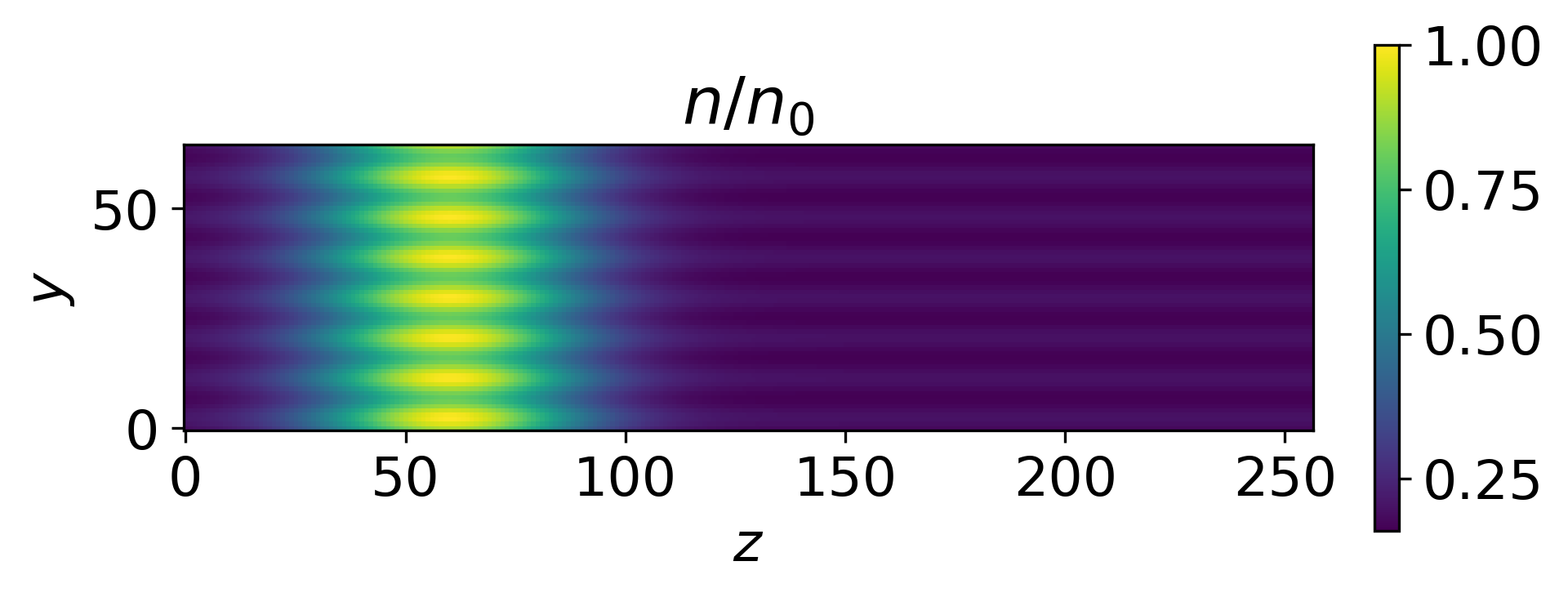}\\
(c)
\includegraphics[width=0.45\textwidth]{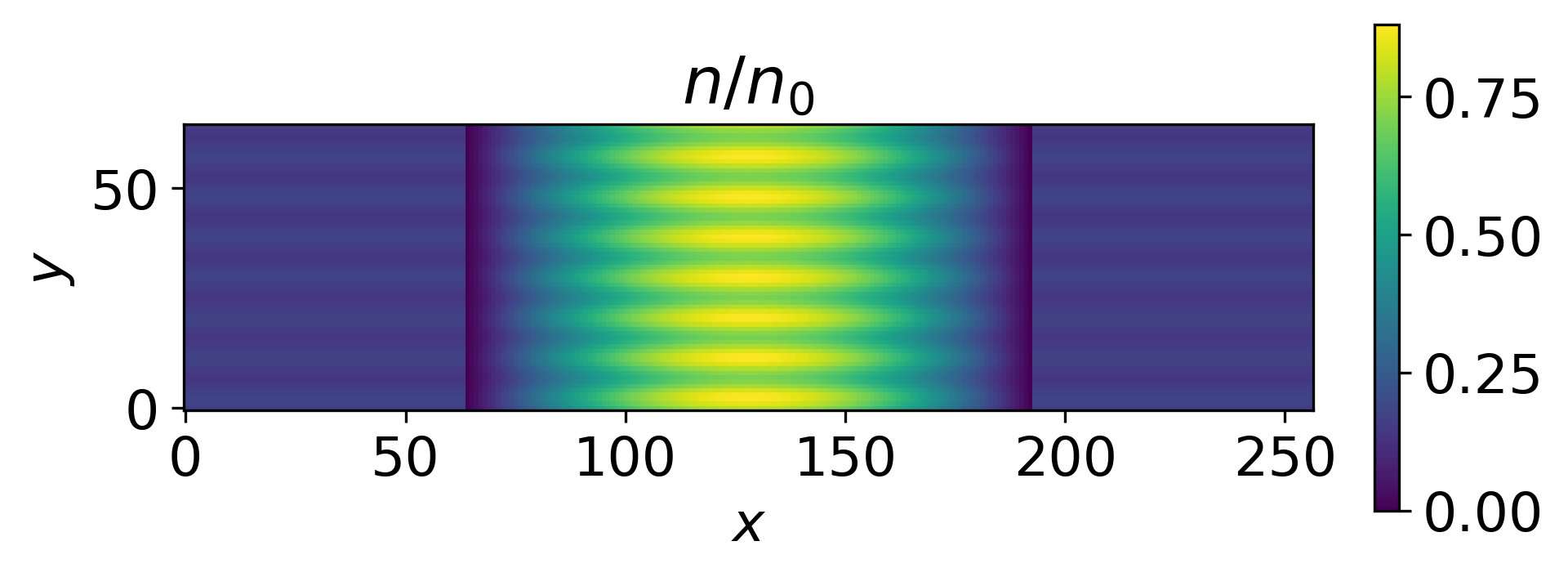}\\
\caption{
Plasma density initialization profiles
in the $y=32$ plane (a),
$x=128$ plane (b),
and $z=72$ plane (c).
}
\label{fig:nn}
\end{figure}

Following the suggestion in
\cite{Xie_2024},
a non-uniform plasma initialization is applied.
The normalized initial plasma density functions
along three directions are defined as,
\begin{equation}
    \hat{n}_x (x) =
    \sin{\left( 2 \pi \dfrac{x-x_m}{l_x}
    \right)},
\end{equation}
\begin{equation}
    \hat{n}_y (y) =
    A_y \sin{\left( 2 \pi \dfrac{y}{l_y}
    \right)} + 1 - A_y,
\end{equation}
\begin{equation}
    \hat{n}_z (z) =
    (1-n_{min}) \exp{
    \left[ -\dfrac{(z-z_p)^2}{l_z^2} \right]} +
    n_{min},
\end{equation}
where $x$, $y$, $z$ values represent
grid indices, $l_x=256$,
$x_m=64$,
$A_y=0.1$, $l_y=64/7$,
$n_{min}=0.2$,
$z_p=60$,
$l_z=30$.
The 3D initial density is thus
$n_e = n_i = n_0 \hat{n}_x \hat{n}_y \hat{n}_z$,
and for those regions with negative
normalized density values, the
minimum $n_{min} n_0$ is used instead.
The initial density profiles are
shown in Fig.\ref{fig:nn}.
Note that the
two lines with nearly zero
values
as shown in Fig.\ref{fig:nn} (a) and (c)
do not matter much,
because the plasma will quickly evolve to fill
these gaps.
Given the 3D profile,
particle positions can be sampled
using a simple accept-reject method.

In addition,
it is found that the initial
ion axial velocity
affects the convergent speed
of the simulation,
we thus provide a more reasonable
ion axial velocity profile
as initialization.
\begin{equation}
    v_{iz} (z) =
    a + \dfrac{b-a}{1+(z/c)^d},
\end{equation}
where
$a = 8287.5$ m/s,
$b = -760.4$ m/s,
$c = 0.0091$ m,
$d = 4.4528$.
The fitting function
and the coefficients
are obtained based on
\cite{Xie_2024}.
Besides the ion axial distribution,
ions are sampled according to
Maxwellian velocity distribution
with temperature 0.5 eV,
and electrons are sampled
with temperature 10 eV.

\subsection{The Cathode Electron Injection}
\label{sec:cathode}

When an analytical ionization source is used,
new pairs of electrons and ions are generated
continuously regardless of existing electrons,
thus a quasi-neutral cathode boundary condition can be applied
to maintain the neutrality of the plasma in the simulation domain.
However, when MCC is used instead,
the ionization is triggered by energetic electrons,
such that
a cathode is needed to continuously supply electrons
like real Hall thrusters.
Therefore, we set the plane $x=246$ to be the cathode boundary,
shifted by 10 grids from the maximum $x=256$ boundary,
which continuously and uniformly injects electrons into the simulation domain.
The injection current is chosen to be
$I_0 \approx 0.3$ A,
which is found to be able to
sustain the discharge.
The temperature of the injected electrons
is 10 eV,
but an additional drifting velocity
along $-x$ is added
with 3 times of the corresponding thermal velocity.
As we can see,
there are a number of free parameters of the cathode
boundary that may affect the simulation,
which is however beyond the scope of this paper,
and will be left as a future work to be
studied in more detail.

\subsection{Neutral Gas Fluid Solver}
\label{sec:neutral}

As our first step
of coupling the ionization using MCC
with self-consistent calculation of the neutral gas density,
only the continuity equation
along the axial direction is considered and solved numerically
with a constant axial fluid velocity $u_z$,
\begin{equation}
    % \dfrac{\partial n_a}{\partial t}
    % + \bm{\nabla} \cdot (n_a \bm{u_a}) = - S_i,
    \dfrac{\partial n_a}{\partial t}
    +  u_z \dfrac{\partial n_a}{\partial z} = - S_i,
\end{equation}
where $n_a$ is the neutral gas density,
$S_i$ is the ionization rate,
the subscript ``a'' is used to denote the neutral atom.

\begin{figure}[!ht]
\centering
\includegraphics[width=0.45\textwidth]{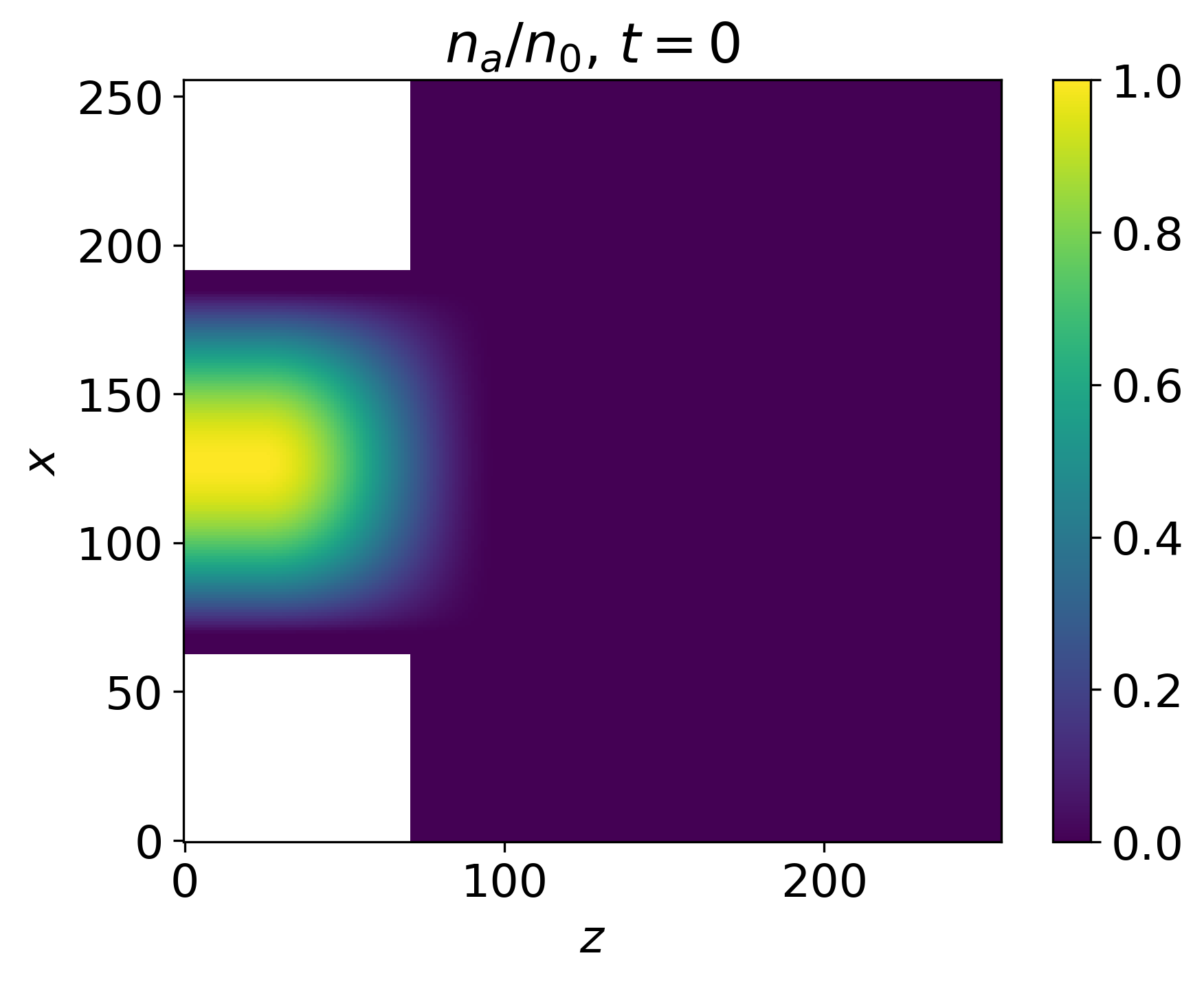}
\caption{
Initial neutral gas density profile.
The $z$ and $x$ values represent grid indices.
}
\label{fig:den_n_zx}
\end{figure}

To set a reasonable initial neutral gas density profile,
we consider the steady state,
where $\partial n_a / \partial t = 0$,
and the ionization source function
that is commonly used in the literature
\cite{Charoy2019,Villafana2021,Xie_2024}.
\begin{equation}
    S_i(z) = S_0 \cos{\left( 
    \pi \dfrac{z - z_m}{z_2 - z_1}
    \right)},
\end{equation}
if $z\in [z_1,z_2]$,
otherwise zero.
Analytically solve the steady state
continuity equation leads to
\begin{equation}
    % n_a(z) = n_0 - \dfrac{S_0}{u_z}\dfrac{z_2-z_1}{\pi} 
    % \left[
    % 1 +
    % \sin{\left(
    % \pi \dfrac{z-z_m}{z_2-z_1}
    % \right)}
    % \right],
    n_a(z) = \dfrac{n_0}{2}
    \left[
    1 -
    \sin{\left(
    \pi \dfrac{z-z_m}{z_2-z_1}
    \right)}
    \right],
\end{equation}
for $z\in [z_1,z_2]$,
where $n_{a0}$ is the peak density
value $n_a(z_1)=n_0$.
For $z < z_1$,
we set $n_a(z) = n_0$,
and for $z>z_2$,
we set $n_a(z) = 0$,
thus a relation between $n_0$ and $u_z$
can be obtained,
\begin{equation}
    n_{a0} = \dfrac{2 S_0}{u_z} \dfrac{z_2-z_1}{\pi}.
\end{equation}
Setting $S_0 = 9.55 \times 10^{23}$ m$^{-3}$s$^{-1}$,
the commonly used value, and
$z_1 = 0.25$ cm,
$z_2 = 1$ cm,
$n_{a0} = 1.5 \times 10^{19}$ m$^{-3}$,
we obtain
$u_z = 304$ m/s.
However, simulations show that
an increasing fluid speed can reduce
the ionization oscillation 
(or the so called breathing mode)
frequency,
which is advantageous for
faster simulation convergence.
We thus apply $u_z = 304 \times 3 = 912$ m/s
with $n_{a0} = 1.5 \times 10^{19}$ m$^{-3}$.

In addition, we notice that
due to the sheath effect,
the plasma density is low
near the discharge channel surface,
thus there would be a gas leak,
which would eventually affect the
ionization in the plume region.
Therefore,
we add an additional cosine
function along the radial $x$ direction,
\begin{align}
    n_a(z,x) = &\dfrac{n_0}{2}
    \Bigg[
    1 -
    \sin{\left(
    \pi \dfrac{z-z_m}{z_2-z_1}
    \right)}
    \Bigg]
    \nonumber \\
    &\times \cos{\left(
    \pi \dfrac{x-x_m}{x_2 - x_1}
    \right)},
\end{align}
for $x\in[x_1,x_2]$,
otherwise zero.
$x_m=1.28$ cm is the center of the channel,
$x_2-x_1 = 1.152$ cm
is equal to 0.9 times of the channel width 1.28 cm.
Of course, along the azimuthal $y$ direction,
$n_a$ is set to be uniform initially.
The initial neutral gas density
profile 
in the $z$-$x$ plane is shown in Fig.\ref{fig:den_n_zx}.

\begin{figure}[!ht]
\centering
(a)
\includegraphics[width=0.45\textwidth]{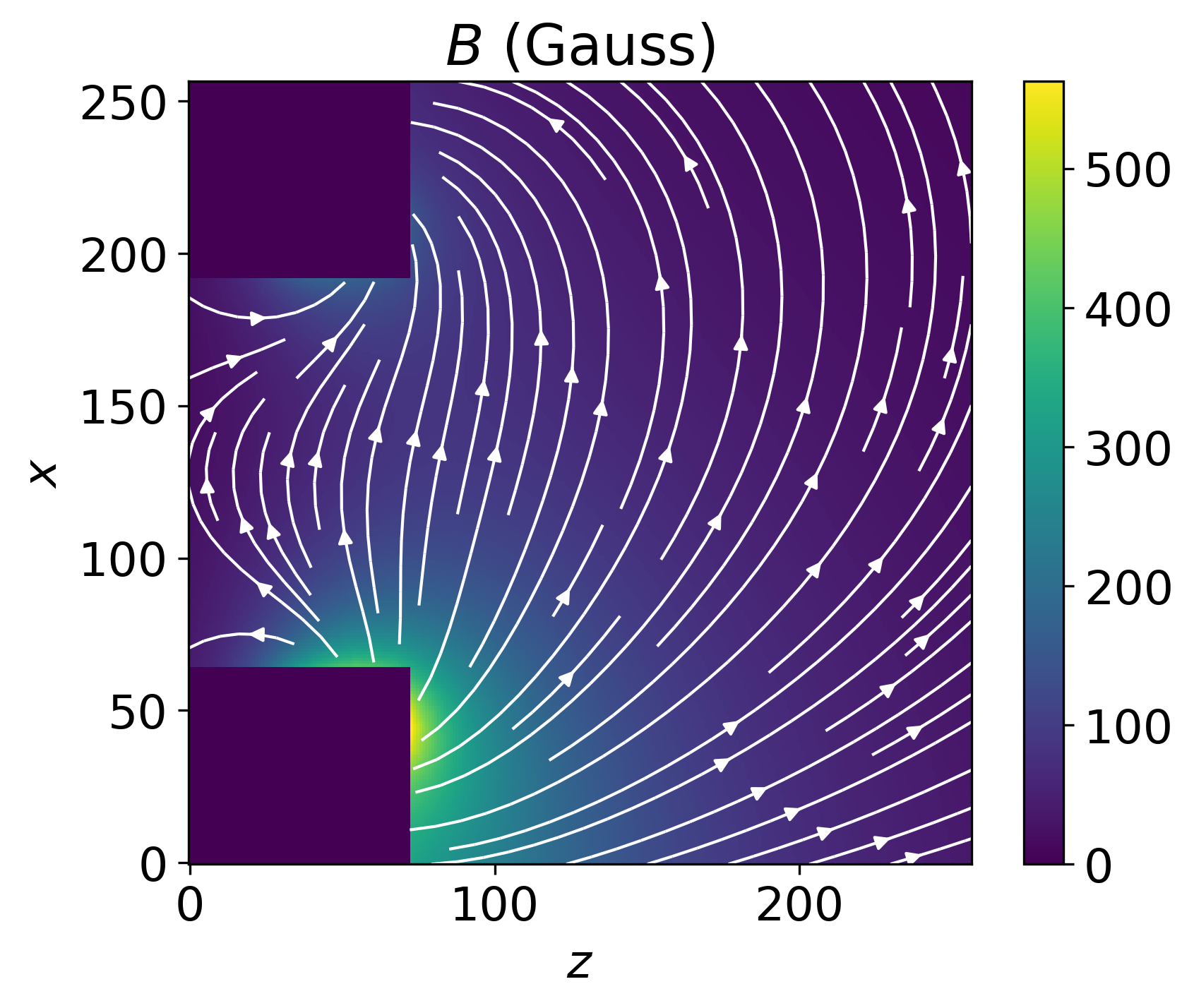}\\
(b)
\includegraphics[width=0.45\textwidth]{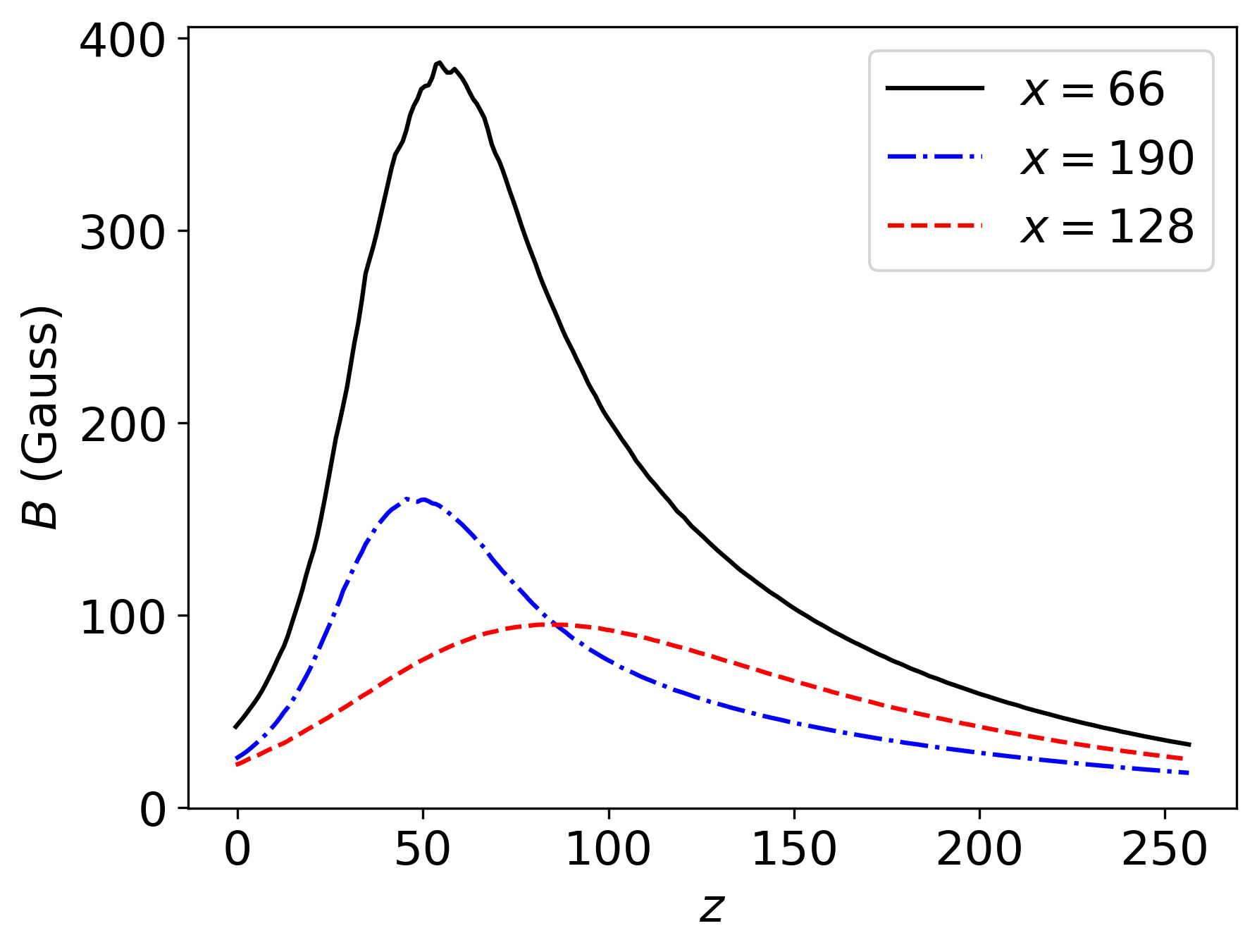}
\caption{
Magnetic field configuration
(a)
and 1D magnetic field strength at
three radial positions along $z$ (b).
The $z$ and $x$ values represent grid indices.
}
\label{fig:B}
\end{figure}

\subsection{Magnetic Field Configuration}
\label{sec:initialization}

The FEMM software
(www.femm.info) is applied to
generate a Hall thruster magnetic field
as shown in Fig.\ref{fig:B} (a),
and the maximum value at the center of the channel
along $z$ is about 100 Gauss,
while that near the inner and outer surfaces are
400 and 180 Gauss, respectively,
as shown in Fig.\ref{fig:B} (b).

Another stronger magnetic field configuration
is also considered,
by increasing the field coil current
to a value such that the
maximum B field of the corresponding $x=128$
curve can reach about 200 Gauss.
Because the field plots are similar
to Fig.\ref{fig:B}
except the values,
we do not present them anymore redundantly.

In addition,
the commonly applied
analytical
magnetic field configuration
is still considered for comparison
\cite{Xie_2024,Charoy2019,Villafana2023},
which has a Gaussian shape
and only the radial component throughout the channel and the plume,
\begin{equation}\label{eq:B}
    B_x(z) = a_k \exp{\left[
    -\dfrac{(z-z_{B_{max}})^2}
    {2 \sigma^2}
    \right]} + b_k,
\end{equation}
where $k=1$ for $z \le z_{B_{max}}$
and $k=2$ for $z>z_{B_{max}}$.
With the given parameters,
$B_A = B(z=0)=60$ Gauss,
$B_C = B(z=2.56\text{cm}) = 10$ Gauss,
$B_{max} = B(z=z_{B_{max}})=100$ Gauss,
$z_{B_{max}} = 0.75$ cm,
and $\sigma = 0.625$ cm,
$a_k$ and $b_k$ can be calculated
according to
\begin{align}\label{eq:ak}
    a_1 &= \dfrac{B_{max} - B_A}
    {1-\exp{[-z_{B_{max}}^2/(2\sigma^2)]}}
    \approx 77.9 \nonumber\\
    a_2 &= \dfrac{B_{max} - B_C}
    {1-\exp{[-(2.56\text{cm}-z_{B_{max}})^2/(2\sigma^2)]}}
    \approx 91.4
\end{align}
and $b_1 = B_{max} - a_1$,
$b_2 = B_{max} - a_2$.

\subsection{Summary of Simulation Cases}

In summary,
three simulation cases are considered in this paper,
as listed in Tab.\ref{tab:cases}.
Weak-B case applies the field
described in Fig.\ref{fig:B};
Strong-B has stronger B field
by increasing the field coil current;
Analytic-B is the case using
Eq.(\ref{eq:B}).
In the next section,
effects due to the
magnetic field configuration and strength
are investigated.
The typical runtime of these three
cases is about 18 days,
when 128 MPI processes are used
for parallelization
on a dual AMD EPYC 9754 CPU
computing node.

\begin{table}[!ht]
\centering
\begin{tabular}{rr}
\hline
Label &
Maximum B on the centerline \\
\hline
Strong-B & FEMM $B_{max}\approx 200$ Gauss \\
Weak-B & FEMM $B_{max}\approx 100$ Gauss \\
Analytic-B & Analytic $B_{max}\approx 100$ Gauss \\
\hline
\end{tabular}
\caption{
List of simulation cases.
}
\label{tab:cases}
\end{table}

\section{Simulation Results}
\label{sec:3}
\begin{figure}[!ht]
\centering
\includegraphics[width=0.42\textwidth]{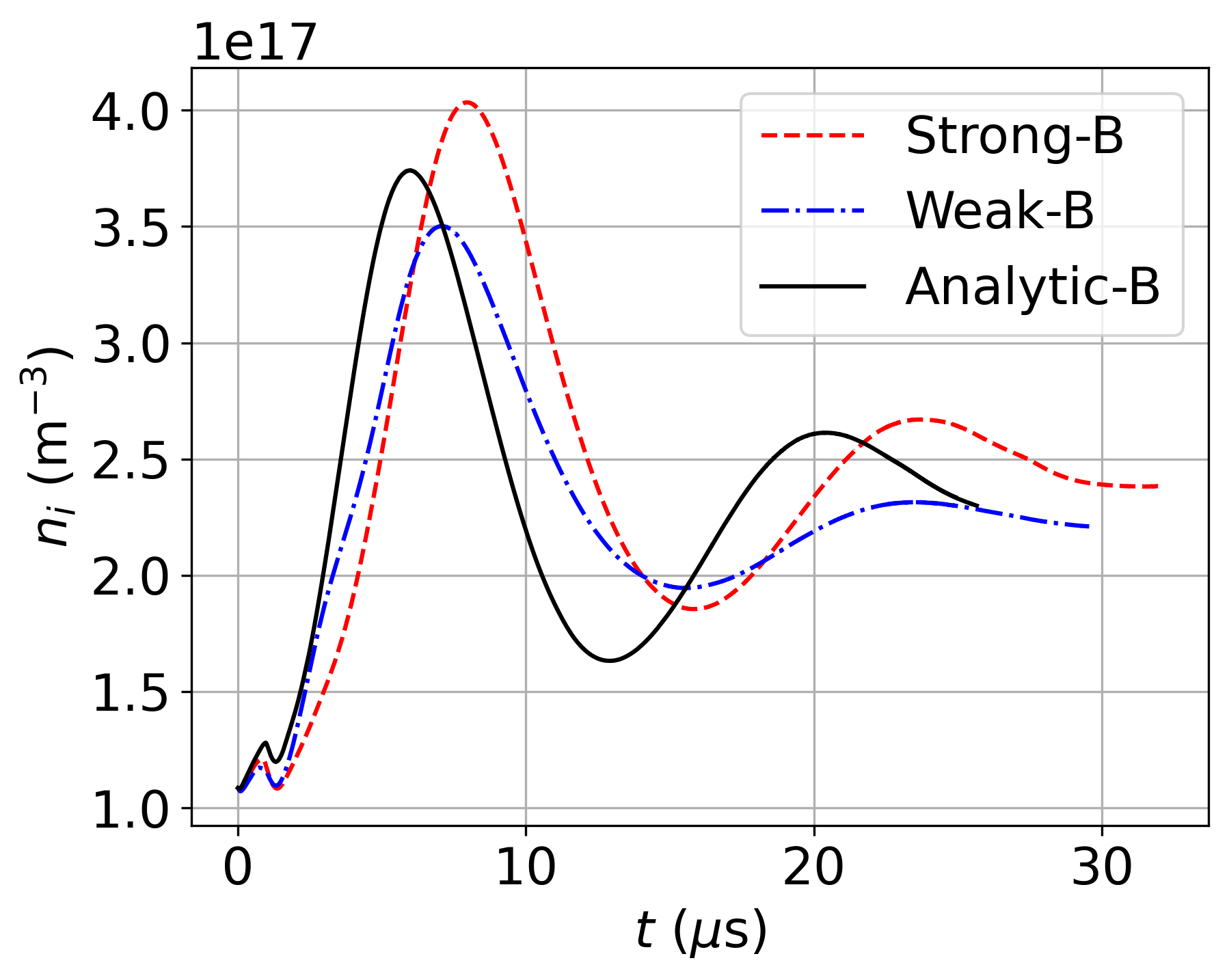}\\
\includegraphics[width=0.42\textwidth]{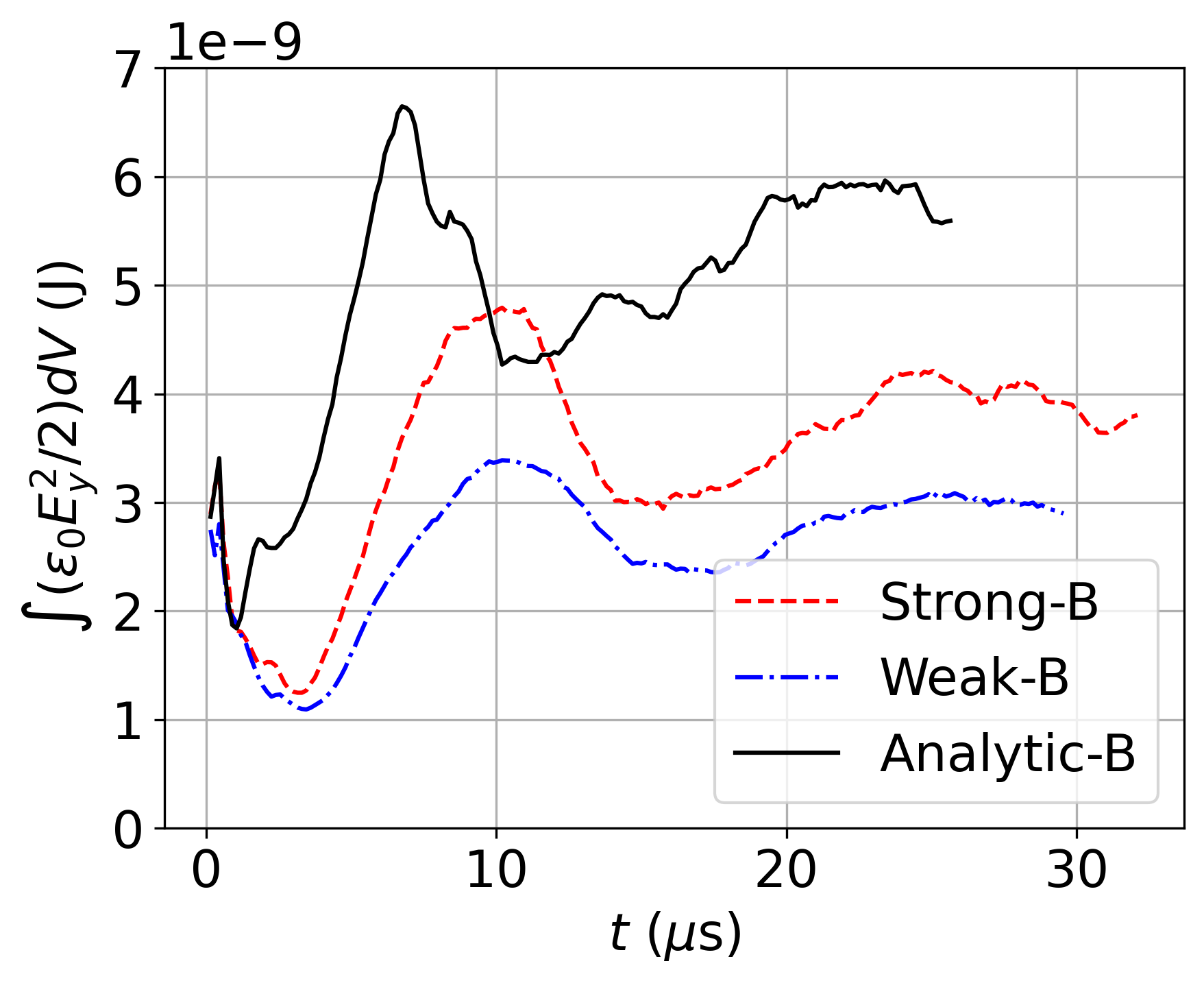}
\caption{
Top: spatially averaged ion number density over time.
Bottom: total azimuthal electric field energy
of the whole simulation domain over time.
}
\label{fig:pn}
\end{figure}

\begin{figure*}[!ht]
\centering
\includegraphics[width=0.32\textwidth]
{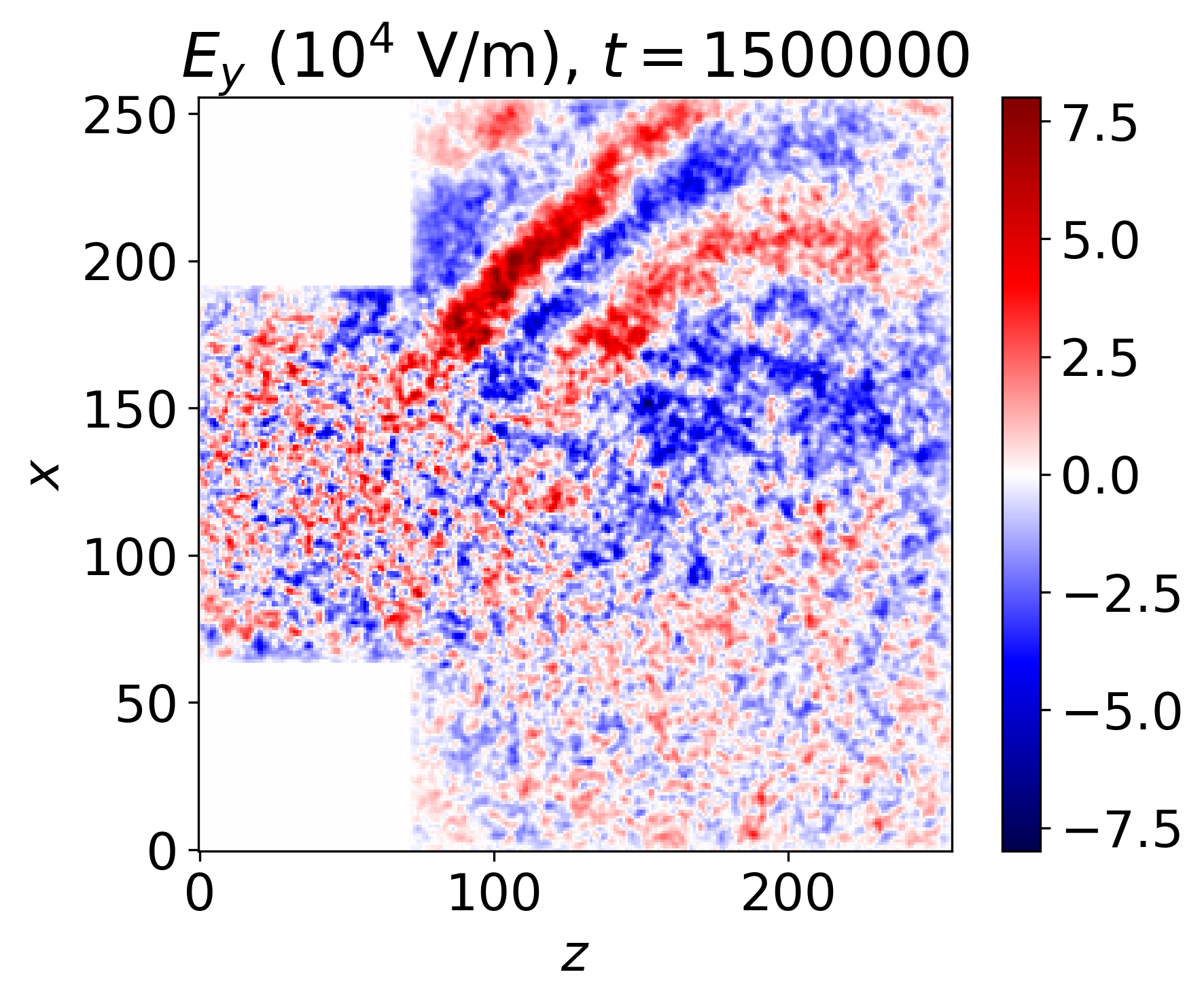}
\includegraphics[width=0.32\textwidth]
{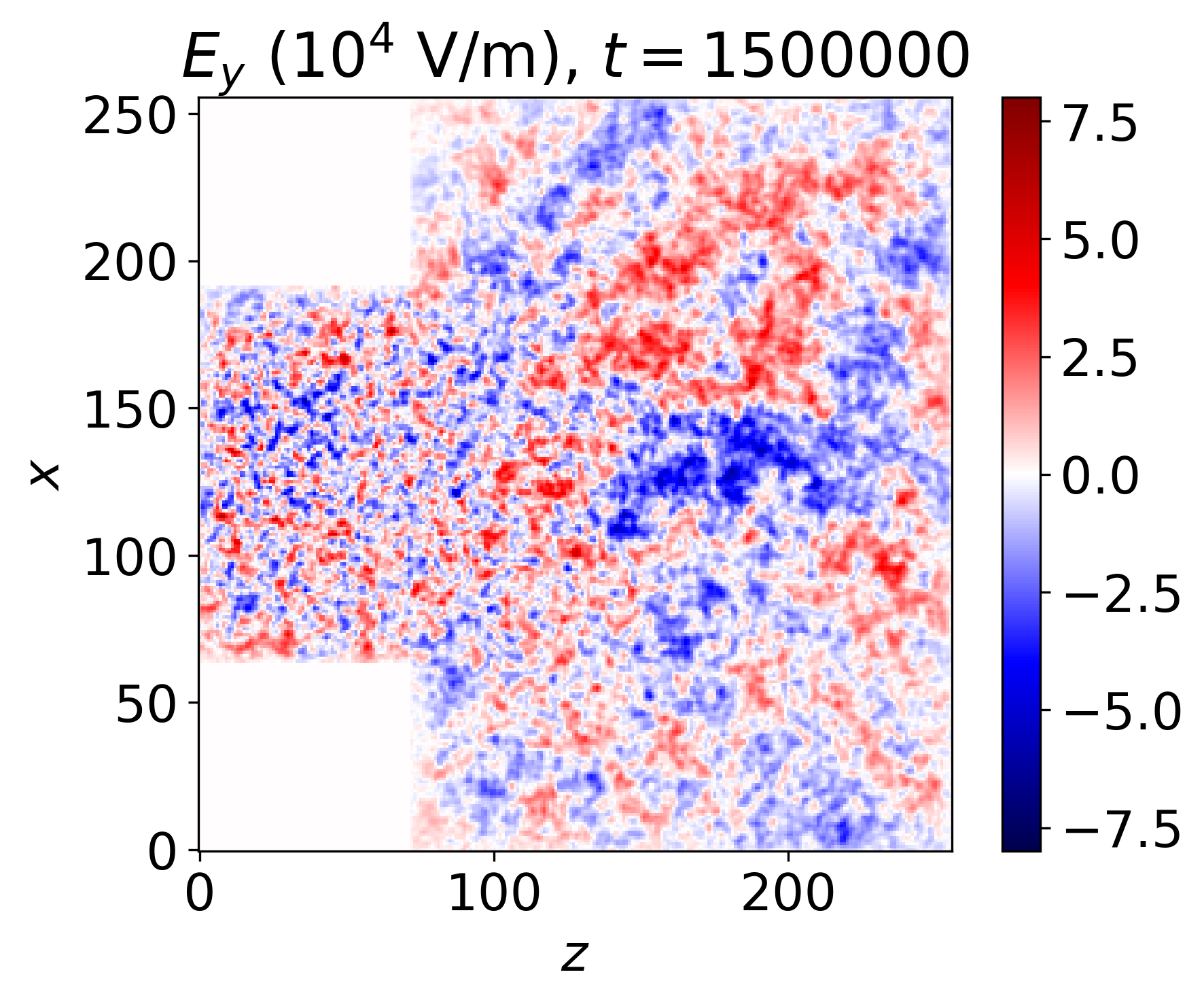}
\includegraphics[width=0.32\textwidth]
{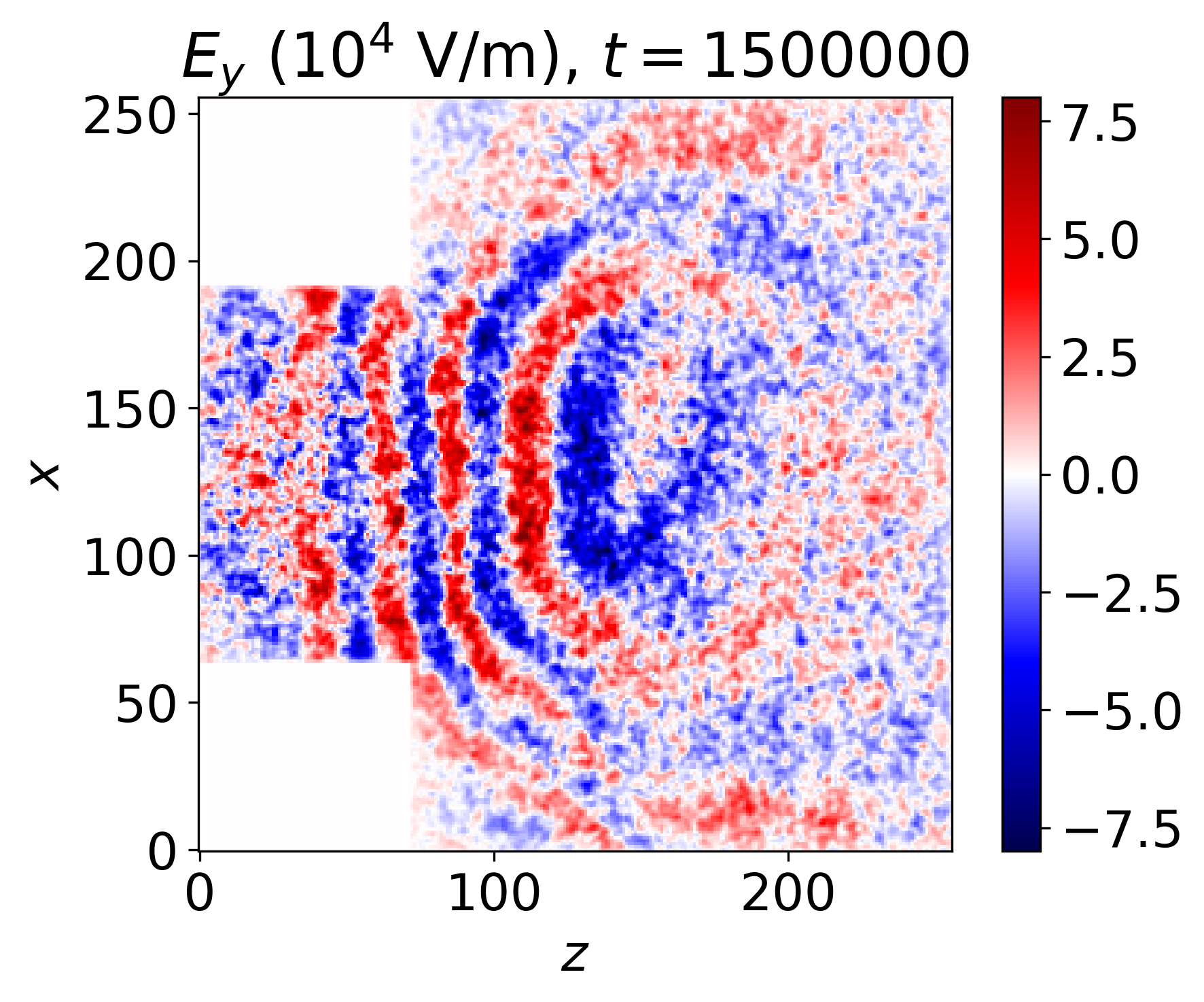}\\
\includegraphics[width=0.32\textwidth]
{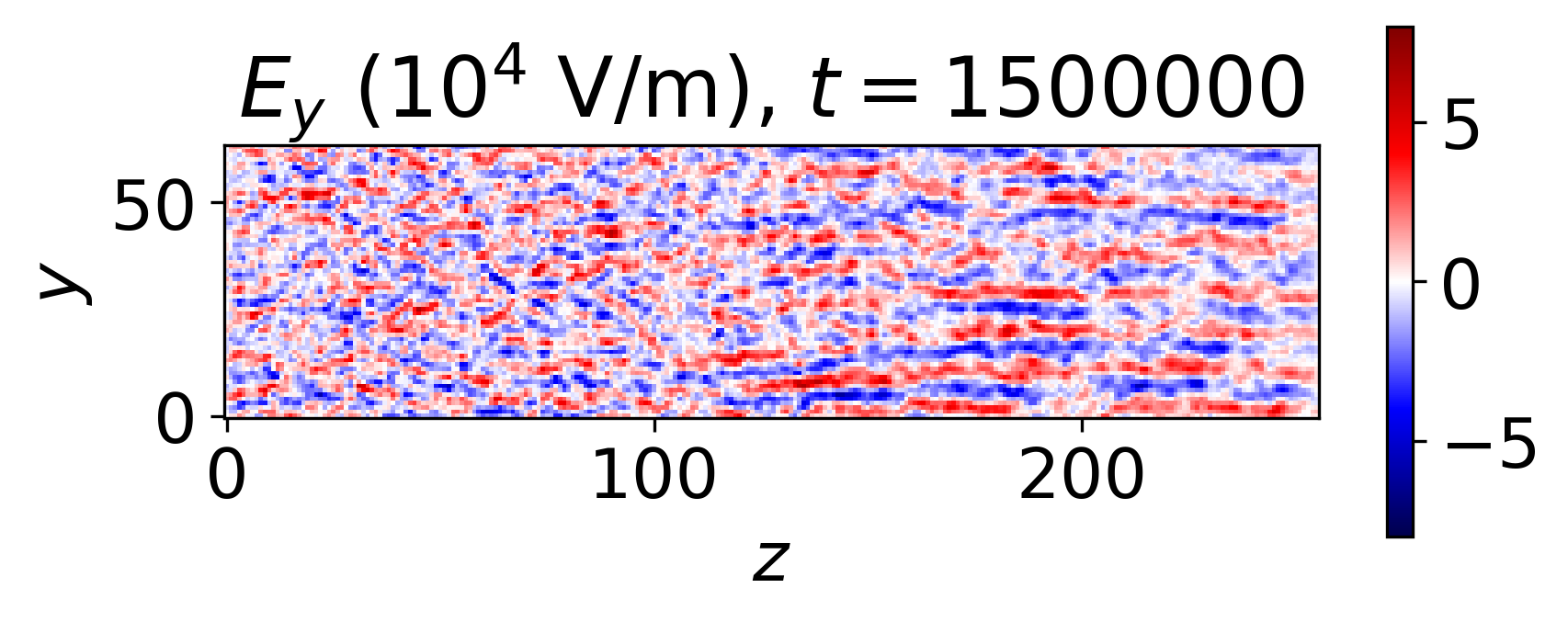}
\includegraphics[width=0.32\textwidth]
{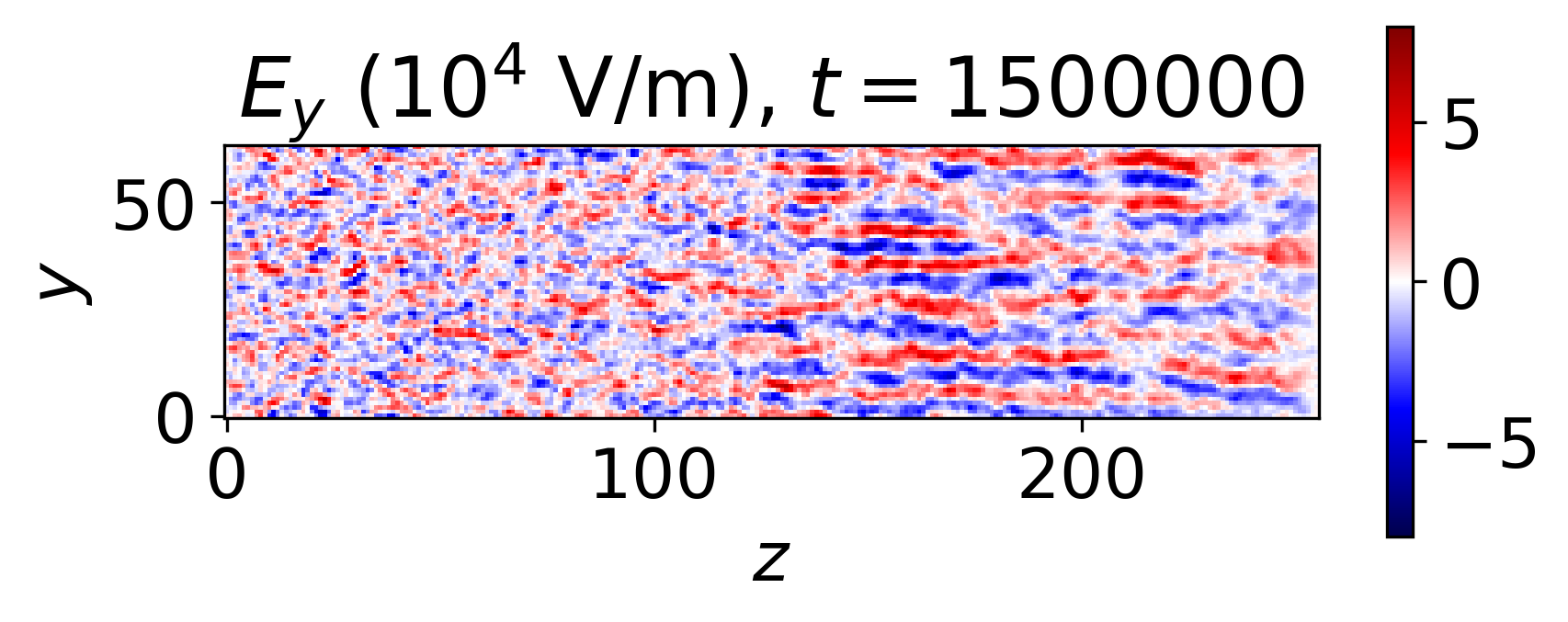}
\includegraphics[width=0.32\textwidth]
{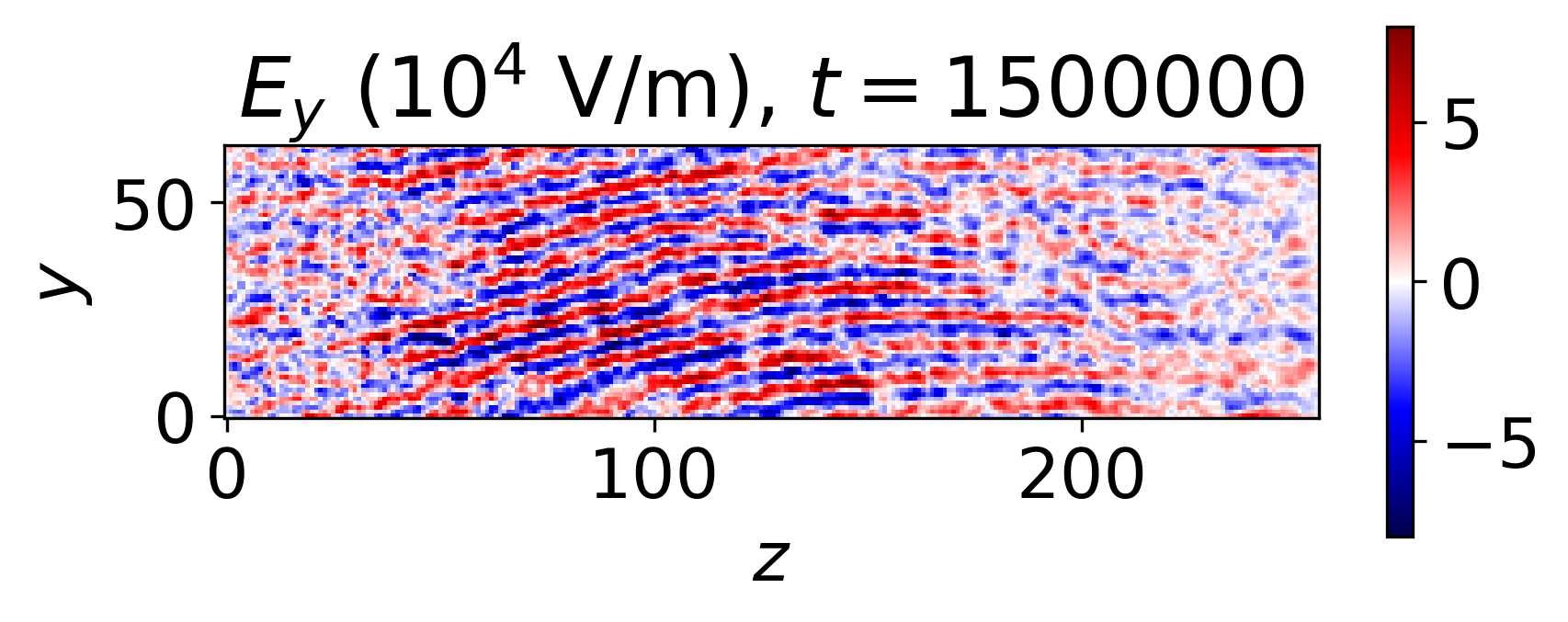}
\caption{
Azimuthal electric field $E_y$
at the middle plane of $y$ (top row) and $x$
(bottom row)
of cases Strong-B (left), Weak-B (middle), and
Analytic-B (right)
at time 22.5 $\mu$s.
}
\label{fig:Ey}
\end{figure*}

\begin{figure*}[!ht]
\centering
\includegraphics[width=0.32\textwidth]
{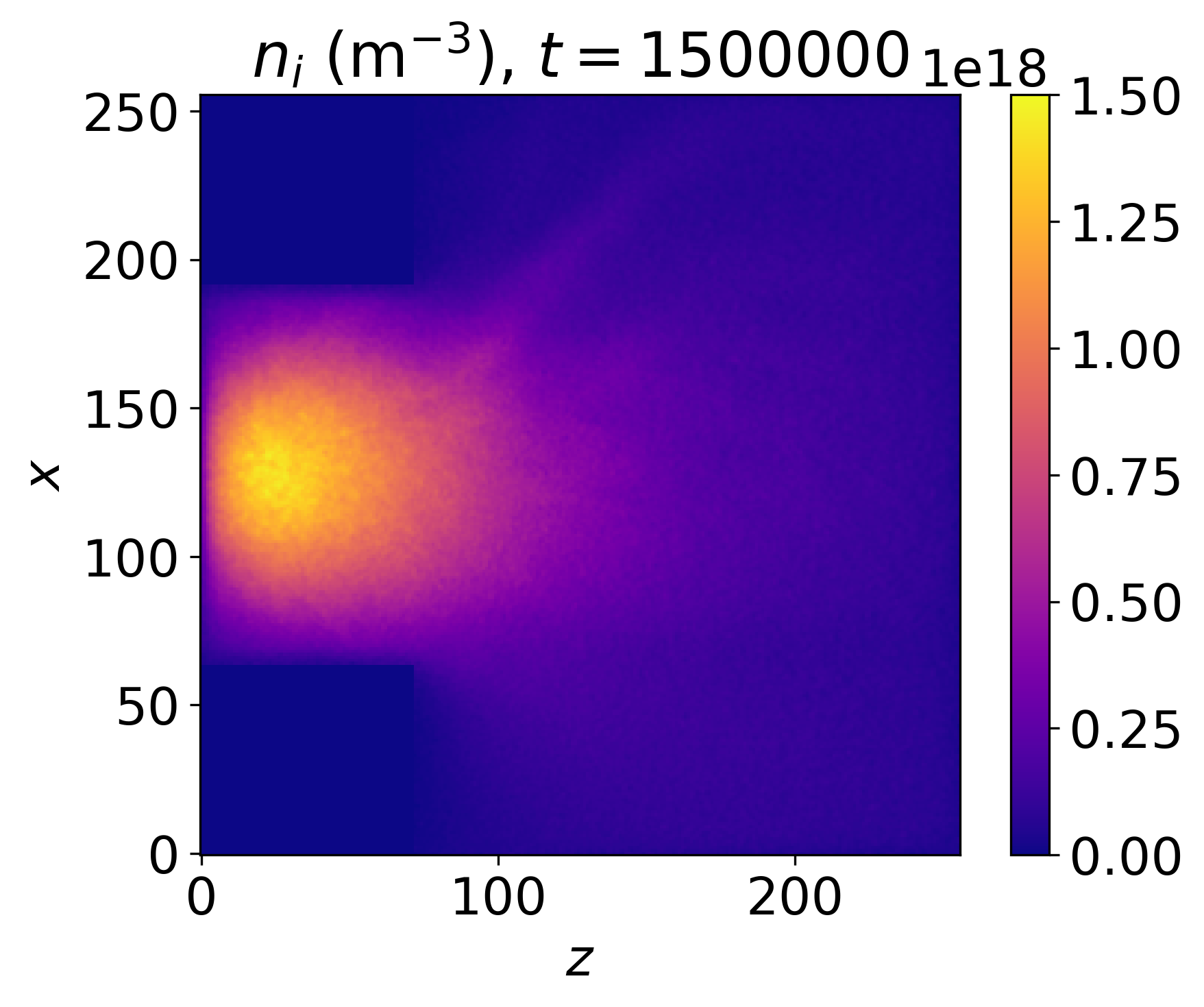}
\includegraphics[width=0.32\textwidth]
{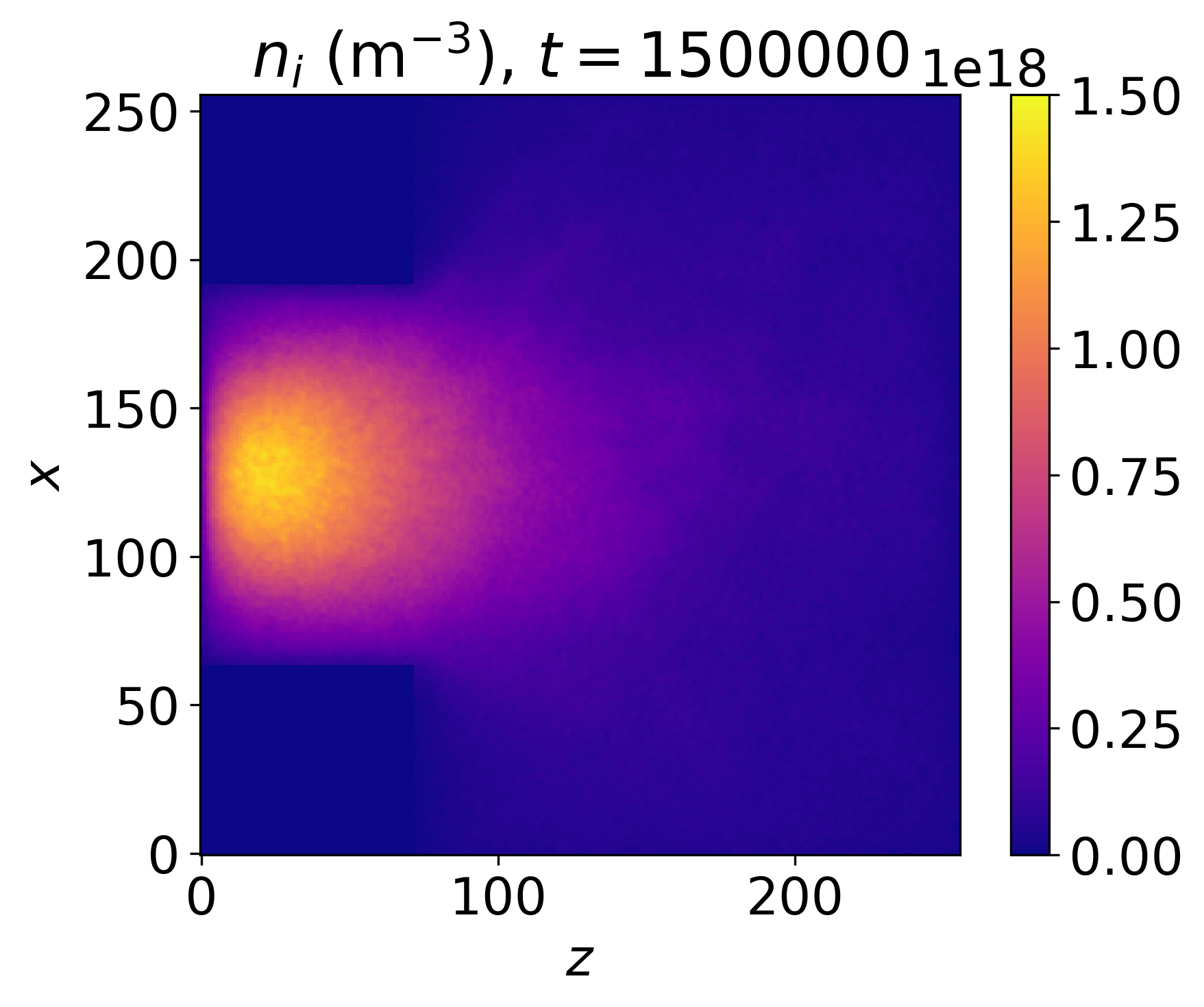}
\includegraphics[width=0.32\textwidth]
{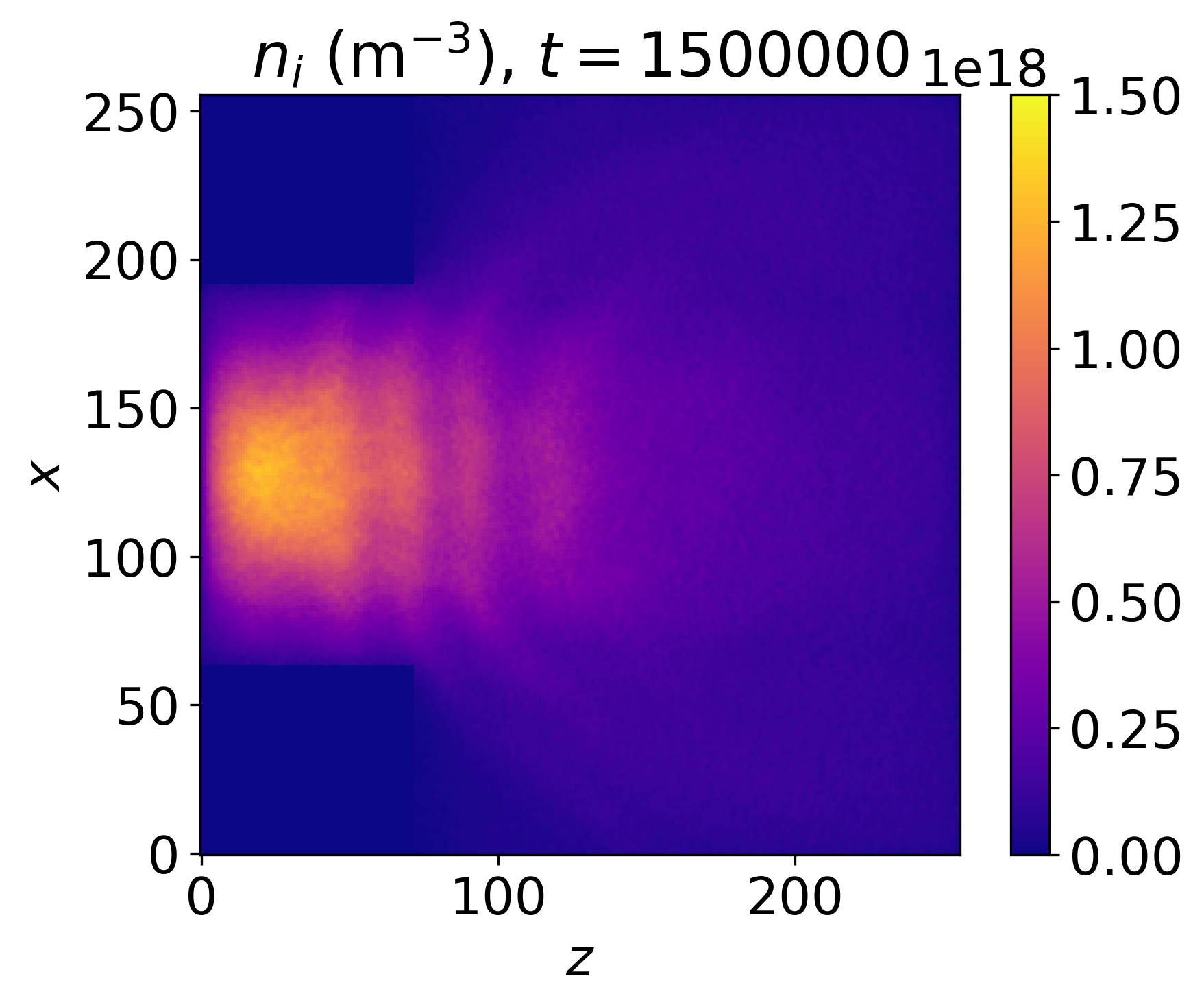}\\
\includegraphics[width=0.32\textwidth]
{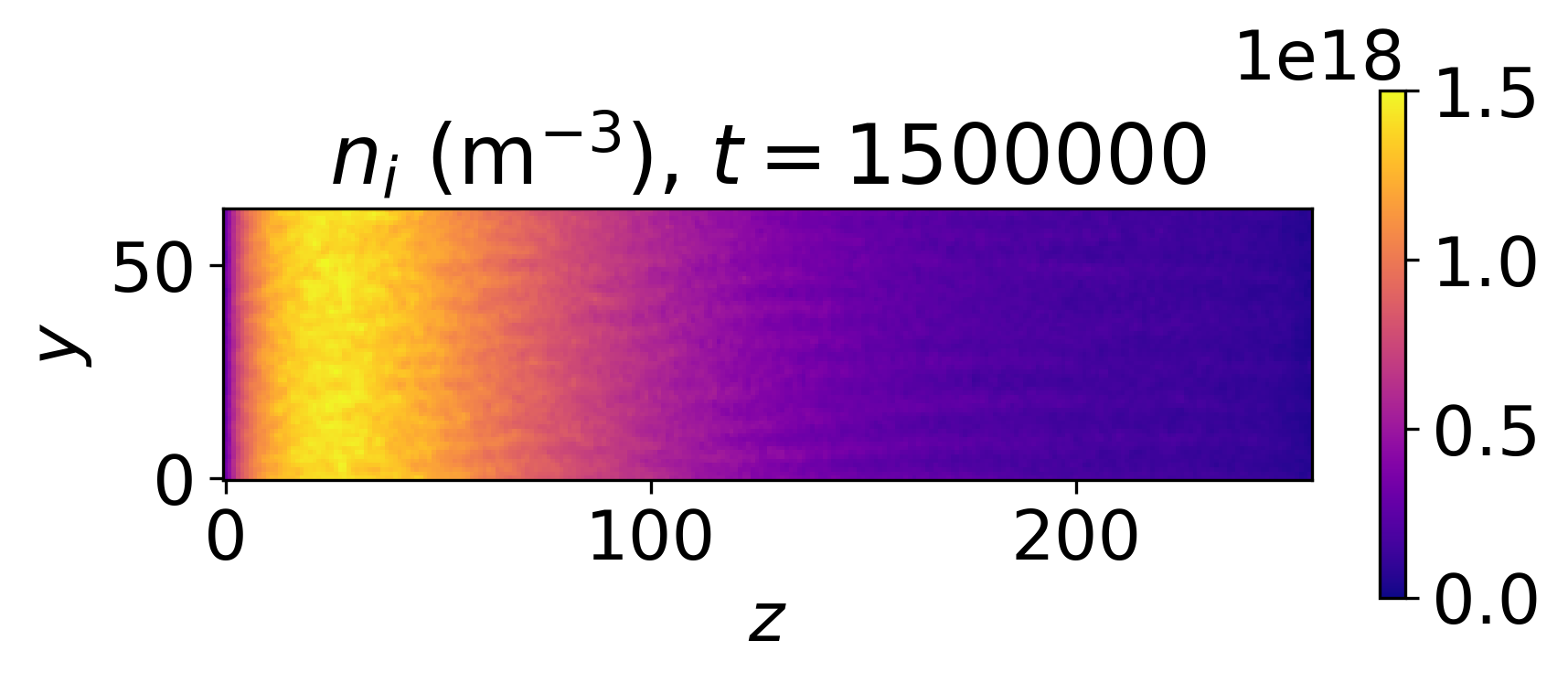}
\includegraphics[width=0.32\textwidth]
{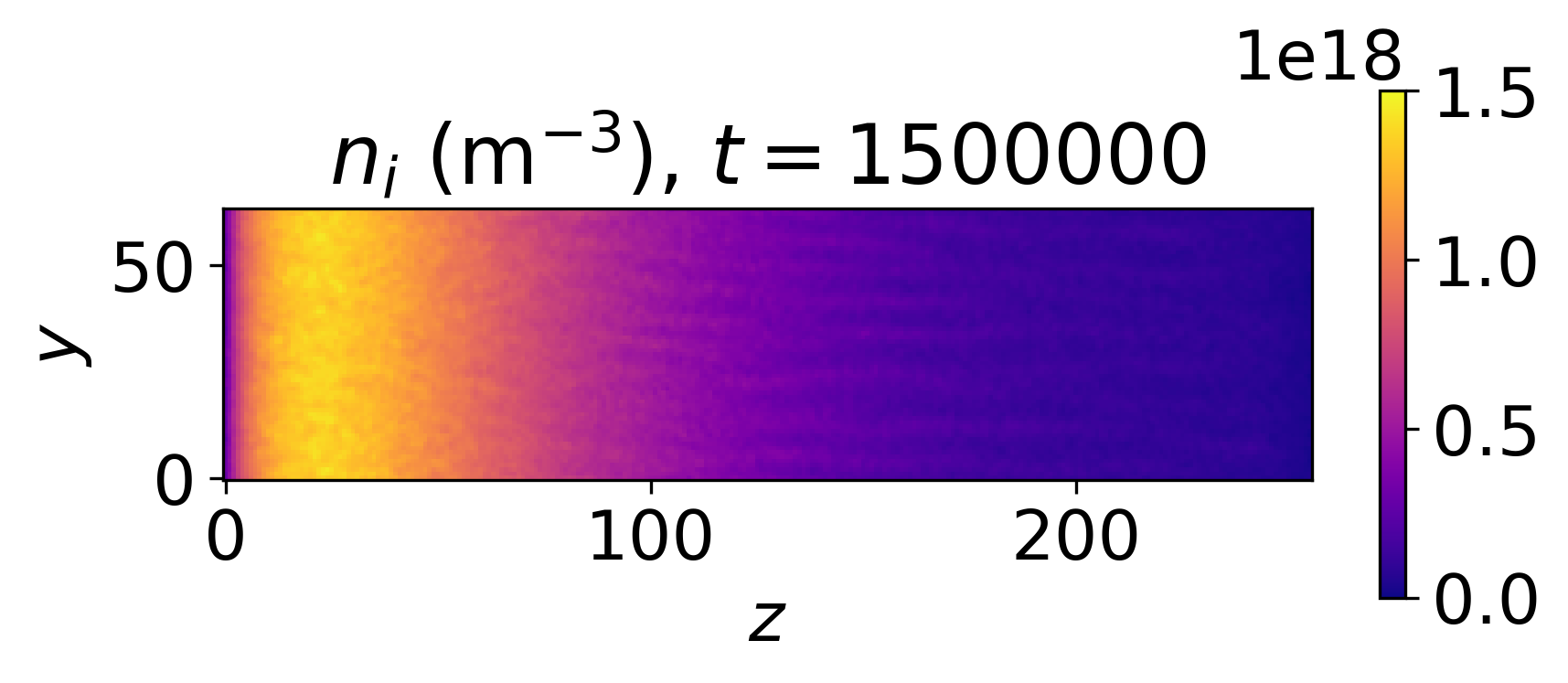}
\includegraphics[width=0.32\textwidth]
{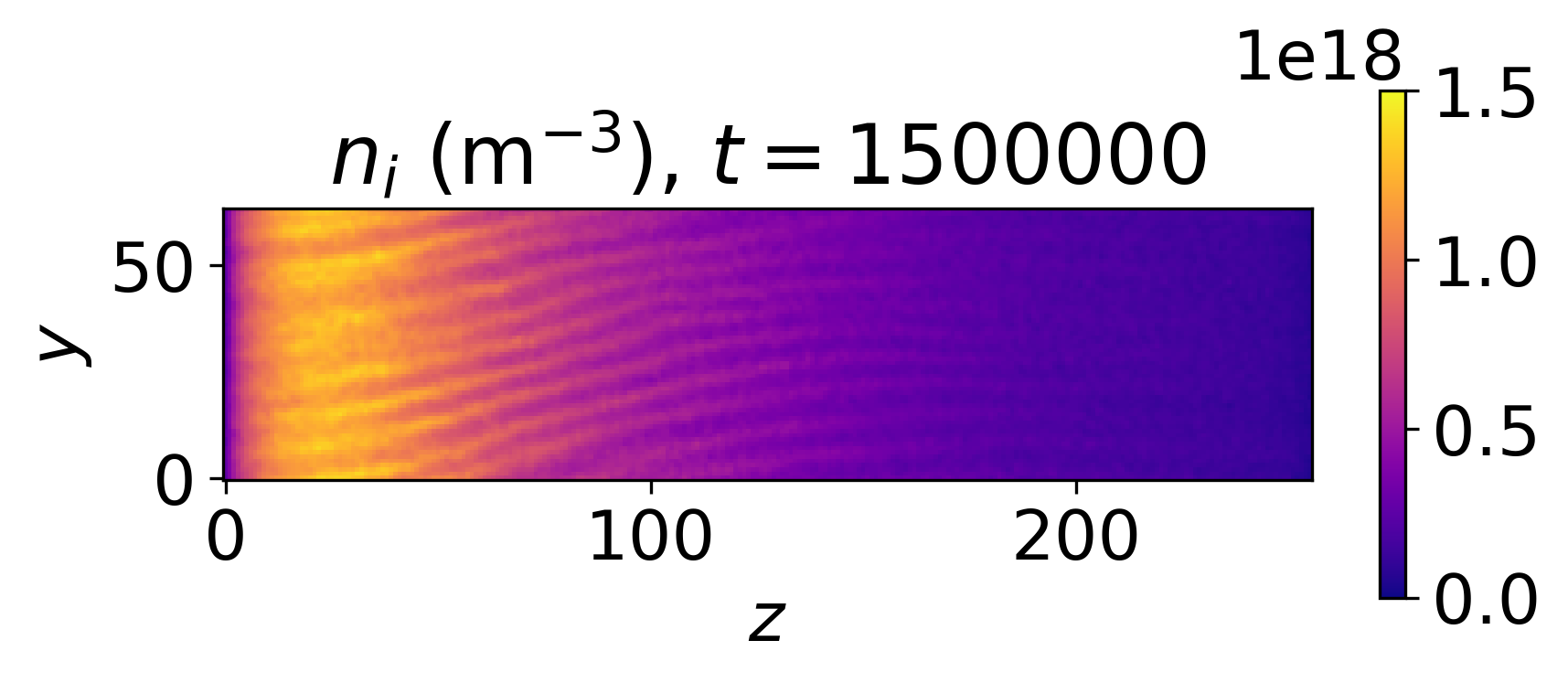}
\caption{
Ion number density $n_i$
at the middle plane of $y$ (top row) and $x$
(bottom row)
of cases Strong-B (left), Weak-B (middle), and
Analytic-B (right)
at time 22.5 $\mu$s.
}
\label{fig:ni}
\end{figure*}

\begin{figure*}[!ht]
\centering
\includegraphics[width=0.32\textwidth]
{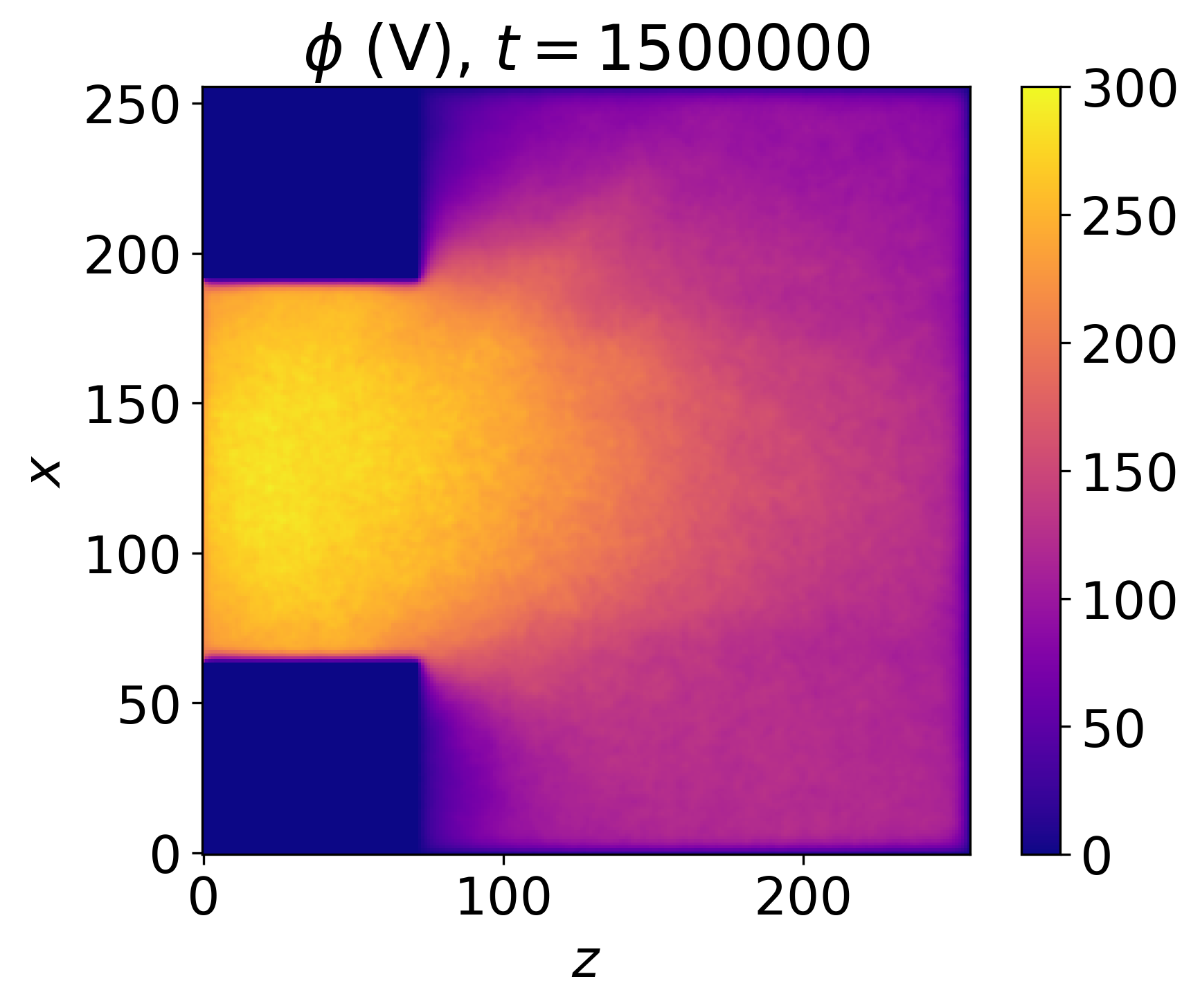}
\includegraphics[width=0.32\textwidth]
{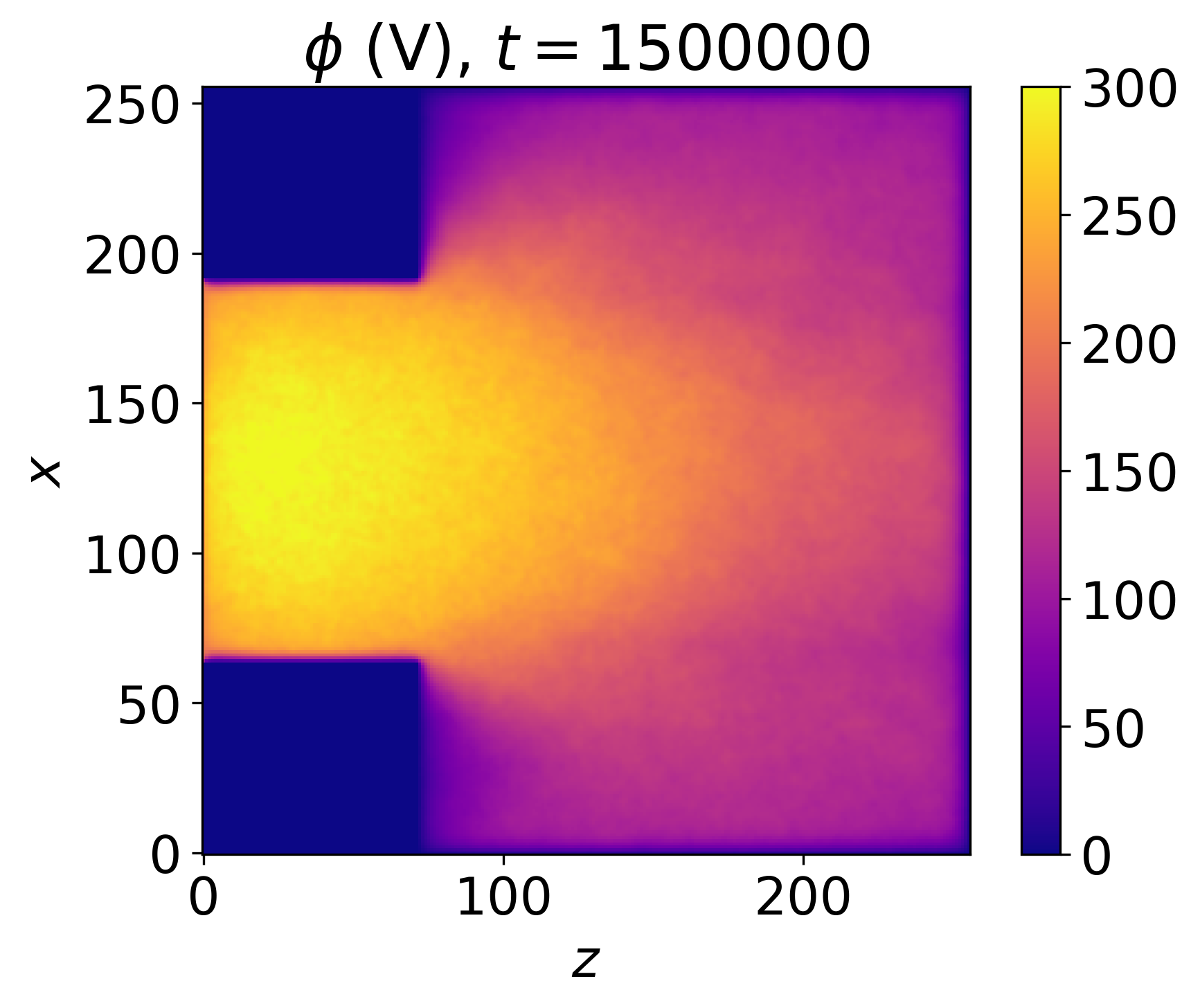}
\includegraphics[width=0.32\textwidth]
{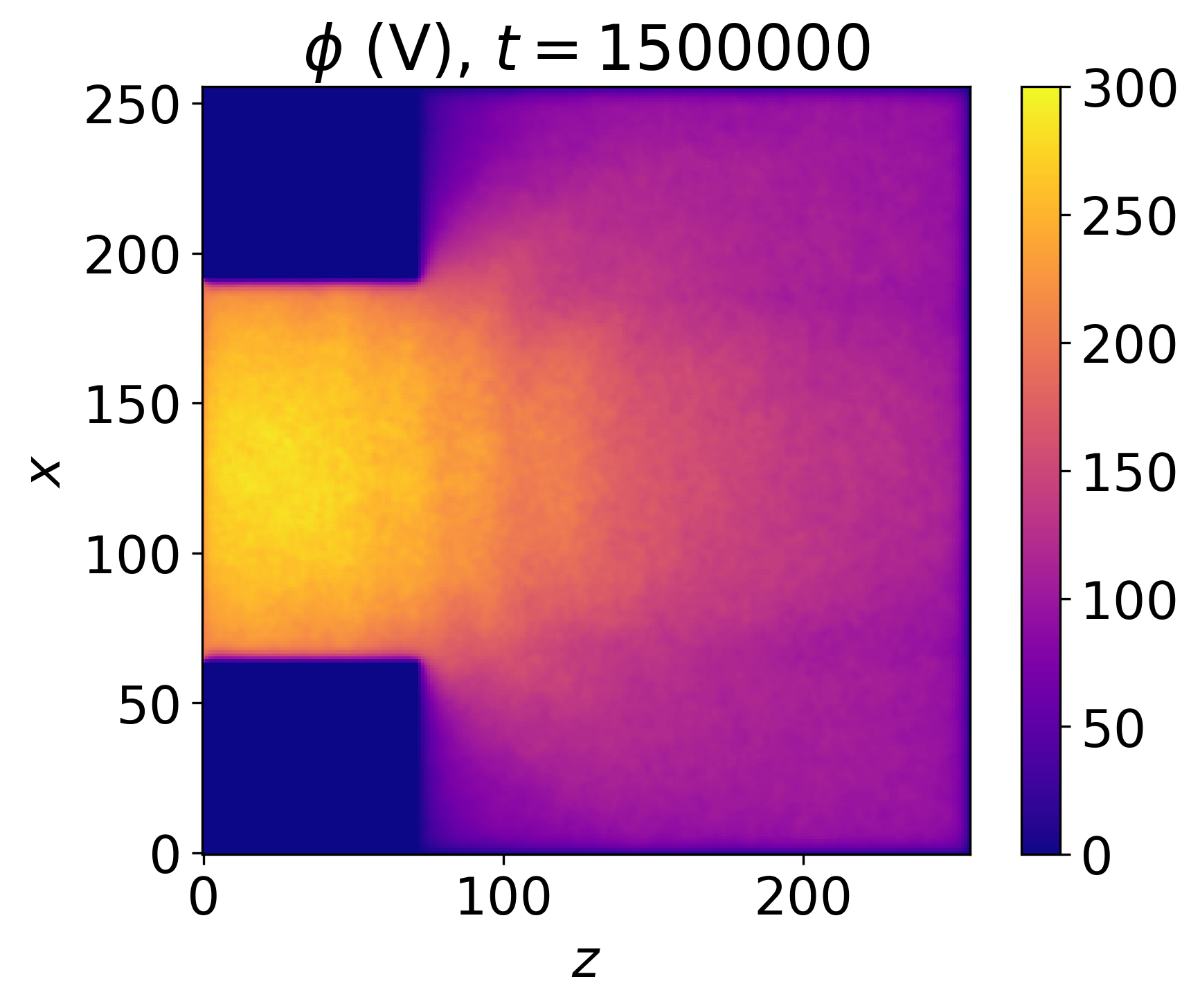}\\
\includegraphics[width=0.32\textwidth]
{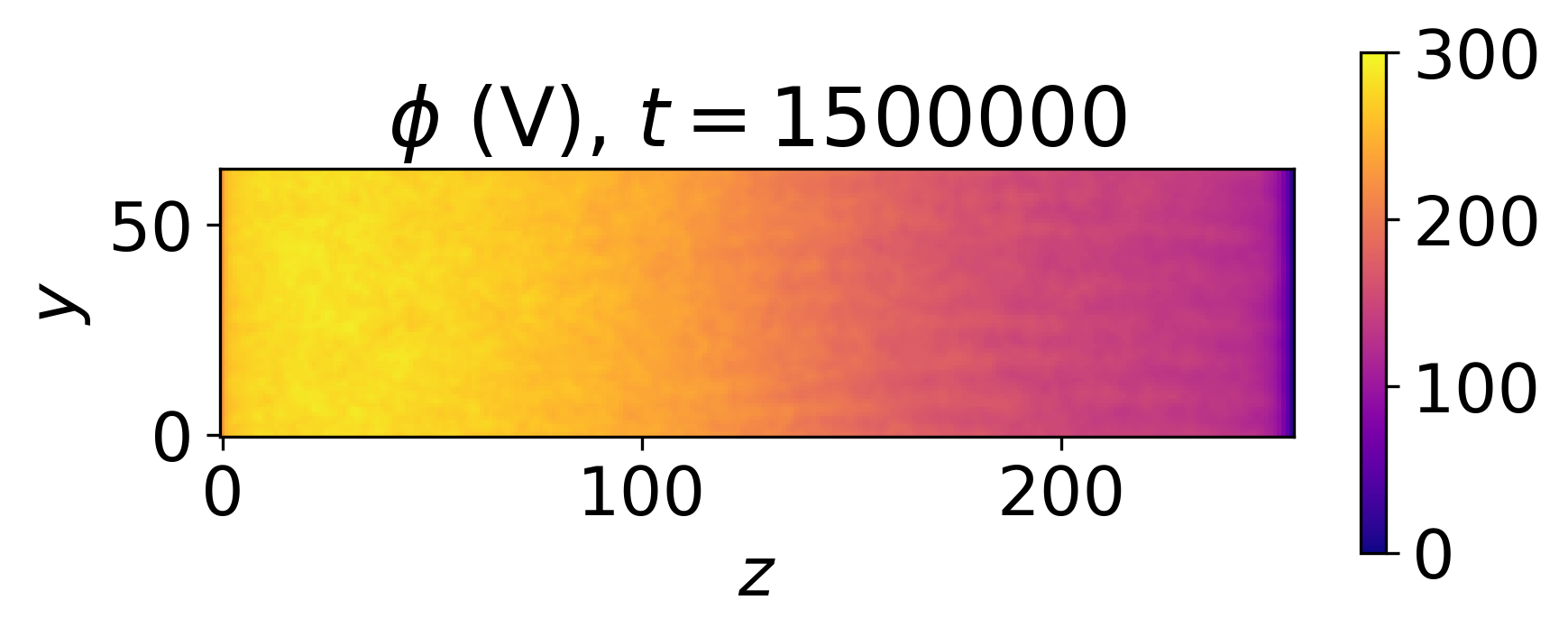}
\includegraphics[width=0.32\textwidth]
{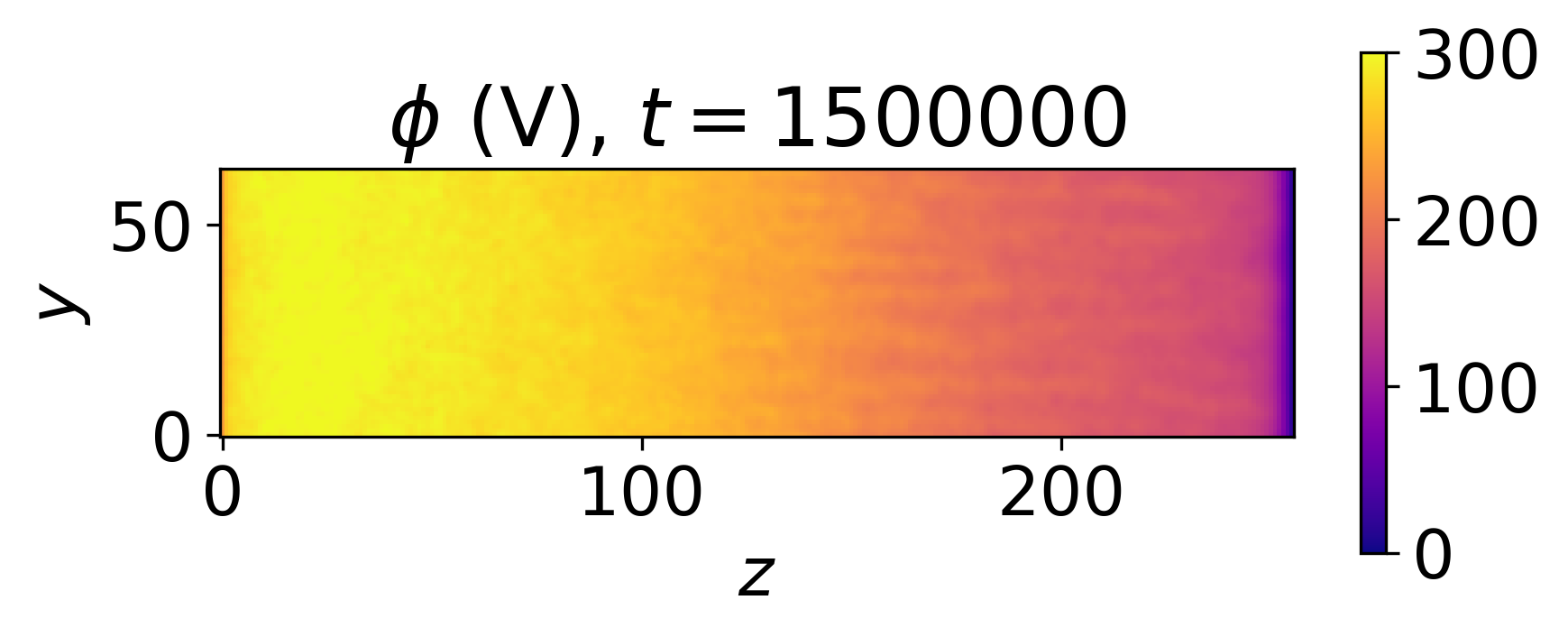}
\includegraphics[width=0.32\textwidth]
{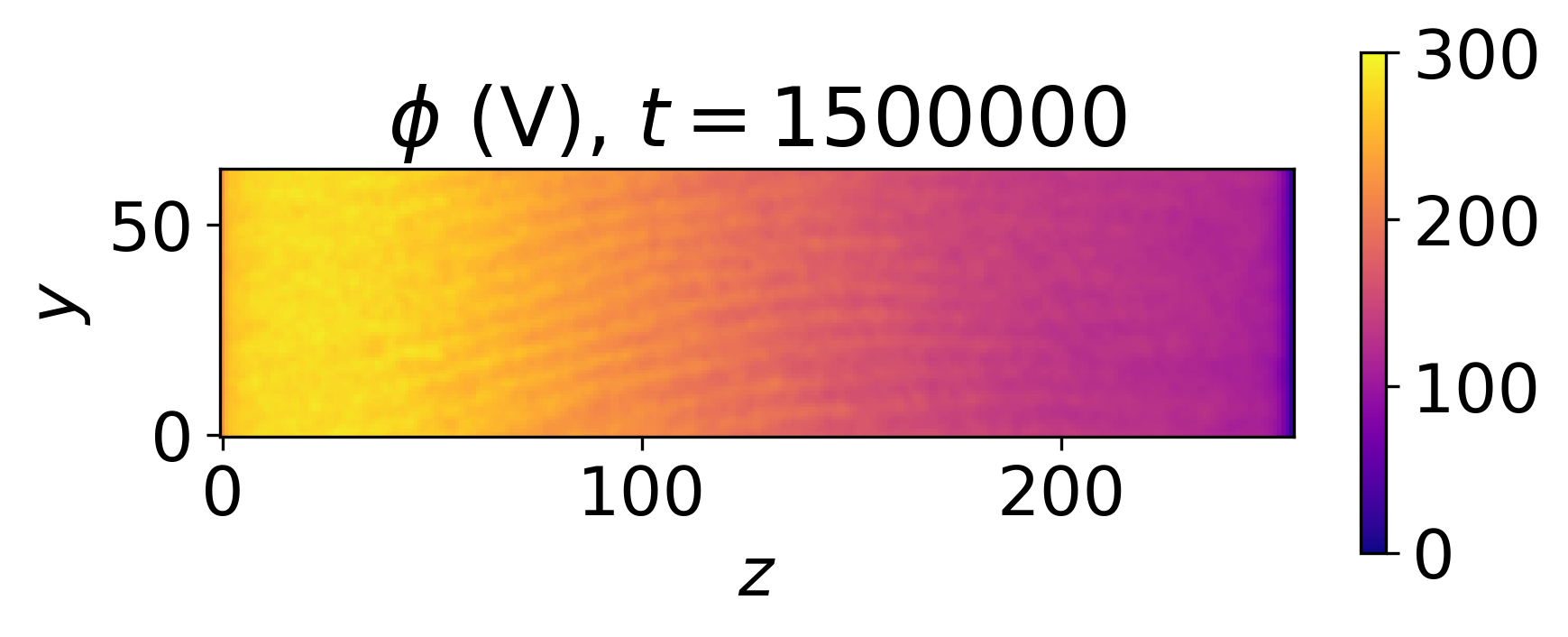}
\caption{
Potential $\phi$
at the middle plane of $y$ (top row) and $x$
(bottom row)
of cases Strong-B (left), Weak-B (middle), and
Analytic-B (right)
at time 22.5 $\mu$s.
}
\label{fig:phi}
\end{figure*}

\begin{figure*}[!ht]
\centering
\includegraphics[width=0.32\textwidth]
{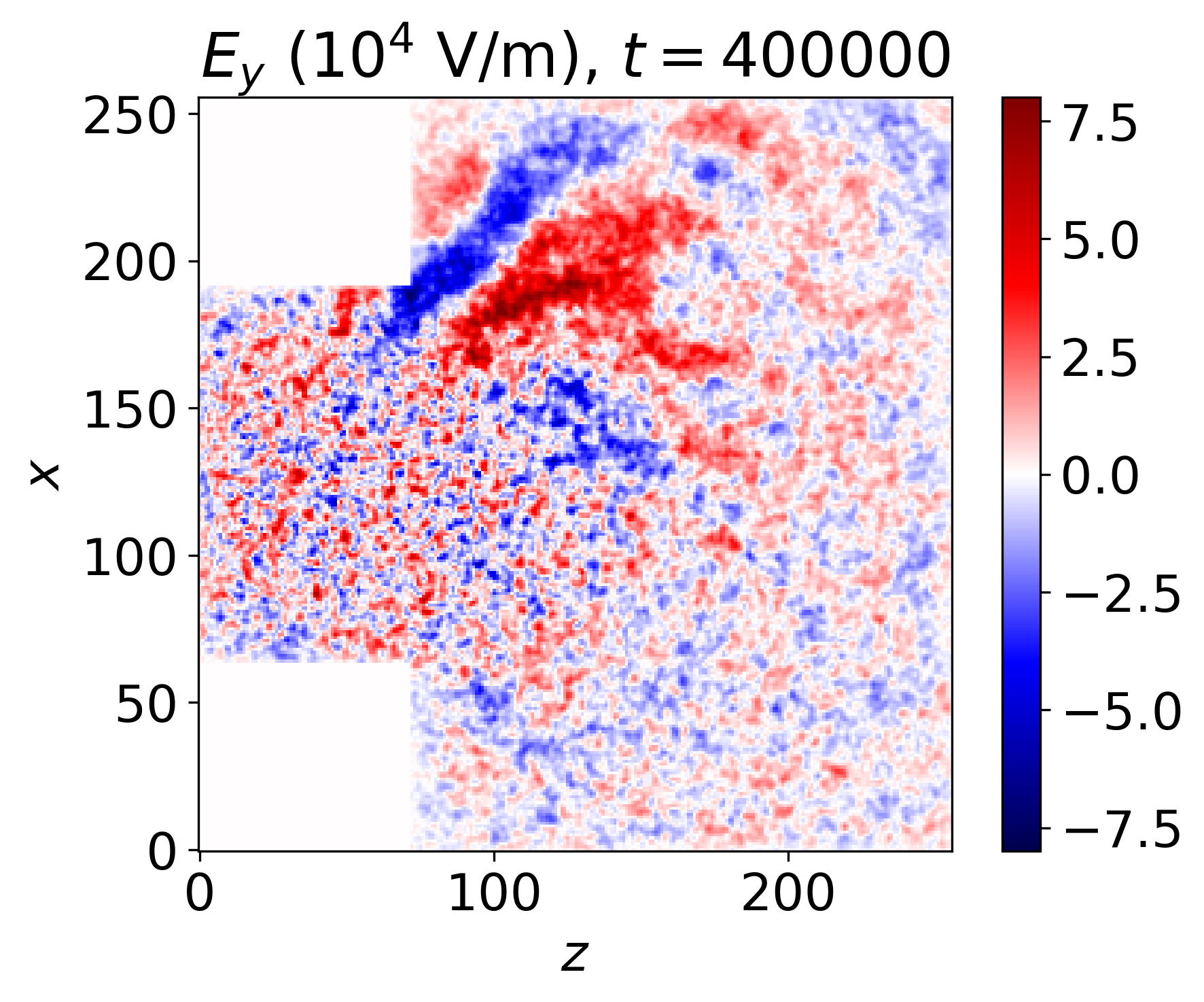}
\includegraphics[width=0.32\textwidth]
{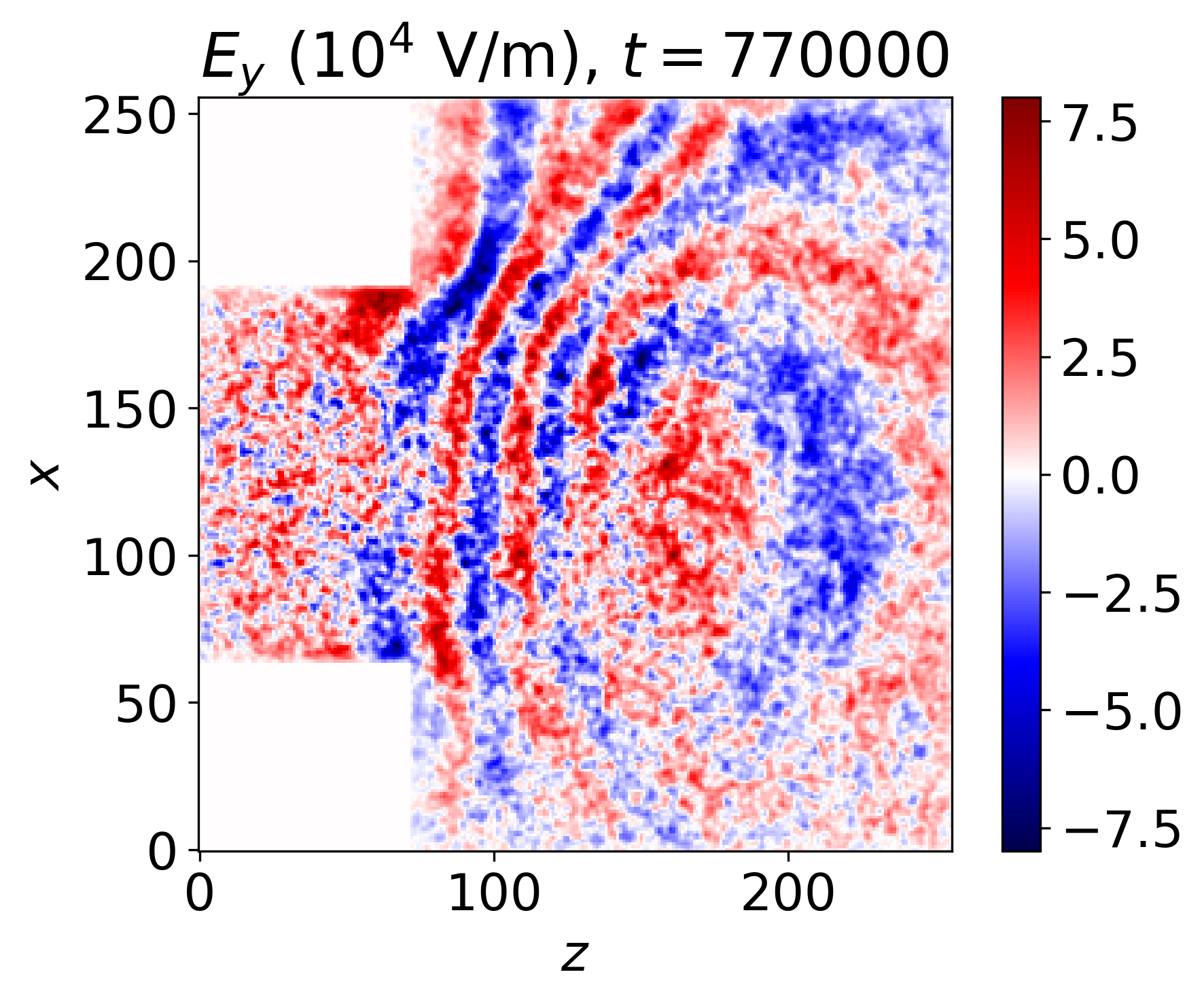}
\includegraphics[width=0.32\textwidth]
{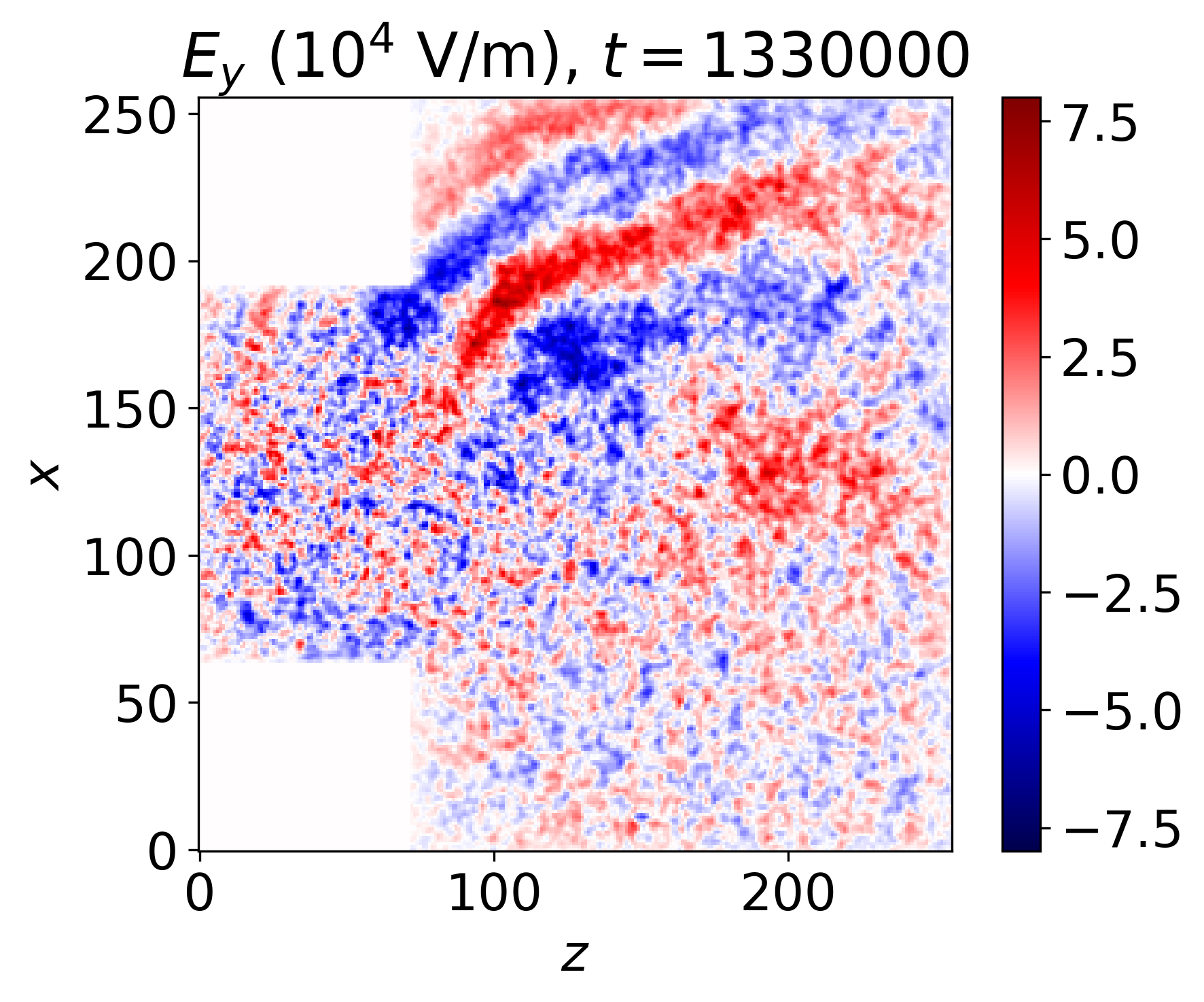}\\
\includegraphics[width=0.32\textwidth]
{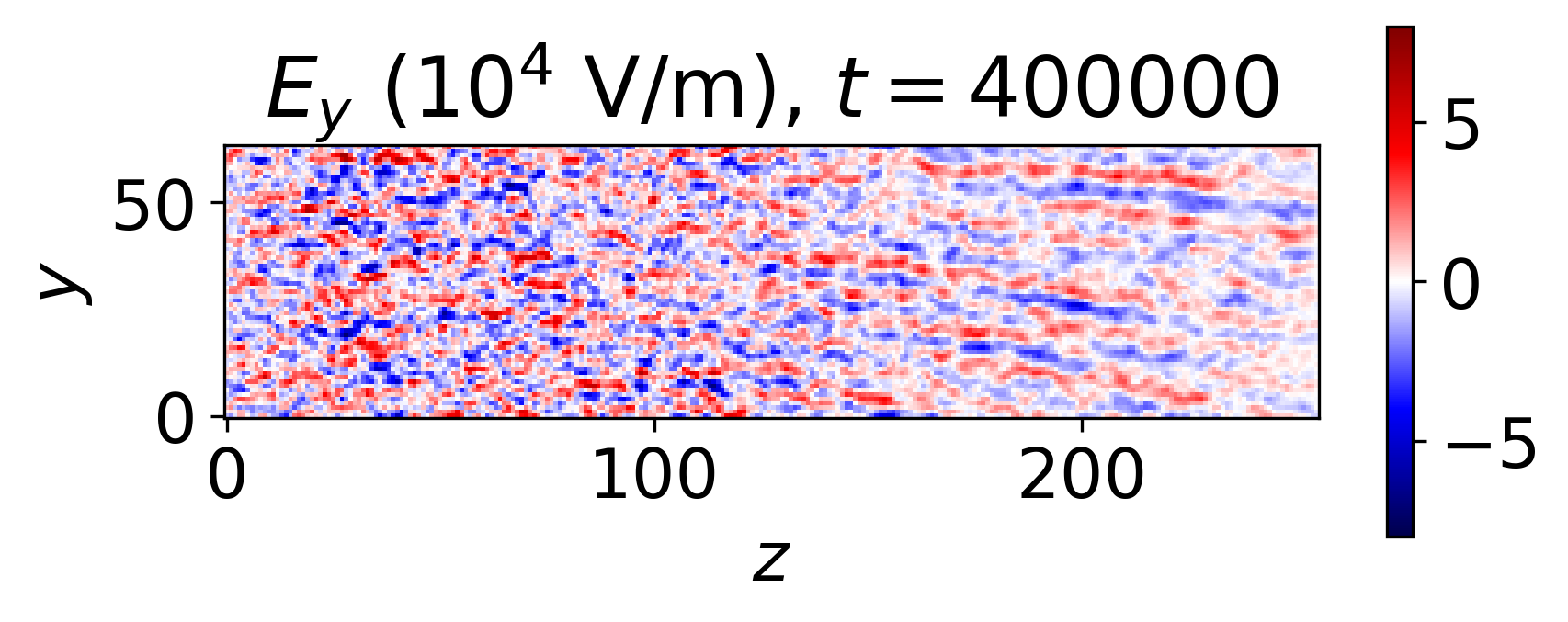}
\includegraphics[width=0.32\textwidth]
{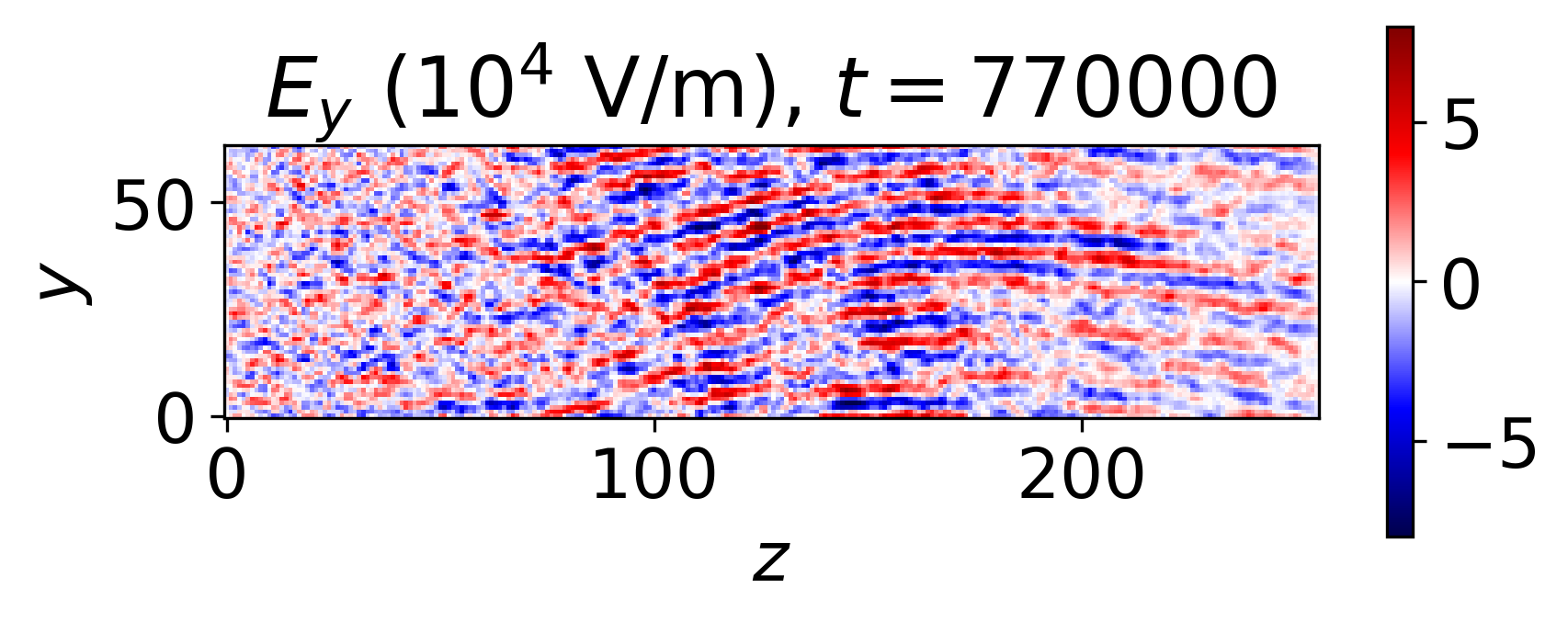}
\includegraphics[width=0.32\textwidth]
{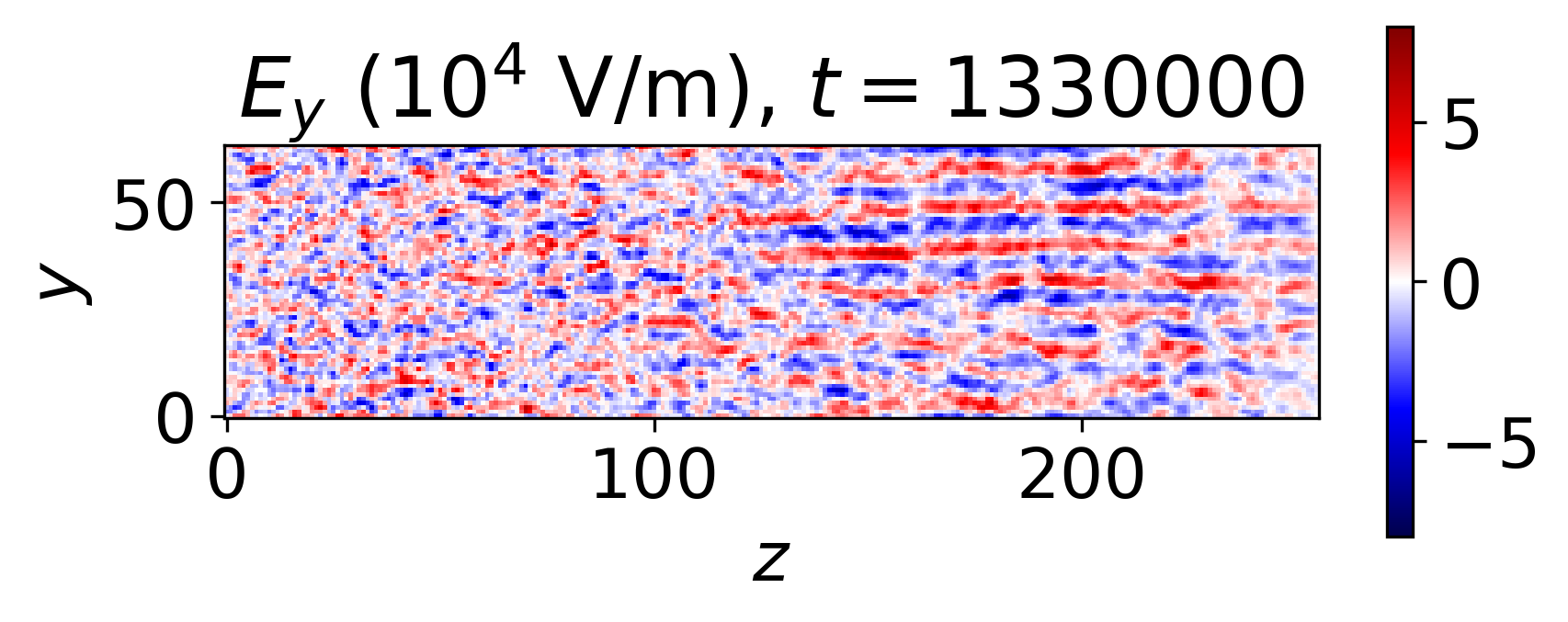}
\caption{
Azimuthal electric field $E_y$
at the middle plane of $y$ (top row) and $x$
(bottom row)
of case Strong-B
at time 6,
11.55,
and 19.95 $\mu$s.
}
\label{fig:Ey_Strong-B}
\end{figure*}

\begin{figure*}[!ht]
\centering
\includegraphics[width=0.32\textwidth]
{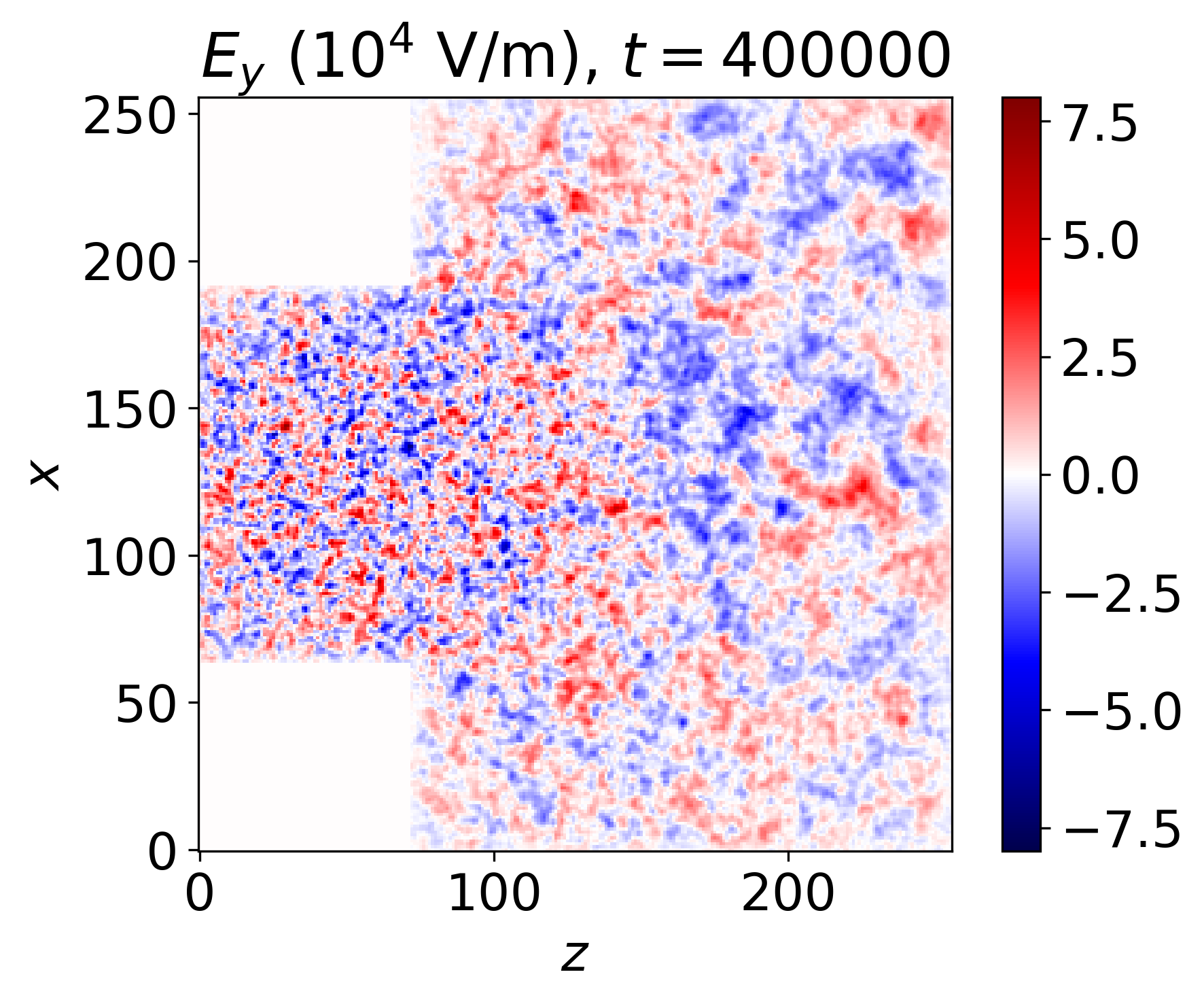}
\includegraphics[width=0.32\textwidth]
{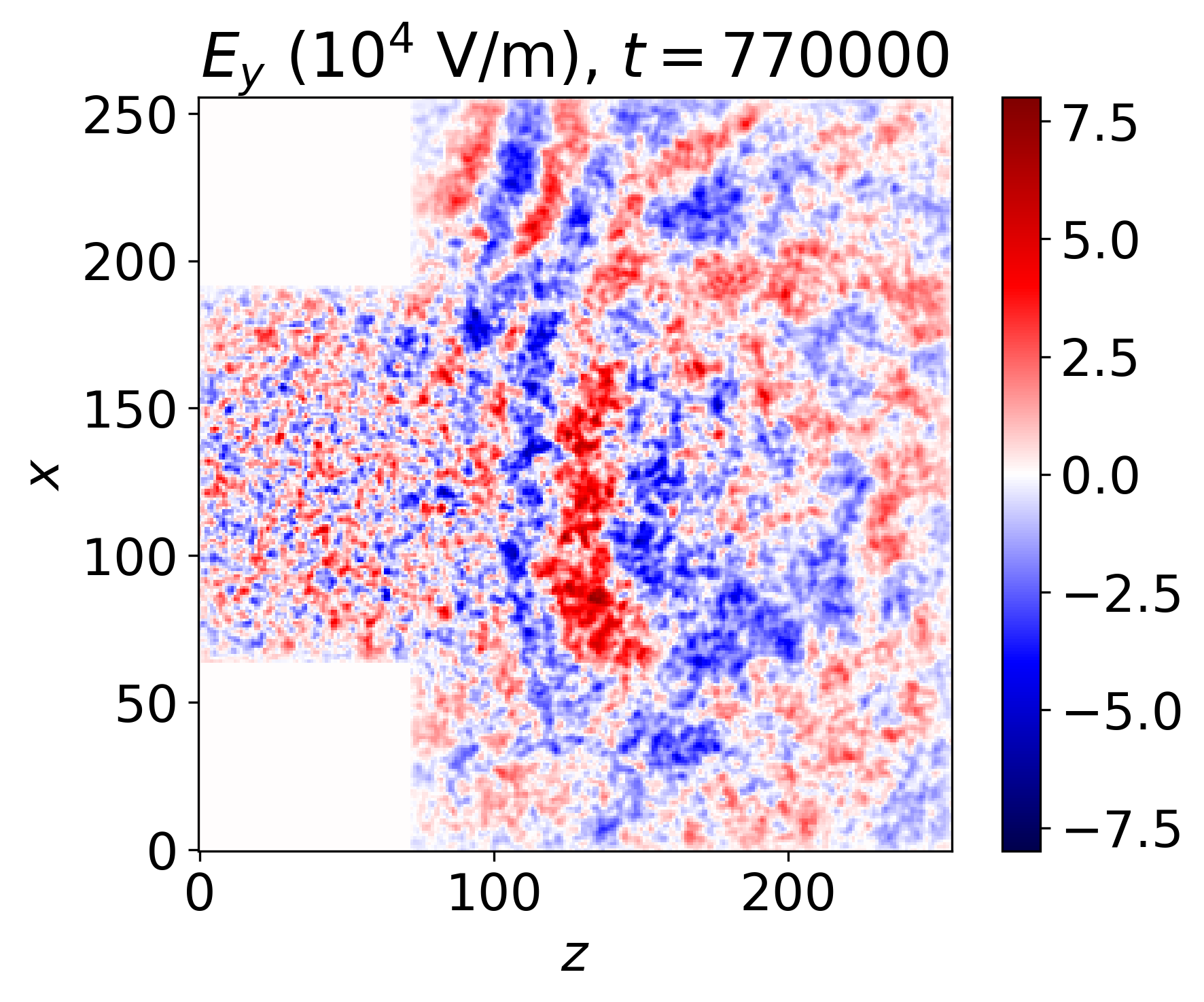}
\includegraphics[width=0.32\textwidth]
{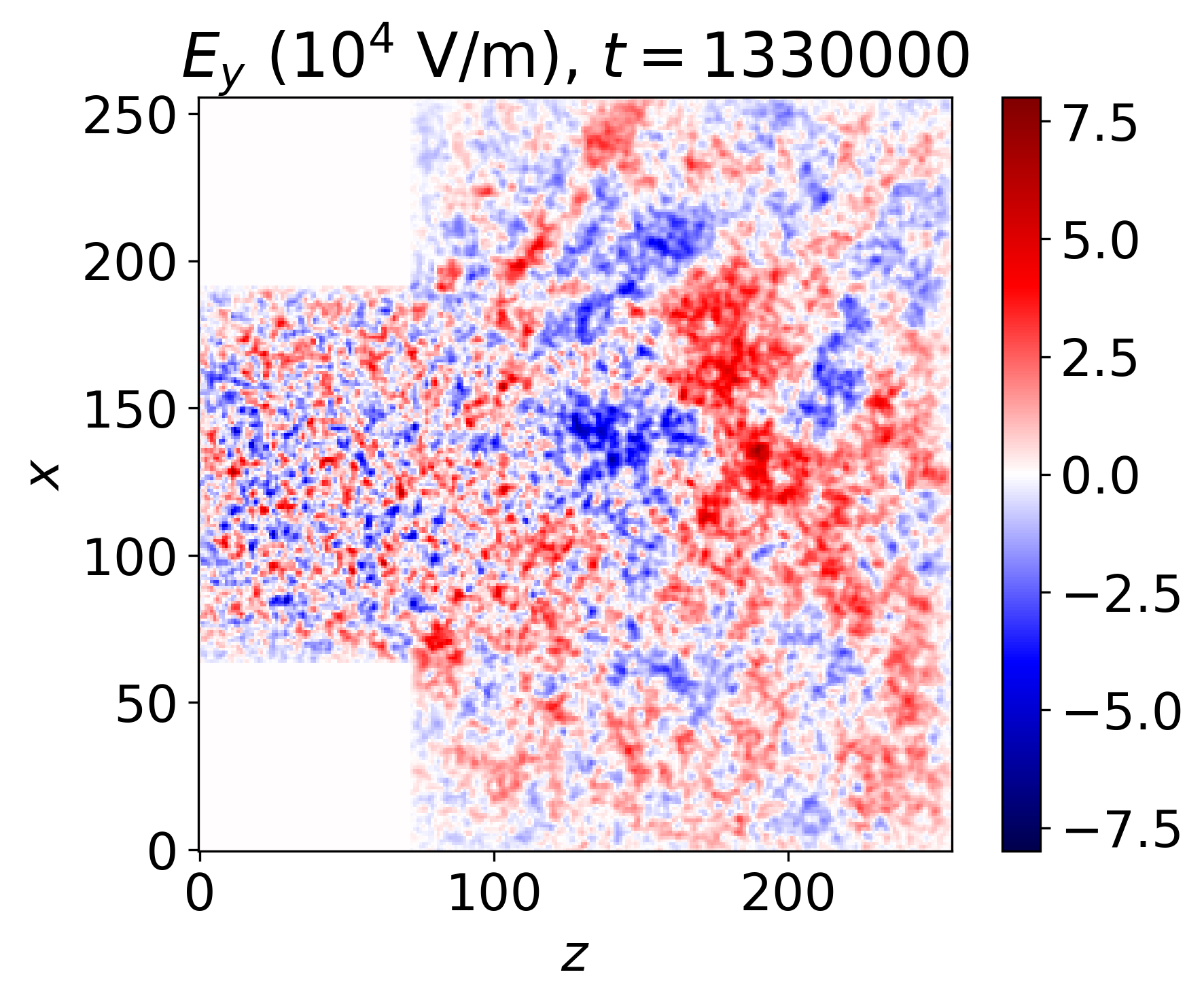}\\
\includegraphics[width=0.32\textwidth]
{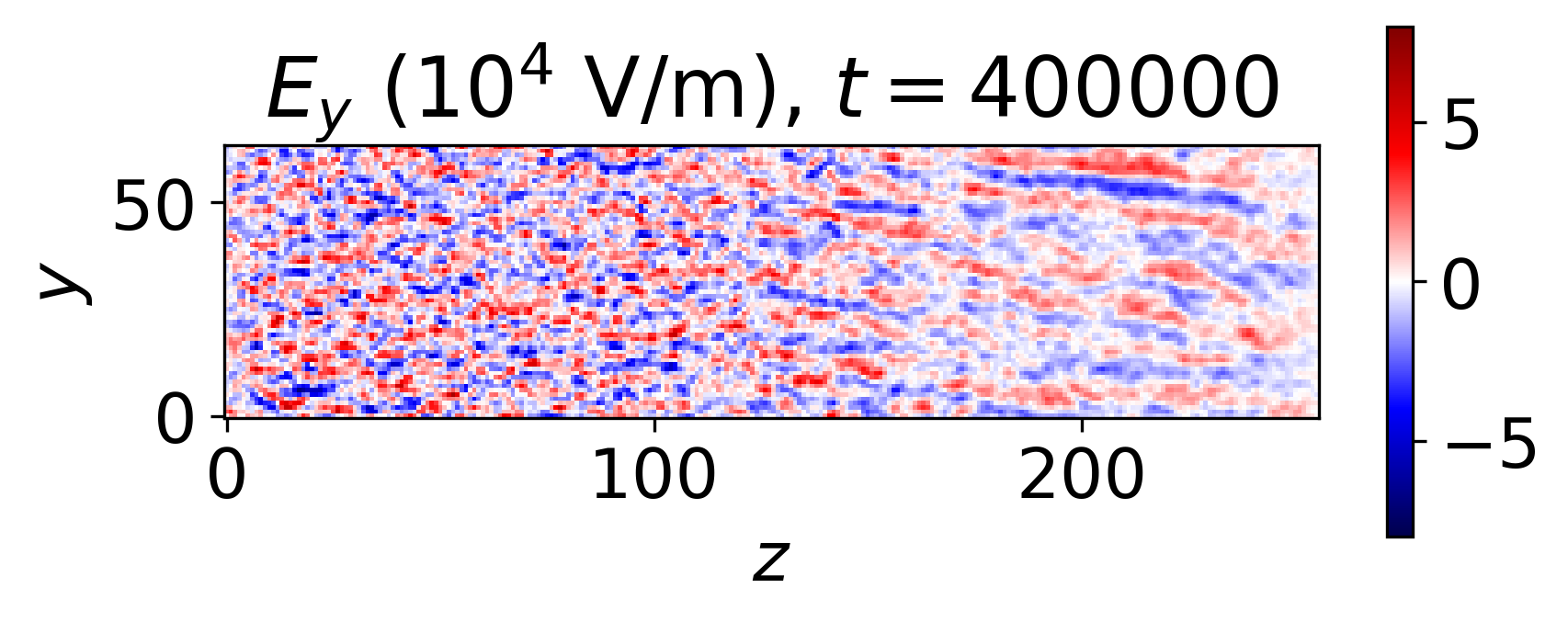}
\includegraphics[width=0.32\textwidth]
{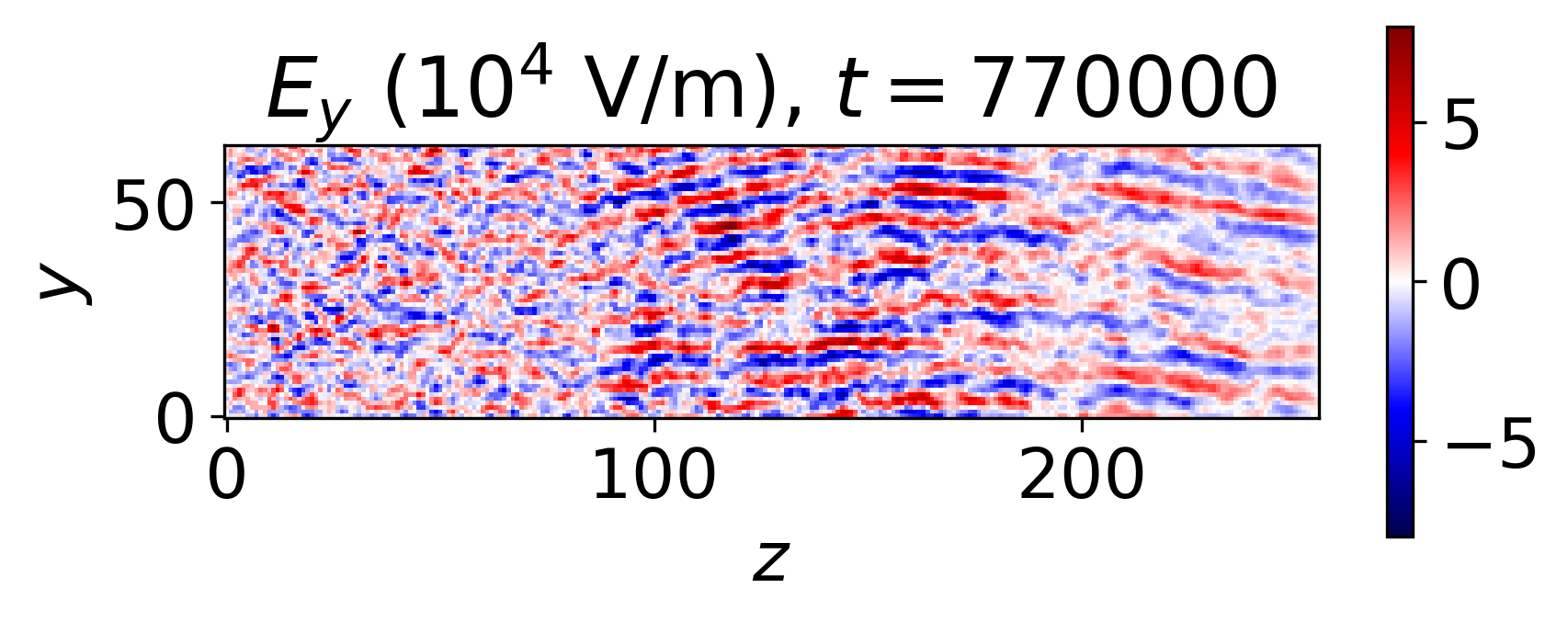}
\includegraphics[width=0.32\textwidth]
{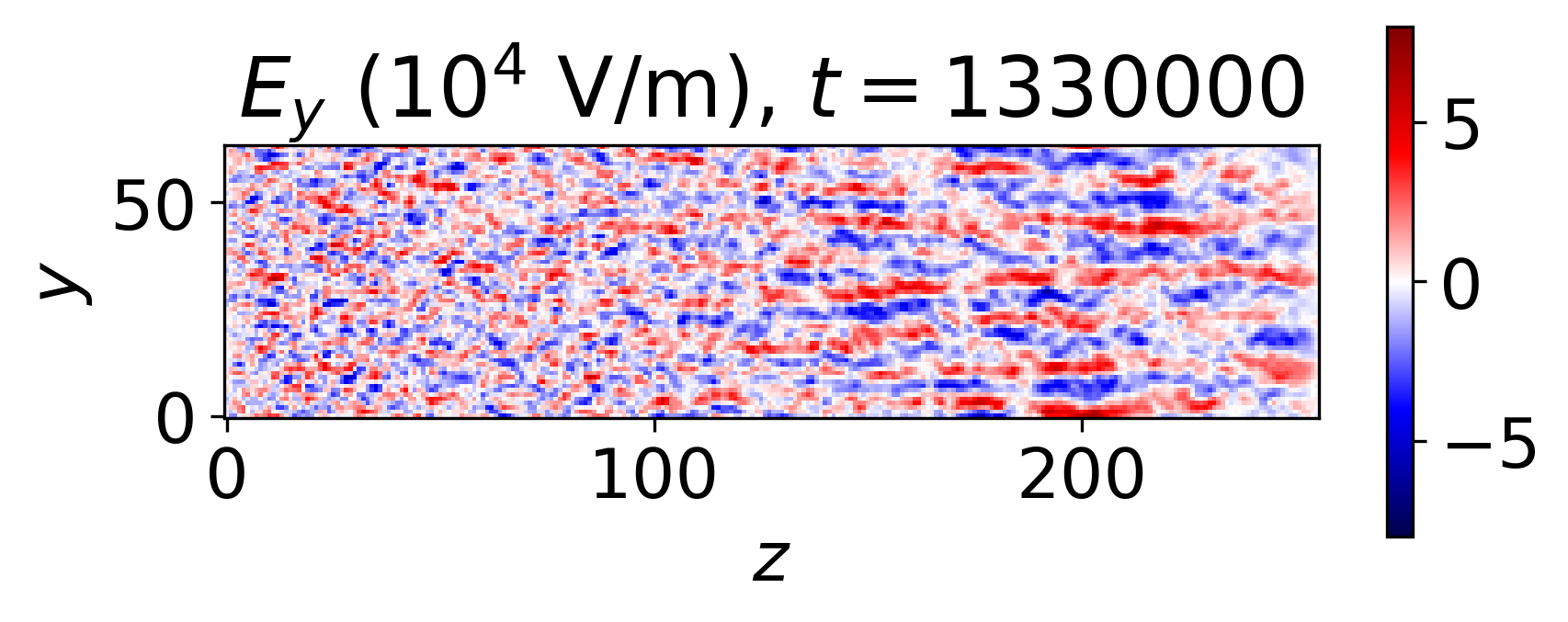}
\caption{
Azimuthal electric field $E_y$
at the middle plane of $y$ (top row) and $x$
(bottom row)
of case Weak-B
at time 6,
11.55,
and 19.95 $\mu$s.
}
\label{fig:Ey_Weak-B}
\end{figure*}

\begin{figure*}[!ht]
\centering
\includegraphics[width=0.32\textwidth]
{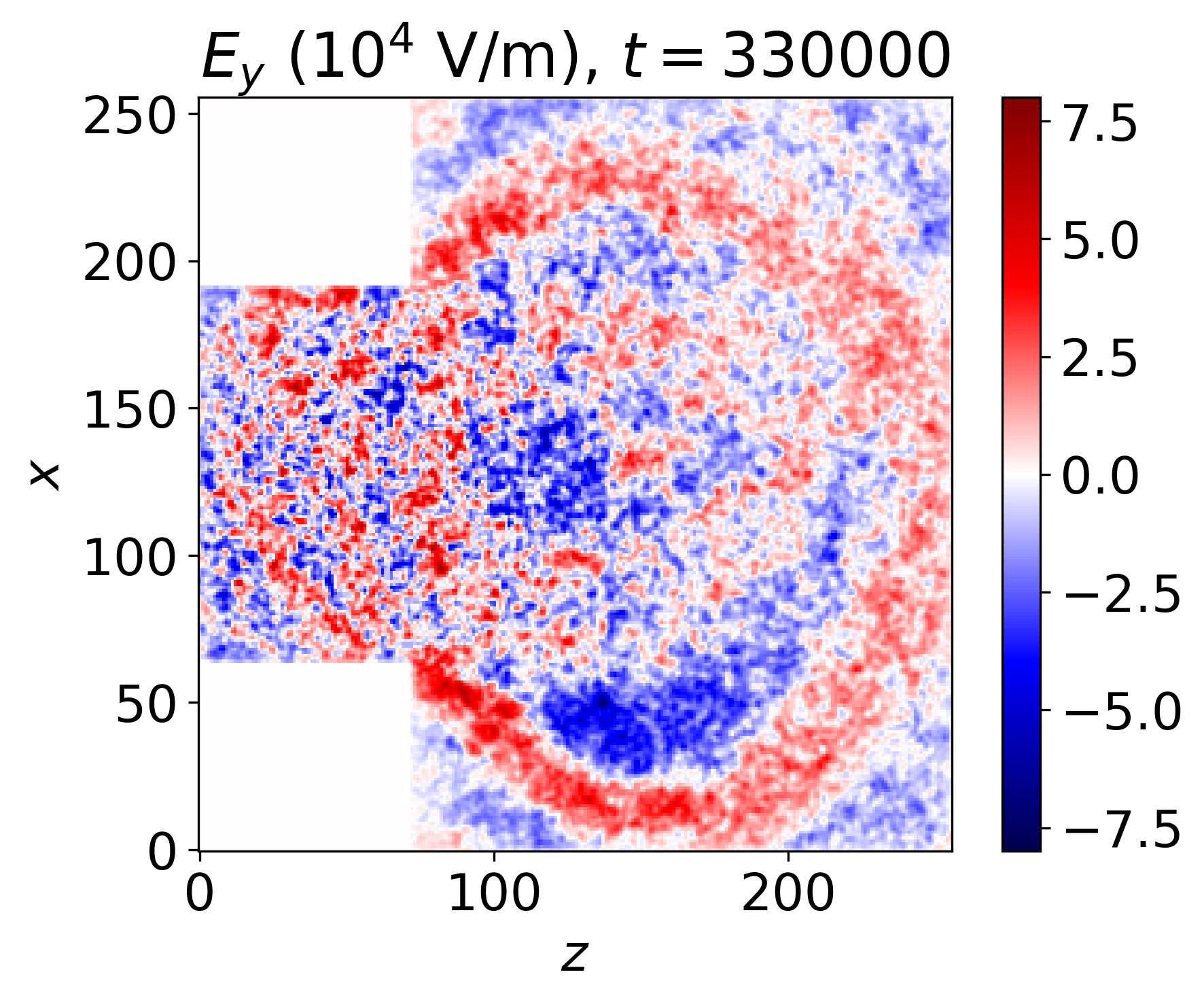}
\includegraphics[width=0.32\textwidth]
{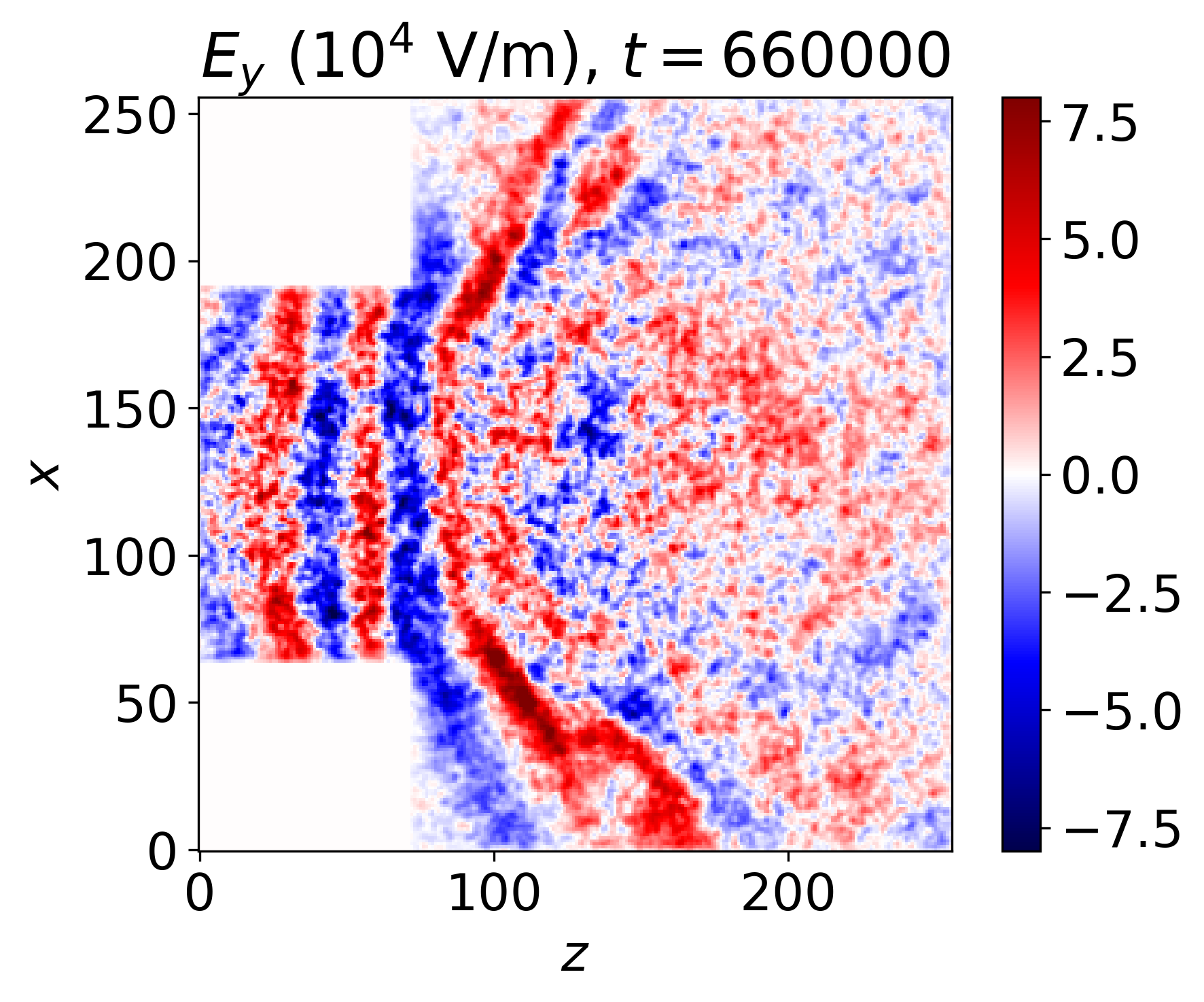}
\includegraphics[width=0.32\textwidth]
{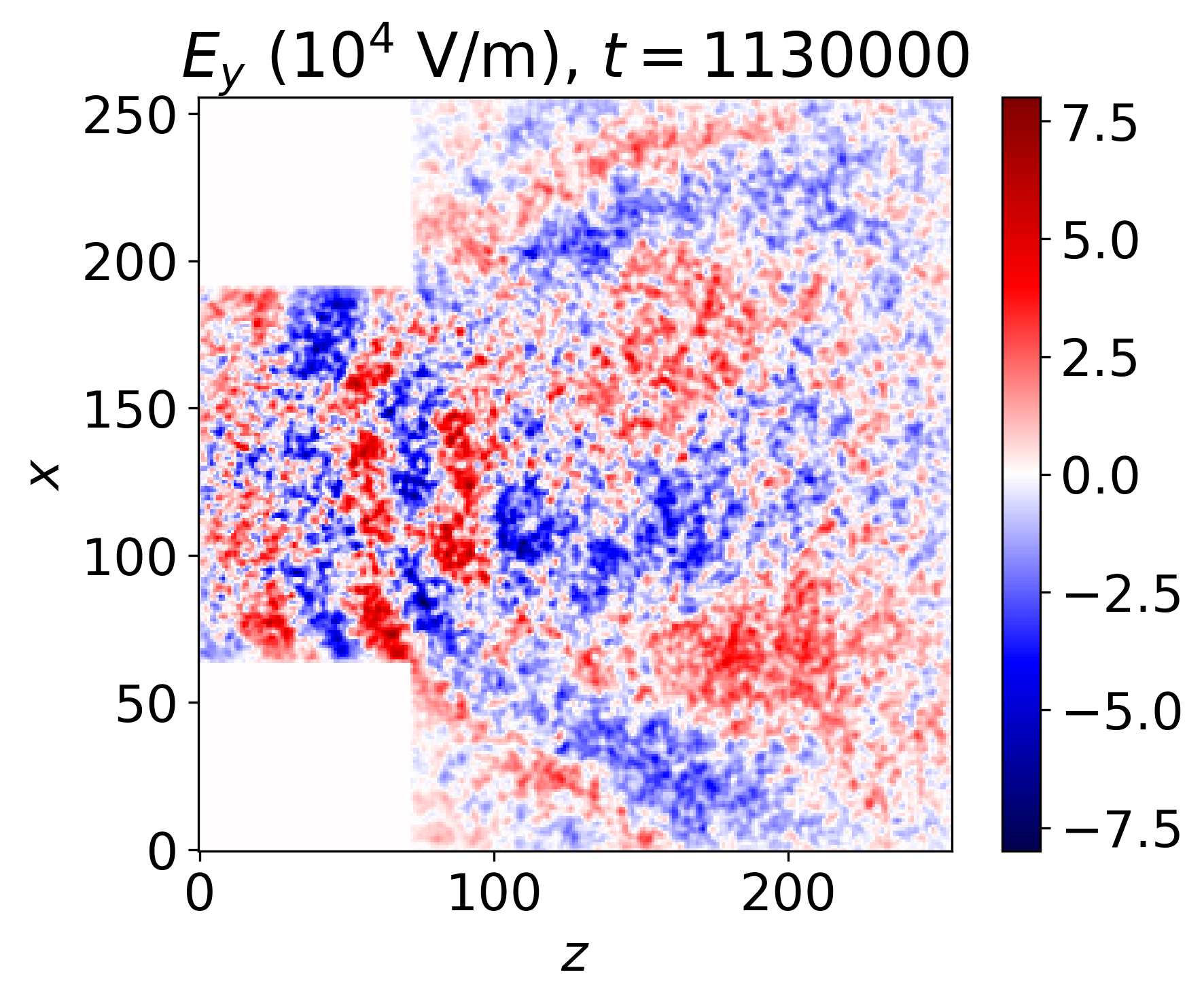}\\
\includegraphics[width=0.32\textwidth]
{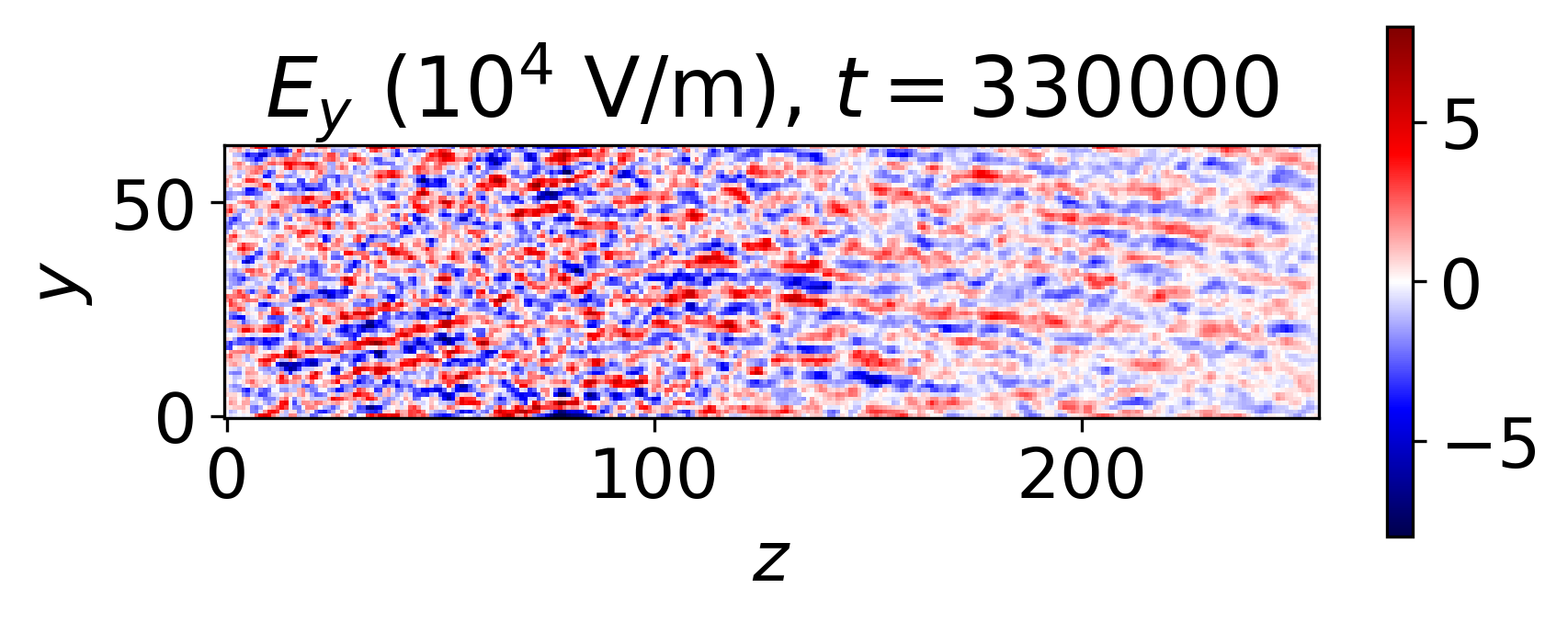}
\includegraphics[width=0.32\textwidth]
{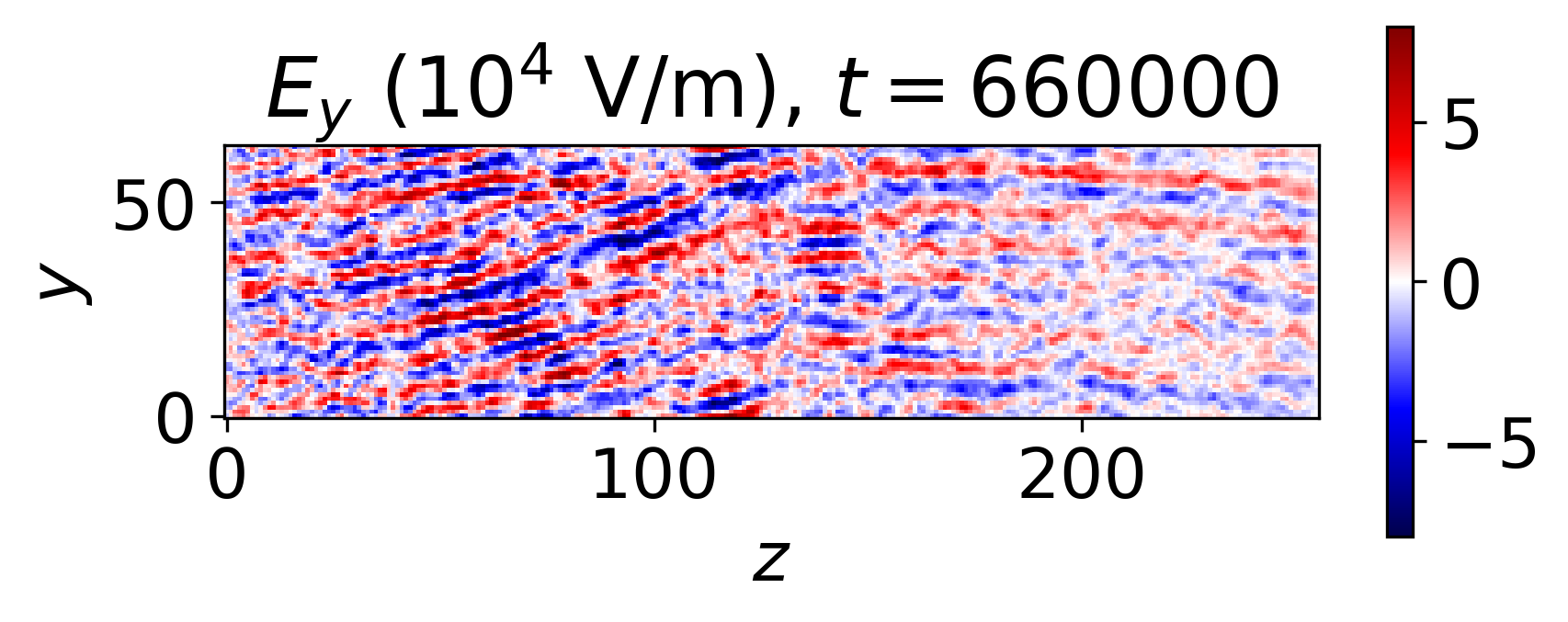}
\includegraphics[width=0.32\textwidth]
{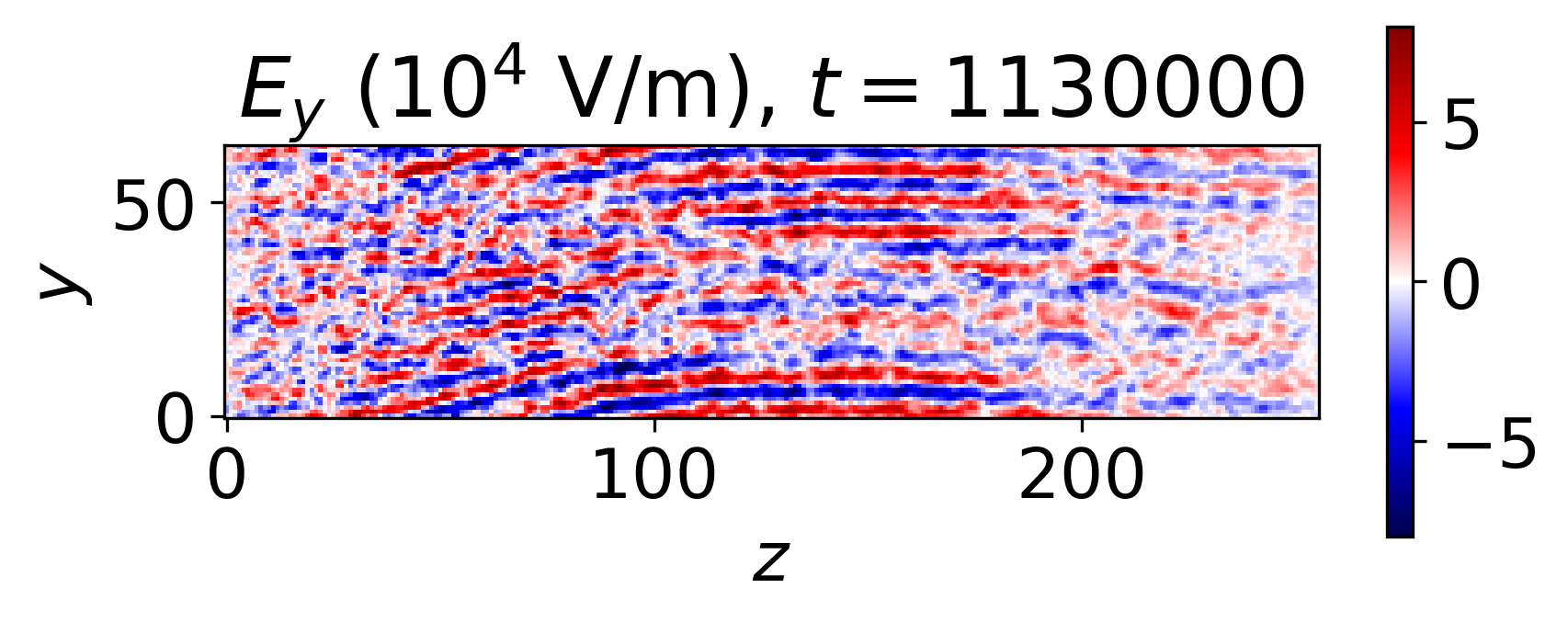}
\caption{
Azimuthal electric field $E_y$
at the middle plane of $y$ (top row) and $x$
(bottom row)
of case Analytic-B
at time 4.95,
9.9,
and 16.95 $\mu$s.
}
\label{fig:Ey_Analytic-B}
\end{figure*}

\begin{figure*}[!ht]
\centering
\includegraphics[width=0.32\textwidth]
{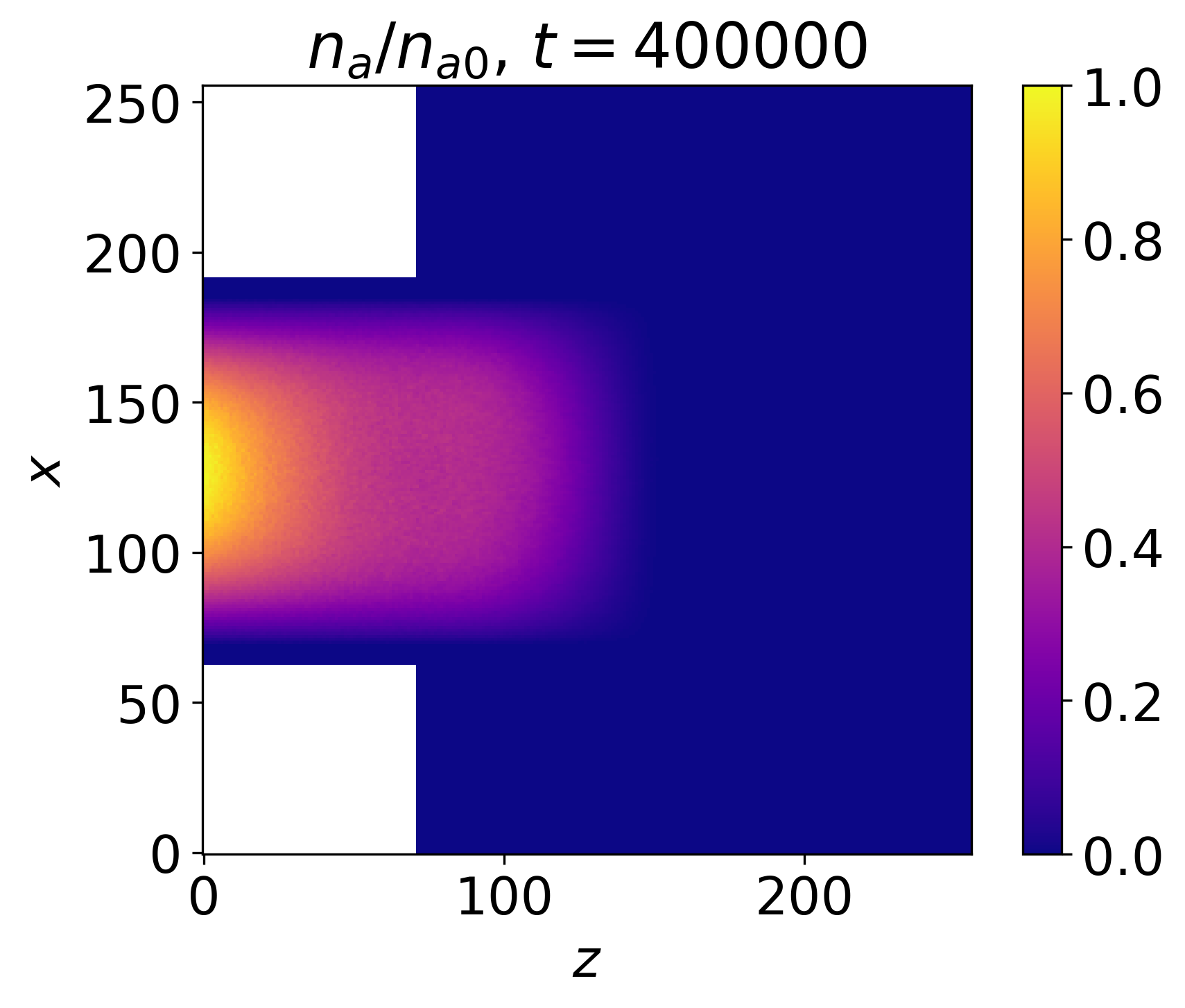}
\includegraphics[width=0.32\textwidth]
{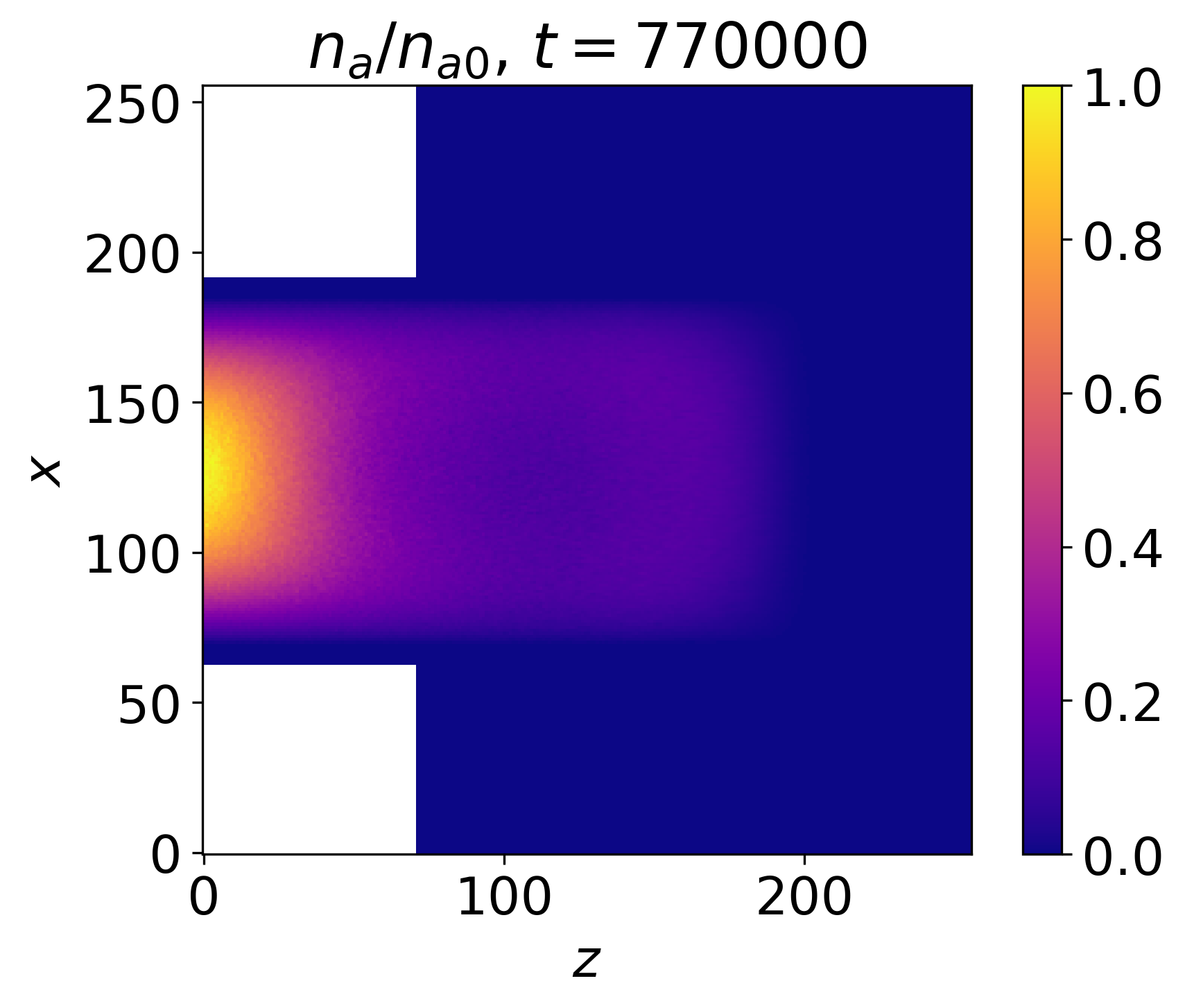}
\includegraphics[width=0.32\textwidth]
{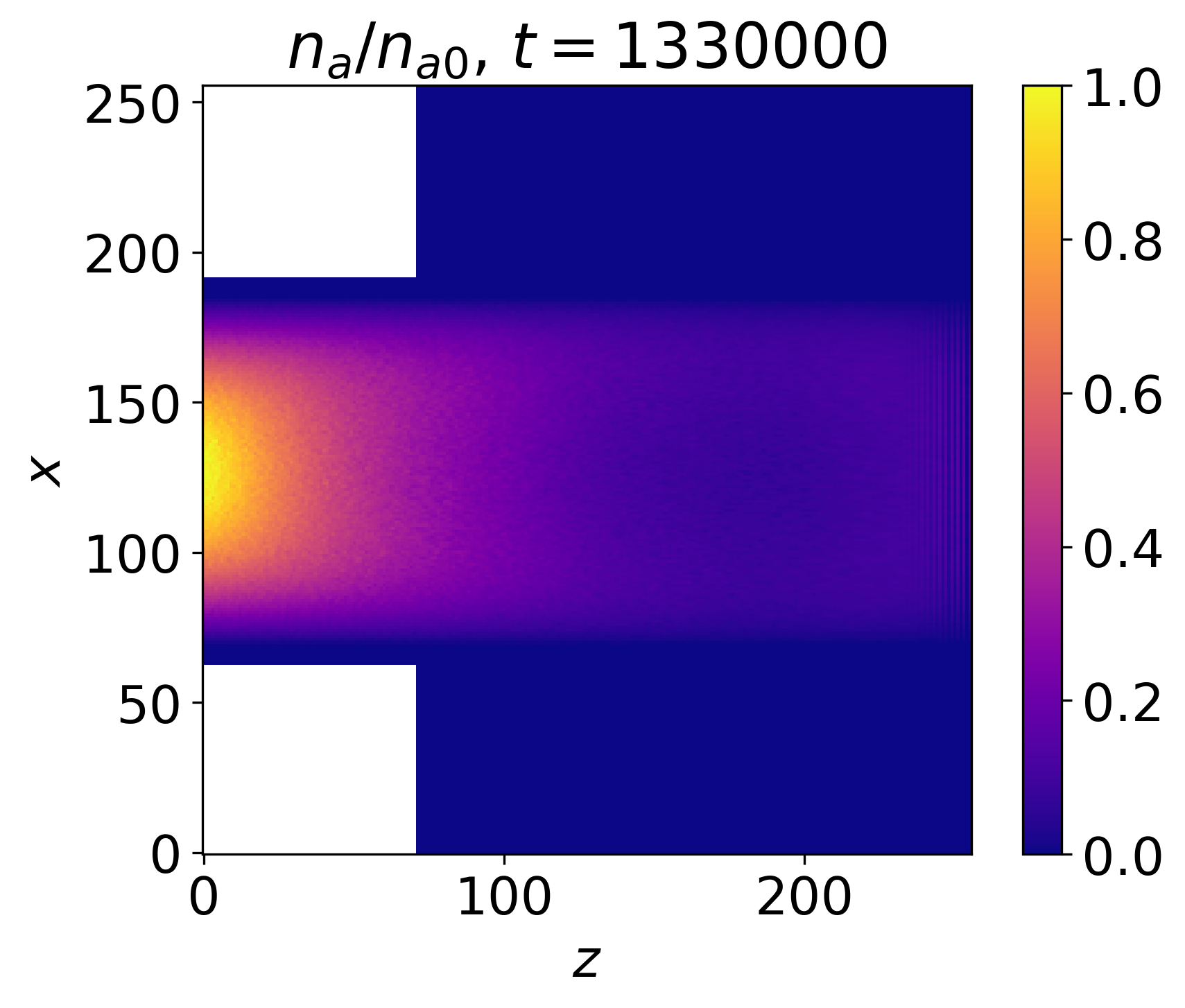}
\caption{
Neutral gas density $n_a$
at the middle plane of $y$ (top row) and $x$
(bottom row)
of case Analytic-B
at time 6,
11.55,
and 19.95 $\mu$s.
}
\label{fig:na_Strong-B}
\end{figure*}

The spatially averaged ion number density
over time of the three cases
is shown in Fig.\ref{fig:pn} (top).
As we can see,
$n_i$ increases to a peak value about $4 \times 10^{17}$
m$^{-3}$
at about 7 $\mu$s,
then drops and oscillates around
$2.3 \times 10^{17}$ m$^{-3}$,
indicating the convergence of the simulations.
Note that, this low frequency oscillation
of $n_i$
corresponds to the ionization oscillation
or the so-called breathing mode
\cite{Pan_2023},
which only occurs when the neutral gas density
is solved self-consistently along
with the ionization.
Looking at the oscillating period,
the Strong-B case has a period of
about 16 $\mu$s,
the Weak-B case has a period of about 17 $\mu$s,
and the analytic-B case has a period of about
14 $\mu$s.
Note that these periods are already
shrunken by increasing the neutral gas flow
speed to $u_z=912$ m/s,
as mentioned in Sec.\ref{sec:neutral}.
A smaller $u_z$ leads to
even longer breathing mode period,
which makes the simulation more computationally
expensive.

In Fig.\ref{fig:pn} (bottom),
we plot the total azimuthal electric field
energy of the whole simulation domain,
which can reflect the extent of the
azimuthal instability (mostly the electron drift
instability),
because without azimuthal instability,
only noisy $E_y$ should exist
without patterned oscillating fields
that can store more energy.
From the $E_y$ energy point of view,
we can see that Analytic-B case
excites the strongest instability,
and the Weak-B case
leads to the weakest instability
among these three cases.
If we compare the evolution of
the $E_y$ energy and $n_i$,
we can see that overall
higher $E_y$ is coupled with higher $n_i$,
but a phase difference exists,
$E_y$ is slower than $n_i$ by about 2 $\mu$s,
which would be explained by the
influence of ionization
to the instability.
We will discuss this in later text
using snapshots of $E_y$.

Next, we present
$z$-$x$ and $z$-$y$ snapshots of
$E_y$, $n_i$,
and the potential $\phi$
of the three cases at time step
1500000,
i.e., 22.5 $\mu$s,
in Fig.\ref{fig:Ey},
Fig.\ref{fig:ni},
and Fig.\ref{fig:phi},
respectively.
Following the previous discussion
on the instability intensity,
we can see from thes plots of $E_y$ first
in Fig.\ref{fig:Ey} that
indeed Analytic-B exhibits the strongest instability.
The major difference between
Analytic-B and Weak-B is that
Analytic-B only has radial component of
the B field.
Thus it seems like a configuration
with only or mostly the radial component
of the magnetic field
leads to stronger electron drift instability,
or a radial gradient of the B field may play a role.
Comparing Weak-B with Strong-B,
we can see that a stronger B field
leads to an enhanced instability.
Moreover, both Weak-B and Strong-B
have only instability wave patterns in
the plume region,
while Analytic-B has wave patterns throughout
the channel and the plume.
In addition,
we can see that Strong-B has the instability
patterns only in the upper plume region,
where the B field is relatively weaker
compared to the other side,
indicating an asymmetry
due to the asymmetric magnetic field
configuration and strength radially
as shown in Fig.\ref{fig:B}.

Looking at the $n_i$ plots in Fig.\ref{fig:ni},
the wave patterns can be seen in the
Strong-B case,
but can barely be seen in the Weak-B case,
and the patterns are the most obvious and clear
in the Analytic-B case
existing in both the channel and plume.
From the potential plots in Fig.\ref{fig:phi},
other than the same tendency of the
instability intensity,
we can see that the Weak-B case has
slightly higher potential (nearly 300 V)
in the channel,
indicating slightly stronger charge separation.

\begin{figure*}[!ht]
\centering
\includegraphics[width=0.32\textwidth]
{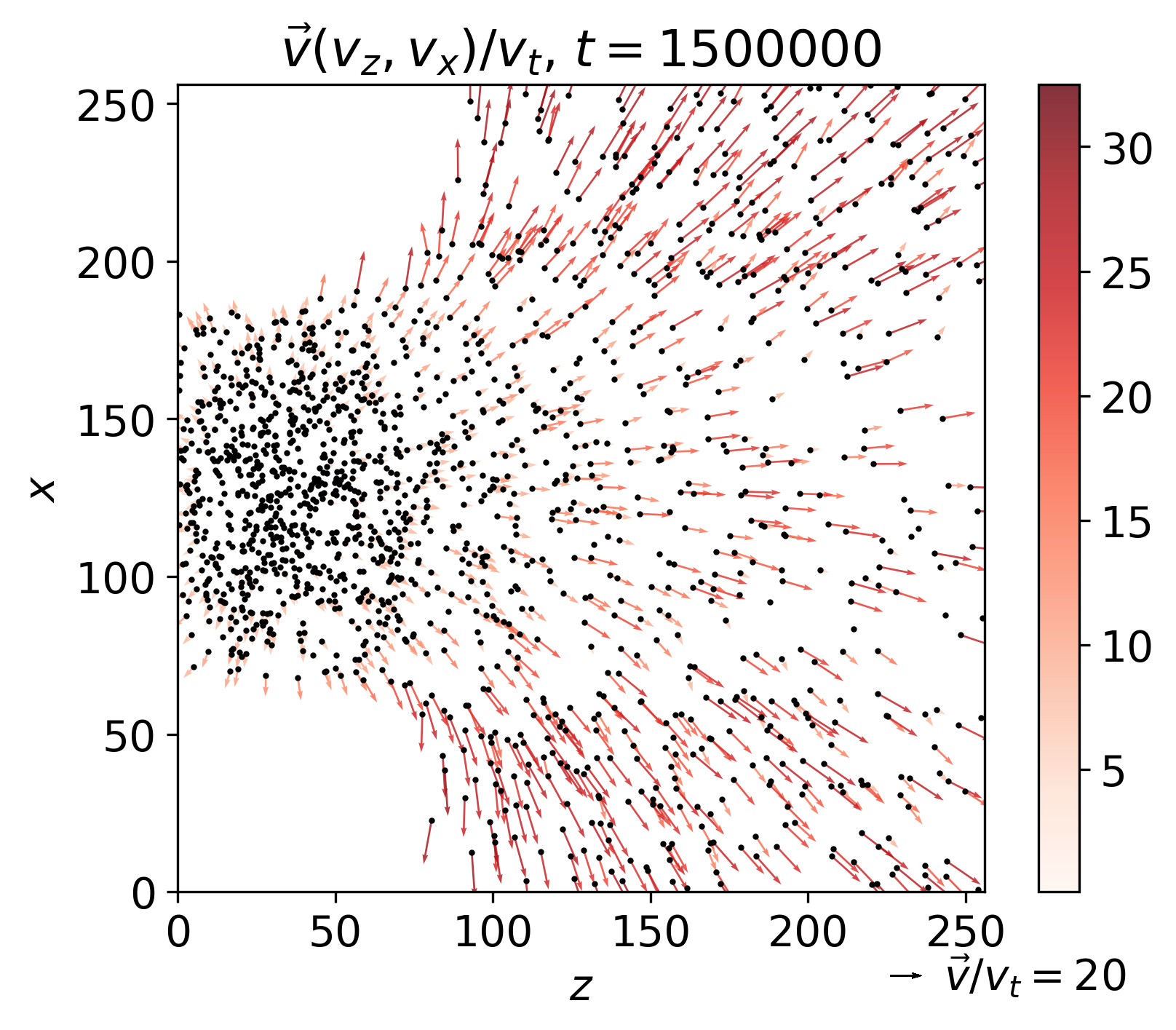}
\includegraphics[width=0.32\textwidth]
{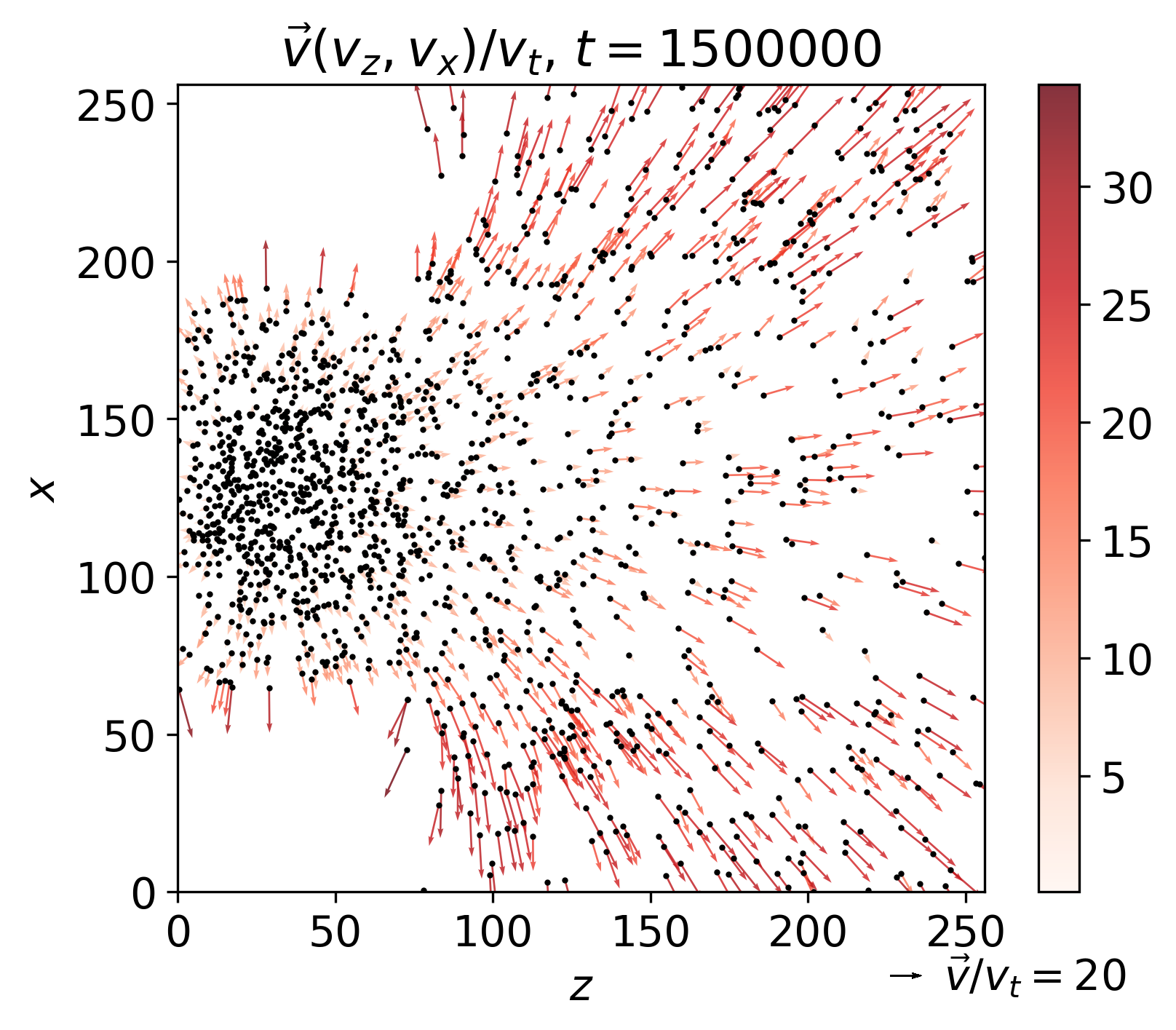}
\includegraphics[width=0.32\textwidth]
{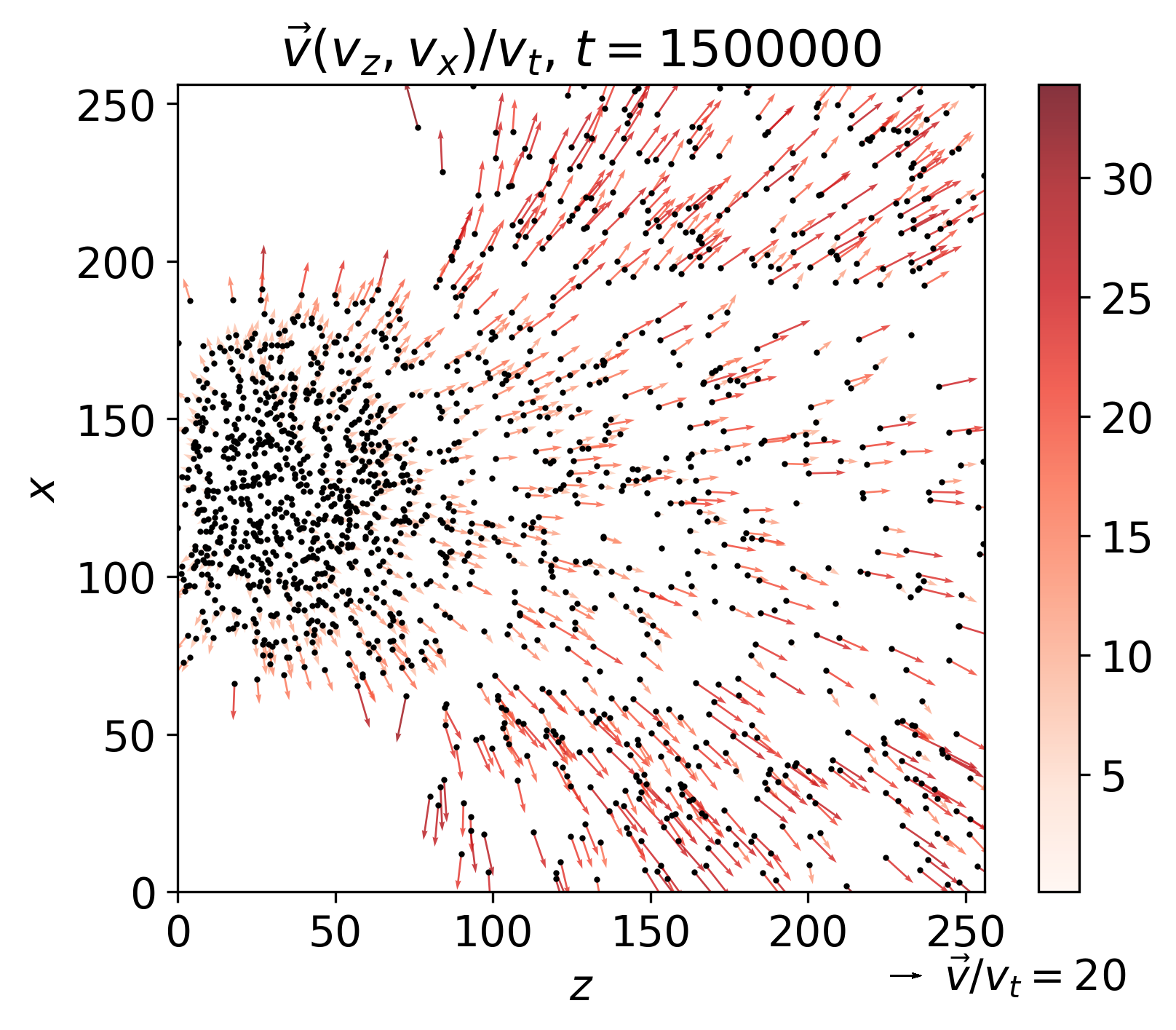}
\caption{
Ion velocity vector plots
at 22.5 $\mu$s
of cases Strong-B (left),
Weak-B (middle),
and Analytic-B (right),
where $v_t$ denotes the ion thermal
velocity corresponding to 0.5 eV.
}
\label{fig:vi}
\end{figure*}

\begin{figure*}[!ht]
\centering
\includegraphics[width=0.24\textwidth]{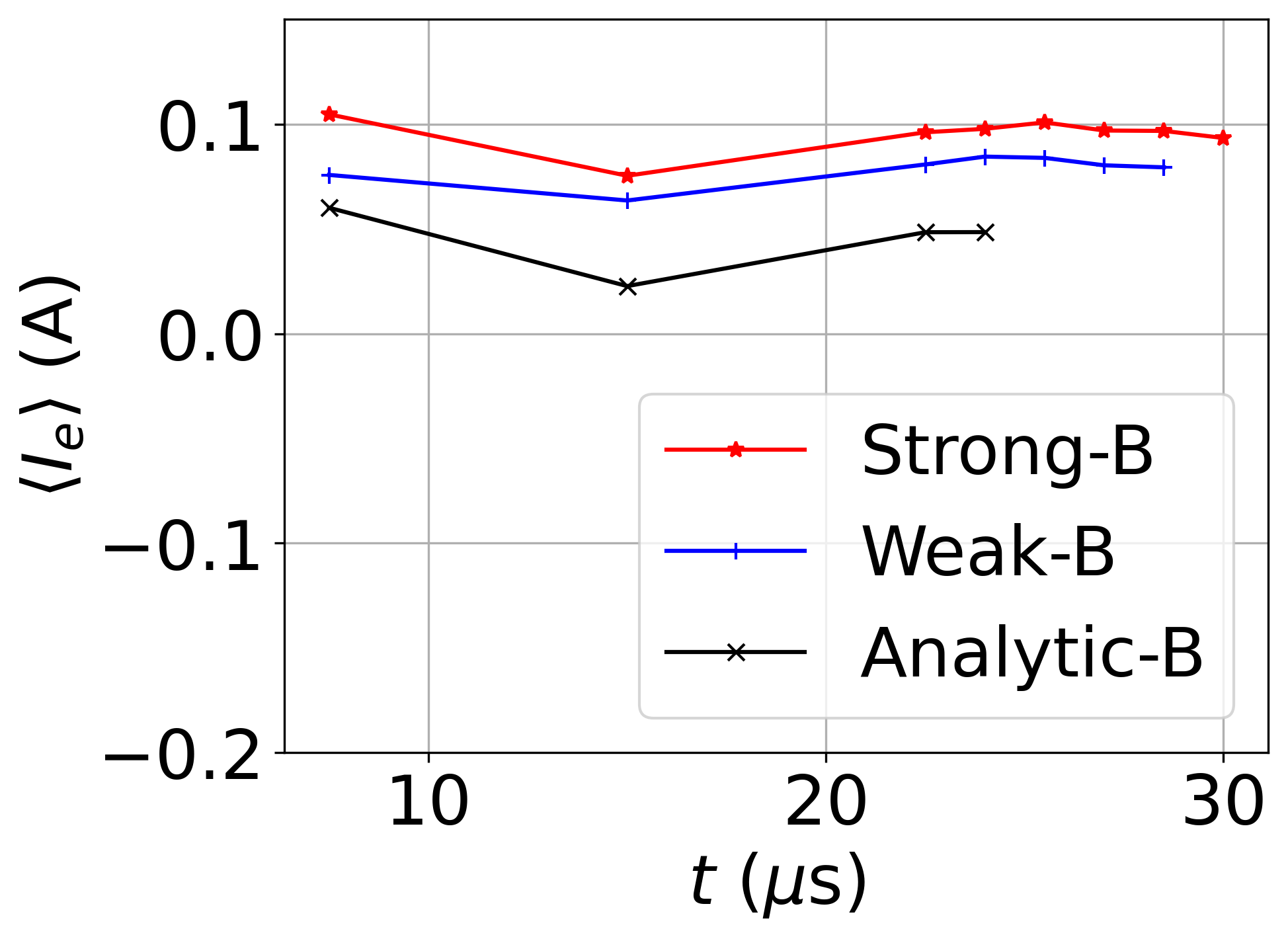}
\includegraphics[width=0.24\textwidth]{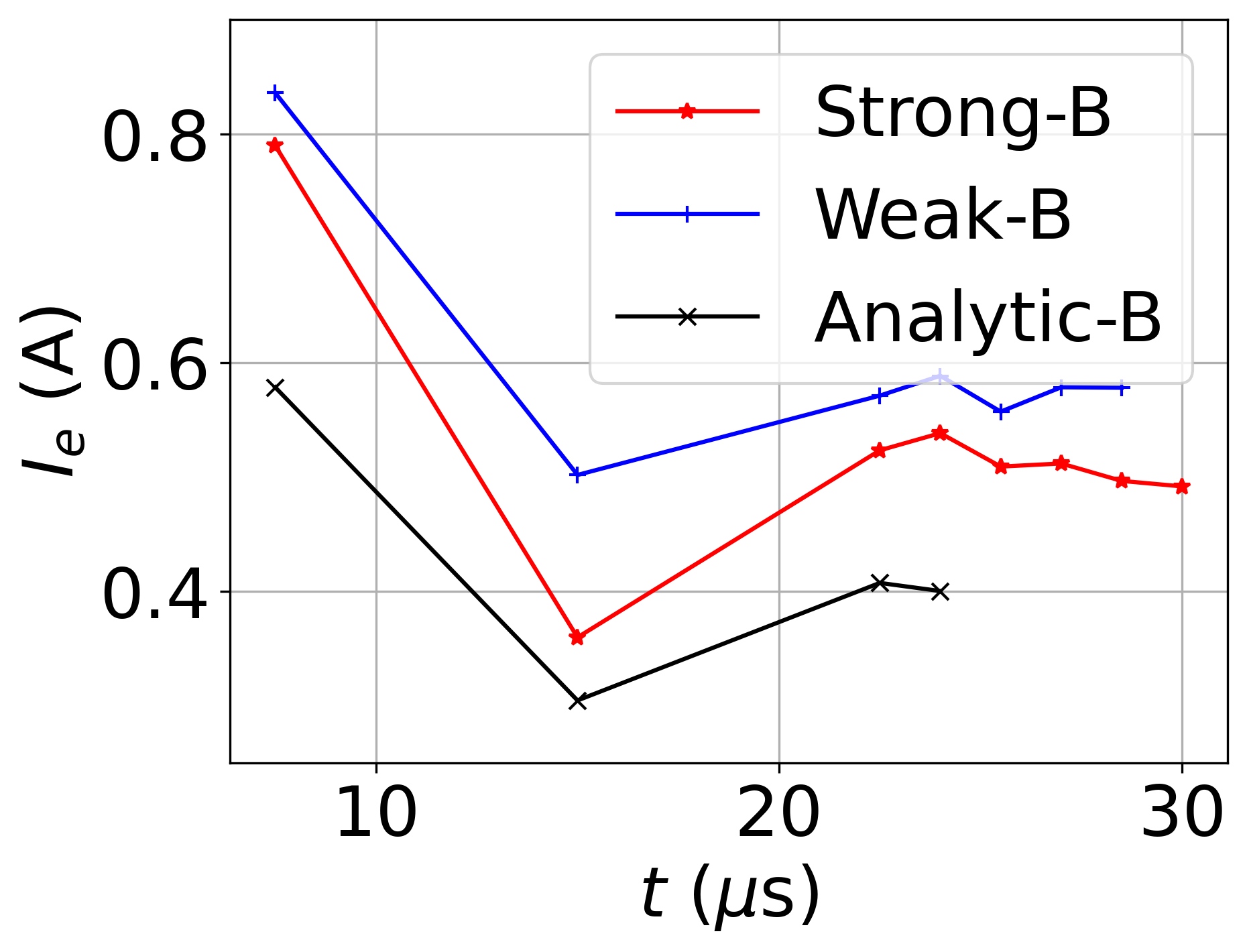}
\includegraphics[width=0.24\textwidth]{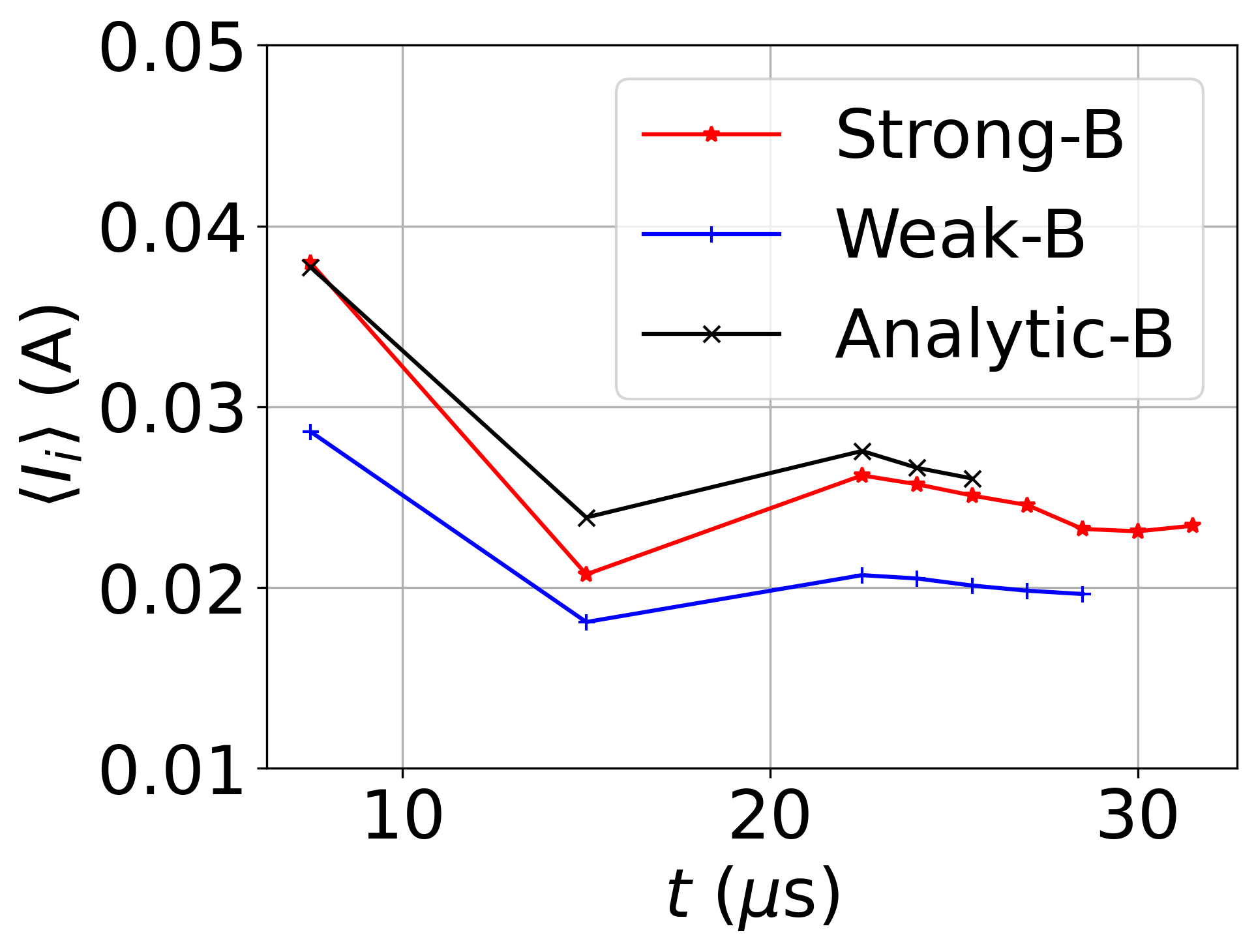}
\includegraphics[width=0.24\textwidth]{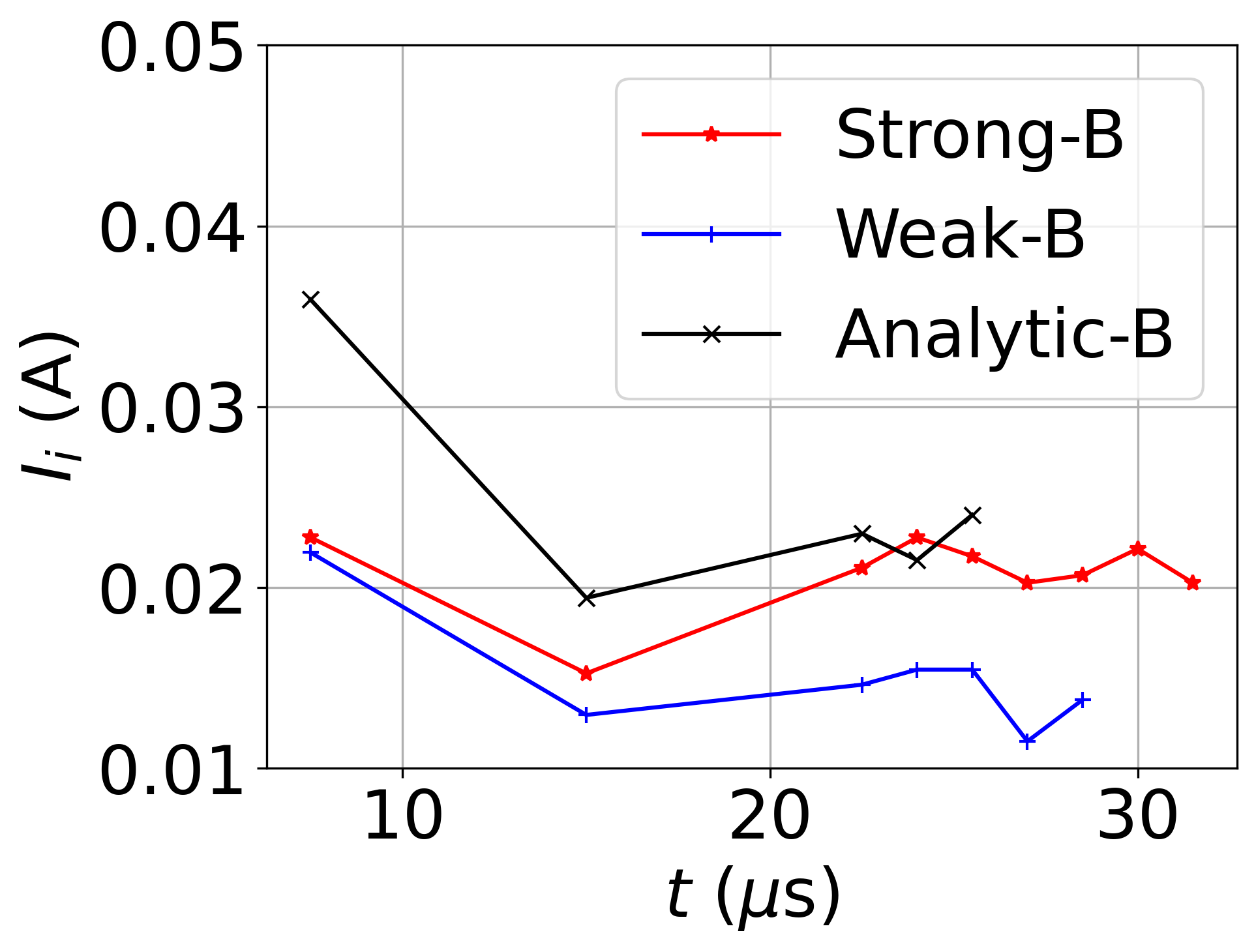}
\caption{
From left to right panels:
the time evolution of
averaged electron current over
the whole domain
$\langle I_e \rangle$;
electron current at anode
$I_e$;
averaged ion current over
the whole domain
$\langle I_i \rangle$;
ion current at the $z=256$ plane
$I_i$.
}
\label{fig:plot_v}
\end{figure*}

Following the previous results,
we now investigate how ionization affects
the instability.
In Fig.\ref{fig:Ey_Strong-B},
Fig.\ref{fig:Ey_Weak-B}, and
Fig.\ref{fig:Ey_Analytic-B},
we present $E_y$ at different times
of the three cases.
For the Strong-B case as shown in
Fig.\ref{fig:Ey_Strong-B},
the first column is at time 6 $\mu$s,
corresponding to the initial strong
ionization stage
as indicated by
the positive large slope of $n_i$
in Fig.\ref{fig:pn} (top).
During this stage,
because many electrons are losing energy
to ionize neutral atoms,
electrons would be hard to
participate in the instability establishment.
In the meanwhile, many newly ionized
ion-electron pairs would dim the
formation of the instability too.
By contrast, the second column
in Fig.\ref{fig:Ey_Strong-B} is at time 11.55 $\mu$s,
corresponding to the (nearly) non-ionization stage,
as indicated by the negative large slope of $n_i$
in Fig.\ref{fig:pn} (top).
During this stage, the neutral gas density
has become low and is restoring gradually,
thus electrons do not lose energy
due to many ionization collisions,
and rarely newly ionized ion-electron pairs
can dim the instability formation.
Therefore,
Comparing the left and the middle
columns in Fig.\ref{fig:Ey_Strong-B},
we can see that the instability
is indeed much stronger in
the non-ionization stage (the middle one),
and that is found to be the only stage,
within which the wave patterns can
develop to cover the whole plume region,
while only the upper plume region exhibits
wave patterns during other time stages.
The third column in Fig.\ref{fig:Ey_Strong-B}
is at time 19.95 $\mu$s,
corresponding to another ionization stage,
thus the instability becomes weaker again.
Similar effect can be seen in
the Weak-B case
(Fig.\ref{fig:Ey_Weak-B}) and the Analytic-B case
(Fig.\ref{fig:Ey_Analytic-B}) as well.

For completeness,
the neutral gas density $n_a$ snapshots are also presented in
Fig.\ref{fig:na_Strong-B}
of the case Strong-B,
at the corresponding times of
6,
11.55,
and 19.95 $\mu$s,
as those shown in Fig.\ref{fig:Ey_Strong-B}.
We can see the ionization stage at 6 $\mu$s,
the non-ionization stage at 11.55 $\mu$s,
and another ionization stage at 19.95 $\mu$s.
We have to mention that a wrong boundary condition was
implemented in the $n_a$ solver,
such that there are some tiny reflecting waves
shown near $z=256$ in the right panel at 19.95 $\mu$s
if one looks closely,
but they do not affect the simulation,
because both the neutral gas density and the electron density
are too low there to trigger any ionization.
By the way, it has been tested
that using a Neumann boundary condition can easily fix this bug,
but since the computation is too massive, we
do not rerun the simulations any more.

In addition, because the magnetic field configuration
is varied,
one would be interested in
seeing the difference of the ion plume angle.
We present three snapshots of
the ion velocity vector plots
of the three cases in Fig.\ref{fig:vi}.
However, it is found that although we applied different
magnetic field configuration and strength,
the plume angle is not affected much.

At last,
the electron and ion currents
are diagnosed to partially reflect the overall
performance of the simulated Hall thruster
under the three cases.
As shown in Fig.\ref{fig:plot_v},
$I_e$ denotes the electron current
at the anode;
$\langle I_e \rangle \equiv
e \langle v_{ez} \rangle
\langle n_e \rangle A_{anode}$
denotes the averaged electron current over the
whole domain,
where $\langle v_{ez} \rangle$ is the
averaged electron axial velocity,
$\langle n_e \rangle$
is the averaged electron number density,
and $A_{anode}$ is the anode surface area.
Similarly,
$I_i$ denotes the ion current
at the $z=256$ plane;
$\langle I_i \rangle \equiv
e \langle v_{iz} \rangle
\langle n_i \rangle A_{plume}$
denotes the averaged electron current over the
whole domain,
where $\langle v_{iz} \rangle$ is the
averaged ion axial velocity,
$\langle n_i \rangle$
is the averaged ion number density,
and $A_{plume}$ is the $z=256$ surface area.

Looking at the Analytic-B case first,
we can see that
it has the smallest
electron current
for both $I_e$ and $\langle I_e \rangle$,
however, we have seen in previous discussion
that the Analytic-B case exhibits the strongest
electron drift instability.
The common sense is that
stronger instability would
help electrons to escape the
confinement of the magnetic field,
resulting in higher electron current,
but the simulated results are against this
common sense.
Possible explanations include that
the simulated instability can somehow
reduce the electron conduction,
or more likely,
other mechanisms also play a significant role.
For example,
the different magnetic field configuration
and strength near the anode
between the two cases may contribute.

Then, we notice that
the order of $I_e$ and $\langle I_e \rangle$
is changed
for the Strong-B case
and the Weak-B case,
indicating that
although Strong-B has a smaller anode
electron current $I_e$
than that of Weak-B,
mostly due to stronger magnetic confinement,
there are other places in the domain
that electrons are easier to conduct
in Strong-B than in Weak-B,
such that Strong-B has a higher
averaged $\langle I_e \rangle$
than that of Weak-B.
For example,
in the plume region near the thruster exit,
stronger instability presented in
Strong-B case than that in Weak-B case
may cause a stronger electron conduction.

Then the trend and order of
$I_i$ and $\langle I_i \rangle$
are the same for the three cases
as shown in Fig.\ref{fig:plot_v},
where
the case of Analytic-B
has the highest ion current,
while the Weak-B case has the
smallest ion current.
In addition,
note that the values of ion
currents are one order of magnitude
smaller than that of the
anode electron current,
because as we can see in
Fig.\ref{fig:vi}
that the plume angle is large,
such that a great amount of ion
acceleration energy is wasted
in the radial direction,
which do not contribute to
the axial ion current diagnosis.

\section{Conclusion and Discussion\label{sec:4}}

Nearly all PIC simulation studies
on the electron drift instability
(involving the azimuthal direction)
in Hall thrusters
in the literature
apply a simplified analytic magnetic field
with only the radial component
\cite{Charoy2019,Villafana2021,Villafana2023,Xie_2024}.
In this paper,
for the first time we
implemented much more realistic B field
configuration
in 3D PIC simulations
by importing field data obtained from
the FEMM software.
Three simulation cases are studied
including one with a stronger B field,
a weaker B field,
and the commonly used analytic B field.
Major conclusions are summarized as follows.

First, it is found that
applying the analytic B field
leads to an obviously stronger
electron drift instability
in both the channel and the plume,
compared to the other two cases
applying the FEMM B field,
in which the instability only occurs
in the plume region.
Thus, a magnetic field configuration with only
radial component and without
radial gradient tends to enhance the
instability.

Second,
comparing the two FEMM B field cases,
it is found that
a stronger instability is presented
in the stronger B field case,
and due to the asymmetric radial B field strength,
the instability is more likely
to be established in the plume region
with weaker B field.

Third,
because the neutral gas density is
solved self-consistently,
the breathing mode is captured
in the simulations.
It is found that
for all three cases
the instability can be best established
in the non-ionization time stage,
during which the ionization collisions
are rare due to less neutral gas density,
such that electrons do not lose energy
for ionization.
By contrast,
in the ionization time stage,
electrons need to spend energy
to ionize atoms, therefore
strong instability is hard to be established,
or is dimmed due to many newly generated
ion-electron pairs.

Fourth,
from the point view of
the axial ion and electron currents,
surprisingly it is found that
the analytic B field case with
the strongest instability
leads to the highest ion current
and the smallest electron current.
For the FEMM B field cases,
it is found that the strong B case
exhibits a higher ion current
(both at the plume and on average),
and a higher electron current at the anode,
but lower averaged electron current,
than the weak B case.

In this paper,
because relatively less number of
macro-particles are applied
to make the massive 3D PIC simulations more feasible,
the noise of the results is relatively large,
which for example impedes further
spectral analysis that requires higher resolution.
In the future, a large number of macro-particles
would be used to carry out simulations,
and the magnetic field effects would be
investigated in more detail.

\section*{Supplementary Material}

Some animations of
$E_y$,
$n_i$,
and $n_a$
are made available through
\url{https://www.bilibili.com/video/BV18wdUYwEpL}.

\section*{Acknowledgment}

The authors acknowledge the support from National Natural Science Foundation of China (Grant No. 5247120164).

\section*{Data Availability}

The data that support the findings of this study is available from
the corresponding author upon reasonable request.

\renewcommand{\bibsection}{\section*{References}}
\bibliographystyle{unsrturl}
\bibliography{reference}

\begin{thebibliography}{1}

\bibitem{Xie_2024}
Lihuan Xie, Xin Luo, Zhijun Zhou, and Yinjian Zhao.
\newblock Effect of plasma initialization on 3d pic simulation of hall thruster
  azimuthal instability.
\newblock {\em Physica Scripta}, 99(9):095602, aug 2024.
\newblock URL: \url{https://dx.doi.org/10.1088/1402-4896/ad69e5}.

\bibitem{Lafleur1}
T.~Lafleur, S.~D. Baalrud, and P.~Chabert.
\newblock Theory for the anomalous electron transport in hall effect thrusters.
  i. insights from particle-in-cell simulations.
\newblock {\em Physics of Plasmas}, 23(5):053502, 05 2016.
\newblock \href {https://doi.org/10.1063/1.4948495}
  {\path{doi:10.1063/1.4948495}}.

\bibitem{Reza_2023}
M~Reza, F~Faraji, A~Knoll, A~Piragino, T~Andreussi, and T~Misuri.
\newblock Reduced-order particle-in-cell simulations of a high-power
  magnetically shielded hall thruster.
\newblock {\em Plasma Sources Science and Technology}, 32(6):065016, jun 2023.
\newblock \href {https://doi.org/10.1088/1361-6595/acdea3}
  {\path{doi:10.1088/1361-6595/acdea3}}.

\bibitem{R2}
M.~Reza, F.~Faraji, and A.~Knoll.
\newblock Parametric investigation...
\newblock {\em Journal of Applied Physics}, 133(12):123301, mar 2023.
\newblock \href {https://doi.org/10.1063/5.0138223}
  {\path{doi:10.1063/5.0138223}}.

\bibitem{VAHEDI1995179}
V.~Vahedi and M.~Surendra.
\newblock A monte carlo collision model for the particle-in-cell method:
  applications to argon and oxygen discharges.
\newblock {\em Computer Physics Communications}, 87(1):179--198, 1995.
\newblock Particle Simulation Methods.
\newblock \href {https://doi.org/10.1016/0010-4655(94)00171-W}
  {\path{doi:10.1016/0010-4655(94)00171-W}}.

\bibitem{Charoy2019}
T~Charoy, J~P Boeuf, A~Bourdon, J~A Carlsson, P~Chabert, B~Cuenot, D~Eremin,
  L~Garrigues, K~Hara, I~D Kaganovich, A~T Powis, A~Smolyakov, D~Sydorenko,
  A~Tavant, O~Vermorel, and W~Villafana.
\newblock 2d axial-azimuthal particle-in-cell benchmark for low-temperature
  partially magnetized plasmas.
\newblock {\em Plasma Sources Science and Technology}, 28(10):105010, Oct.
  2019.
\newblock \href {https://doi.org/10.1088/1361-6595/ab46c5}
  {\path{doi:10.1088/1361-6595/ab46c5}}.

\bibitem{Villafana2021}
W~Villafana, F~Petronio, A~C Denig, M~J Jimenez, D~Eremin, L~Garrigues,
  F~Taccogna, A~Alvarez-Laguna, J~P Boeuf, A~Bourdon, P~Chabert, T~Charoy,
  B~Cuenot, K~Hara, F~Pechereau, A~Smolyakov, D~Sydorenko, A~Tavant, and
  O~Vermorel.
\newblock {2D radial-azimuthal particle-in-cell benchmark for E$\times$B
  discharges}.
\newblock {\em Plasma Sources Science and Technology}, 30(7):075002, Jul. 2021.
\newblock \href {https://doi.org/10.1088/1361-6595/ac0a4a}
  {\path{doi:10.1088/1361-6595/ac0a4a}}.

\bibitem{Villafana2023}
W.~Villafana, B.~Cuenot, and O.~Vermorel.
\newblock {3D particle-in-cell study of the electron drift instability in a
  Hall Thruster using unstructured grids}.
\newblock {\em Physics of Plasmas}, 30(3):033503, 03 2023.
\newblock \href {https://doi.org/10.1063/5.0133963}
  {\path{doi:10.1063/5.0133963}}.

\bibitem{Pan_2023}
Ruojian Pan, Junxue Ren, Renfan Mao, and Haibin Tang.
\newblock Practical analysis of different neutral algorithms for particle
  simulation of hall thruster.
\newblock {\em Plasma Sources Science and Technology}, 32(3):034005, mar 2023.
\newblock \href {https://doi.org/10.1088/1361-6595/acc134}
  {\path{doi:10.1088/1361-6595/acc134}}.

\end{thebibliography}

\end{document}